\documentclass[aps,pre,twocolumn,amsmath,amssymb,nofootinbib,superscriptaddress,floatfix]{revtex4-1}

\usepackage{amsmath,amsfonts,amssymb}
\usepackage{amssymb}
\usepackage{amsthm}
\usepackage[dvipdf]{color}
\usepackage{graphicx}
\usepackage{dcolumn}
\usepackage{bm}

\begin{document}
\title{Non-Hermitian Topological Metamaterials with Odd Elasticity}

\author{Di Zhou}
\email[Corresponding author: ]{dzhou90@gatech.edu}
 \affiliation{
School of Physics, Georgia Institute of Technology, Atlanta, GA 30332, USA
 }

 \affiliation{
 Department of Physics,
  University of Michigan, Ann Arbor, 
 MI 48109-1040, USA
 }

\author{Junyi Zhang}
 \affiliation{
Department of Physics, Princeton University, Princeton, 08544, New Jersey, USA
 }

\begin{abstract}
We establish non-Hermitian topological mechanics in one dimensional (1D) and two dimensional (2D) lattices consisting of mass points connected by meta-beams that lead to odd elasticity. Extended from the ``non-Hermitian skin effect" in 1D systems, we demonstrate this effect in 2D lattices in which bulk elastic waves exponentially localize in both lattice directions. We clarify a proper definition of Berry phase in non-Hermitian systems, with which we characterize the lattice topology and show the emergence of topological modes on lattice boundaries. The eigenfrequencies of topological modes are complex due to the breaking of $\mathcal{PT}$-symmetry and the excitations could exponentially grow in time in the damped regime. Besides the bulk modes, additional localized modes arise in the bulk band and they are easily affected by perturbations. These distinguishing features may manifest themselves in various active materials and biological systems. 
\end{abstract}

\maketitle

\section{Introduction}
Recent years have witnessed advances in applying the notion of ``topological protection" to mechanical systems which have led to the blossom of the new field ``topological mechanics"\cite{huber2016topological, susstrunk2015observation, kane2014topological, yang2015topological, mao2018maxwell, barlas2018topological}. Enormous progress not only offer us a plethora of applications of metamaterials\cite{nash2015topological, PhysRevB.101.104106, rocklin2017transformable, meeussen2016geared, paulose2015topological}, but also deepen the understandings of topological protection in aperiodic systems\cite{zhou2018topological, zhou2019topological, bandres2016topological, tran2015topological, varjas2019topological, mitchell2018amorphous}. 


The study of topological band theory has been extended to open systems governed by non-Hermitian Hamiltonians\cite{PhysRevLett.80.5243, heiss2012physics, lee2016anomalous, xu2017weyl, PhysRevB.84.205128, PhysRevLett.118.040401, PhysRevB.97.045106, PhysRevLett.121.026808, yao2018edge, yao2018non, song2019non, shen2018topological, yuce2018edge, PhysRevB.99.201103, zhou2018observation, xiong2018does, okuma2019topological, zhang2019correspondence, lee2019topological, ghatak2019new, tserkovnyak2019exceptional, kawabata2019symmetry, gong2018topological, PhysRevResearch.1.023013, malzard2015topologically, zhao2019topological, yoshida2019non, longhi2019topological, PhysRevB.100.144106, bender2007making, PhysRevA.99.062107, chang2019entanglement, PhysRevB.100.054109, PhysRevB.98.035141, PhysRevB.99.121101, borgnia2019nonhermitian}, which are realized in classical systems subjected to intrinsic gain and/or loss such as optical\cite{makris2008beam, liu2019second, lee2019hybrid, PhysRevB.99.081302, lopez2019multiple, chong2011p, chong2011p, regensburger2012parity, hodaei2014parity, peng2014parity, feng2014single, peng2014loss, jing2015optomechanically, liu2016metrology, kawabata2017information, jing2017high, lu2017exceptional, ashida2017parity, el2018non, zhang2018phonon, lu2014topological, ozawa2019topological, malzard2018bulk, longhi2018parity, zhen2015spawning, el2018non, zhang2018thermal, hofmann2019chiral, PhysRevB.99.201411, xu2016topological, tzortzakakis2019non, liertzer2012pump, bergholtz2019exceptional}, electric\cite{hofmann2019reciprocal, helbig2019observation, hofmann2019reciprocal, ezawa2019electric, xiao2019observation, ezawa2019non, PhysRevLett.124.046401}, and acoustic\cite{PhysRevLett.122.195501, jing2014pt} structures. In contrast to the ordinary topological band theory, non-Hermitian systems exhibit unique features 
such as exceptional points, 
band structure which is sensitive to boundary conditions, and exponentially localized bulk modes (``non-Hermitian skin effect"). Among them the most fascinating subject is the interplay between non-Bloch bulk waves and the characterization of lattice topology. 
It is thus intriguing to ask: can exotic mechanical properties arise in energy non-conserving systems whose Hamiltonian is non-Hermitian? 

In this paper, we study non-Hermitian topological lattices composed of mass points and meta-beams that lead to odd elasticity. 
Odd elasticity originates from active matters with microscopic interactions which do not conserve energy\cite{scheibner2020odd}, and is ubiquitous in a broad range of natural (such as fiber networks with driving\cite{alberts2002drosophila, brangwynne2008cytoplasmic, joanny2009active} and active microtubule networks\cite{foster2015active}) and manmade materials (such as coupled gyroscopes\cite{nash2015topological, mitchell2018amorphous, wang2015topological, mitchell2018realization} and active fluids\cite{engheta2005circuit, souslov2019topological}). Recent progress on the mechanics of active biological surfaces\cite{PhysRevE.96.032404}, such as cell cortex and epithelial tissues, has revealed up-down or chiral symmetry breaking that leads to odd elasticity. Odd elasticity offers unconventional elastostatics and dynamics\cite{scheibner2020odd} which are absent in passive solids, such as horizontal deflection and wave propagation in the overdamped regime. It elucidates non-reciprocal linear responses of active materials.

Up to date, major efforts of non-Hermitian mechanics have been limited to 1D parity-time ($\mathcal{PT}$) symmetric systems\cite{ghatak2019observation, schomerus2019nonreciprocal} whose eigenvalues are real and the eigenmode amplitudes do not grow in time. Our research lifts this $\mathcal{PT}$-symmetry by allowing the eigenvalues to be complex, 
meaning that in the dissipationless limit the lattice is unstable against infinitesimal stimulations. We show that a damping which counteracts the intrinsic energy production naturally stabilizes the lattice.
We first realize the basic two-band models in 1D and 2D rotor lattices, and then in the third honeycomb lattice model we study a four-band non-Hermitian Hamiltonian. In the honeycomb lattice, the eigenvalues of the topological modes are complex due to the breaking of $\mathcal{PT}$-symmetry, and the excitation exponentially increases if the damping is not strong enough to counteract internal energy production. Besides the bulk modes, on lattice boundaries we observe additional localized modes whose eigenvalues are not separated from the bulk band. These non-topological localized modes may be essential for the in-band boundary softness in energy non-conserving systems such as biological structures\cite{RevModPhys.85.1143, toner2005hydrodynamics, geyer2018sounds, petrie2012leading, petrie2016multiple}. We leave non-Hermitian mechanics of amorphous systems in future research.

{This article is organized as follows. In section.\uppercase\expandafter{\romannumeral2} we introduce the general formulation of the lattices composed of mass particles and meta-beams that lead to odd elasticity. In section.\uppercase\expandafter{\romannumeral3} the notion of Berry phase is extended to non-Hermitian systems, and we introduce three different symmetries which quantize the Berry phase. In section.\uppercase\expandafter{\romannumeral4}, the basic non-Hermitian two-band models are realized in the 1D and 2D rotor lattices. The ``non-Hermitian skin effect" is featured by the bulk modes, and the topological edge modes are characterized by the quantized Berry phase. In section.\uppercase\expandafter{\romannumeral5}, we study non-Hermitian topological mechanics in the 2D honeycomb lattice. The complex eigenvalues of the topological modes and the subsequent unconventional mechanical response are summarized at the end of section.\uppercase\expandafter{\romannumeral5}. The main results, extra discussions, and future perspectives are summarized in section.\uppercase\expandafter{\romannumeral6}.}

\section{General formulation of the models}
We model the non-conservative odd-elastic interaction between particles as a pairwise force $\vec F(\vec u)$, where $\vec u$ is the relative displacement away from equilibrium between an interacting pair of particles. According to Ref.\cite{scheibner2020odd}, to linear order, we use the force
\begin{eqnarray}\label{OE}
\vec F(\vec u) = -(k\hat{n}+k^o \hat{\phi})\vec u\cdot\hat{n}, 
\end{eqnarray}
where $\hat{n}$ is the unit vector along the connection orientation, $\hat{\phi}$ is the unit vector rotated from $\hat{n}$ by $90^\circ$ counterclockwise [Fig.\ref{Fig1}(a)], 
$k$ is the spring constant, and $k^o$ represents the strength of the energy non-conserving force (dubbed odd elastic constant). The unit cells of all three lattices are subjected to fixed boundary conditions, and are composed of two mass particles labeled $A$ and $B$ with mass $m$ which are connected by odd-elastic meta-beams. Each cell is labeled by $(n_1, \ldots, n_d)$, and the site displacements are denoted as $\vec u_{A,(n_1,\ldots,n_d)}$ and $\vec u_{B,(n_1,\ldots,n_d)}$, where $d$ is the spatial dimension, $n_i$ ($1\le n_i\le N_i$)  is the cell labeling, and $N_i$ is the lattice length scale. We consider the Newtonian equation of motion and take the ansatz  
\begin{eqnarray}\label{ansatz}
(\vec u_{A,(n_1,\ldots,n_d)}, \vec u_{B,(n_1,\ldots,n_d)}) = \beta_1^{n_1}\ldots \beta_d^{n_d}(\vec u_{A}, \vec u_{B})\qquad
\end{eqnarray}
in every model, where $\beta_i$ is the decay rate of the non-Bloch bulk modes\cite{yao2018edge}. In all three models, $|\beta_i|$ stays a constant while ${\rm Arg\,}\beta_i$ varies from $0$ to $2\pi$. We denote $\vec\beta = (\beta_1, \beta_2, \ldots, \beta_d)$ for simplicity. The equation of motion in every lattice is simplified as an eigenvalue problem
\begin{eqnarray}\label{Newton}
D(\vec\beta) u = \lambda u, 
\end{eqnarray}
where $D(\vec\beta)$ is the dynamical matrix,  $u$ is the unit cell displacement field, and $\lambda$ is the eigenvalue.

\section{Generalized Berry phase in non-Hermitian systems}
We first generalize the Berry phase for the $N$-band non-Hermitian dynamical matrix $D(\vec\beta)$ to characterize the lattice topology. We consider the simple case that all bands are \emph{separable bands}\cite{shen2018topological} (i.e., eigenvalues $\lambda_{n}(\vec\beta)\neq \lambda_{n'}(\vec \beta')$ for all $\vec\beta$, $\vec \beta'$ as long as $1\le n\neq n' \le N$). Thus the system is free of exceptional point. Based on the biorthogonality and completeness[Appendix.A] of the eigenbasis, we define the generalized Berry phase of band $n$ as follows, 
\begin{eqnarray}\label{Berry}
\gamma_i^{(n)} = \oint_{\mathcal{C}_i}  \mathcal{A}_i^{(n)} \,  d{\, \rm Arg\,}\beta_i
\end{eqnarray}
where $\mathcal{C}_i$ is the closed loop trajectory connecting ${\rm Arg\,}\beta_i=0$ and $2\pi$, 
\begin{eqnarray}\label{BerryConnection}
\mathcal{A}_i^{(n)} = i \langle n^L | \partial_{{\rm Arg}\,\beta_i}|n^R\rangle
\end{eqnarray}
is the generalized Berry connection, and $|n^R(\vec\beta)\rangle$ and $|n^L(\vec\beta)\rangle$ are the right and left eigenvectors, respectively. We note that Eq.(\ref{Berry}) is the proper definition for one of the Berry phases of Ref.\cite{shen2018topological}. The generalized Berry phase can be quantized by various symmetries in non-Hermitian systems. In what follows, we discuss three different symmetries that quantize the generalized Berry phase. 

In the first case, we consider a 1D lattice with the dynamical matrix $D(\beta_1)$ subjected to the symmetry property $M_x D(\beta_1) M_x^{-1} = D(\beta_1^*)$, where $M_x$ is the symmetry operator which satisfies $M_x = M_x^{-1}$, and $\beta_1^*$ is the complex conjugate of $\beta_1$. To prove that the generalized Berry phase is quantized by this symmetry, we notice
\begin{eqnarray}\label{reflection}
 D(\beta_1^*)M_x |n^R(\beta_1)\rangle = \lambda_n(\beta_1) M_x |n^R(\beta_1)\rangle,\qquad
\end{eqnarray}
which means 
\begin{eqnarray}
M_x |n^R(\beta_1)\rangle = e^{i\mathcal{R}^{(n)}_x(\beta_1)}|n^R(\beta_1^*)\rangle,
\end{eqnarray}
where $\mathcal{R}^{(n)}_x(\beta_1)$ is the symmetry phase connecting $M_x|n^R(\beta_1)\rangle$ and $|n^R(\beta_1^*)\rangle$. Similarly, the left eigenvector yields
\begin{eqnarray}
\langle n^L(\beta_1)| M_x  = e^{-i\mathcal{L}^{(n)}_x(\beta_1)}\langle n^L(\beta_1^*)|,
\end{eqnarray}
where $\mathcal{L}^{(n)}_x(\beta_1)$ is the symmetry phase factor connecting $\langle n^L(\beta_1)| M_x$ and $\langle n^L(\beta_1^*)|$. Due to the normalized biorthogonality $\langle n^L|n^R\rangle=1$, $\mathcal{L}^{(n)}_x(\beta_1) = \mathcal{R}^{(n)}_x(\beta_1)\mod 2\pi$. At high symmetry points when ${\rm Arg\,}\beta_{1, {\rm hs}}=0, \pi$ (``hs" is short for ``high symmetry"), we have $[M_x, D(\beta_{1,\rm hs})]=0$. Thus, $M_x$ and $D$ share the same eigenvector $|n^R(\beta_{1,\rm hs})\rangle$, with
\begin{eqnarray}
M_x |n^R(\beta_{1,\rm hs})\rangle = \pm |n^R(\beta_{1,\rm hs})\rangle.
\end{eqnarray}
At high symmetry points, $\mathcal{R}^{(n)}_x(\beta_{1,\rm hs})=0$ or $\pi\mod 2\pi$. The generalized Berry phase of band $n$ is quantized:
\begin{eqnarray}
\gamma_1^{(n)} 
= \mathcal{R}^{(n)}_x\big|^{{\rm Arg\,}\beta_1=\pi}_{{\rm Arg\,}\beta_1=0}
=
0 \,\,{\rm or\,\,}\pi \mod 2\pi.
\end{eqnarray}
It is notable that in the Hermitian case, $\beta_1 = e^{iq_1}$ and  $M_x$ is the reflection symmetry operator.

In the second case, we consider a 2D lattice with the dynamical matrix $D(\vec\beta = (\beta_1, \beta_2))$ subjected to the symmetry property $\mathcal{I}D(\vec\beta)\mathcal{I}^{-1}=D(\vec\beta^*)$, where the symmetry operator $\mathcal{I} = \mathcal{I}^{-1}$. To demonstrate that the generalized Berry phase is quantized, we notice
\begin{eqnarray}\label{inversion}
D(\vec\beta^*) \mathcal{I} |n^R(\vec\beta)\rangle = \lambda_n(\vec\beta) \mathcal{I} |n^R(\vec\beta)\rangle,
\end{eqnarray}
which means 
\begin{eqnarray}
\mathcal{I} |n^R(\vec\beta)\rangle = e^{i\mathcal{R}_{\mathcal{I}}^{(n)}(\vec \beta)} |n^R(\vec\beta^*)\rangle,
\end{eqnarray}
At high symmetry points when $({\rm Arg\,}\beta_{1,{\rm hs}}, {\rm Arg\,}\beta_{2,{\rm hs}}) = (0,0), (0,\pi), (\pi,0), (\pi,\pi)$, the right eigenvector yields 
\begin{eqnarray}
\mathcal{I}|n^R(\vec\beta_{\rm hs})\rangle = \pm |n^R(\vec\beta_{\rm hs})\rangle,
\end{eqnarray}
which means $\mathcal{R}_{\mathcal{I}}^{(n)}(\vec \beta_{\rm hs})=0$ or $\pi\mod 2\pi$.
The generalized Berry phases $\gamma_1^{(n)}({\rm Arg\,}\beta_2)$ and $\gamma_1^{(n)}(-{\rm Arg\,}\beta_2)$ are related to one another:
\begin{eqnarray}\label{gammaInv}
\gamma_1^{(n)}({\rm Arg\,}\beta_2) = -\gamma_1^{(n)}(-{\rm Arg\,}\beta_2)\mod 2\pi.
\end{eqnarray}
Although $\gamma_1^{(n)}({\rm Arg\,}\beta_2)$ is not quantized\cite{benalcazar2017electric} at each ${\rm Arg\,}\beta_2$, 
when averaged over ${\rm Arg\,}\beta_2$ the Berry phase $\gamma_1^{(n)}$ is quantized 
\begin{eqnarray}
\gamma_1^{(n)} = \frac{1}{2\pi}\int_{-\pi}^\pi \gamma_1^{(n)}({\rm Arg\,}\beta_2) \,\, d{\rm Arg\,}\beta_2 = 0\,\,{\rm or}\,\, \pi.
\end{eqnarray}
For the bands which are separated from other bands, the summation of generalized Berry phase is also quantized, 
\begin{eqnarray}
\gamma_1 = \sum_n \gamma_1^{(n)} = 0\,\,{\rm or}\,\, \pi \mod 2\pi.
\end{eqnarray}
The above results also hold true for $\gamma_2$. It is notable that $\mathcal{I}$ is the inversion symmetry operator in the Hermitian case when $\vec\beta = (e^{iq_1}, e^{iq_2})$.


In the third case, we consider an $N\times N$ dynamical matrix $D(\vec\beta)$ ($N$ is even) subjected to the symmetry property $\Pi D\Pi^{-1} = -D$, where the symmetry operator $\Pi = \Pi^{-1}$. We prove the quantized Berry phase by noticing that
\begin{eqnarray}\label{particlehole}
D\Pi |n^R\rangle = -\lambda_n \Pi |n^R\rangle,
\end{eqnarray}
which means the eigenvalues come in $\pm\lambda_n$ pairs (we let $\lambda_n>0$). Thus, $N/2$ bands have $+\lambda_n$ eigenvalues and $N/2$ bands have $-\lambda_n$ eigenvalues. 
The eigenstates $|n^R\rangle$ and $|-n^R\rangle$ are related to one another by
\begin{eqnarray}
\Pi |n^R\rangle = e^{i\mathcal{R}^{(n)}_{\Pi}} |-n^R\rangle,
\end{eqnarray}
meaning that the generalized Berry phase $\gamma_1({\rm Arg\,}\beta_2)$ at each ${\rm Arg\,}\beta_2$ is quantized under this symmetry, 
\begin{eqnarray}
\gamma_1({\rm Arg\,}\beta_2)
 & = & 
- \frac{1}{2}\sum_{n=1}^{N/2} \mathcal{R}^{(n)}_{\Pi}({\rm Arg\,}\beta_2)\big|_{{\,\rm Arg\,}\beta_1 =-\pi}^{{\,\rm Arg\,}\beta_1 =+\pi}\nonumber \\
 & = & 0 \,\, {\rm or\,\,}\pi \mod 2\pi,
\end{eqnarray}
where we sum over all $N/2$ bands with $-\lambda_n$ eigenvalues. The above conclusion holds true for $\gamma_2({\rm Arg\,}\beta_1)$ as well. 
In the later discussions it is useful to study the following form of non-Hermitian dynamical matrix:
\begin{eqnarray}\label{Chiral}
D = \sigma_- \otimes h_2(\vec \beta)+\sigma_+ \otimes h_1(\vec\beta),
\end{eqnarray}
where $h_1$, $h_2$ are $\frac{1}{2}N\times \frac{1}{2} N$ invertible matrices which are free of exceptional point, $\sigma_\pm = (\sigma_x\pm i\sigma_y)/2$, and $\sigma_{x,y,z}$ are the Pauli matrices. 
The generalized Berry phase is quantized by the symmetry operator $\Pi=\sigma_z\otimes I_{\frac{1}{2}N\times \frac{1}{2}N}$ such that $\Pi D\Pi^{-1}=-D$ and $\Pi = \Pi^{-1}$. 
Hermiticity constrains $h_1=h_2^\dag$, which is lifted in non-Hermitian systems. We now attempt to express the generalized Berry phase in a simple form in terms of $h_1$ and $h_2$. To this end, we denote $|n^R\rangle = (|n^R_A\rangle^T, |n^R_B\rangle^T)^T$ and $\langle n^L| = (\langle n^L_A|, \langle n^L_B|)$. 
We have $\langle n_A^L|n_A^R\rangle = \langle n_B^L|n_B^R\rangle = 1/2$ according to the normalized biorthogonality $\langle n^L|n^R\rangle=1$. As an intermediate step, the Berry phase of the $\frac{1}{2}N$ bands with $-\lambda_n$ eigenvalues is simplified as  
\begin{eqnarray}\label{chiral1}
\gamma_i & = & 
2i \sum_{n=1}^{\frac{1}{2}N} \oint_{\mathcal{C}_i} d{\,\rm Arg\,}\beta_i \langle n_{A}^L| \partial_{{\rm Arg\,}\beta_i} |n_{A}^R\rangle\nonumber \\
 & {} & +
i \sum_{n=1}^{\frac{1}{2}N} \oint_{\mathcal{C}_i} d{\,\rm Arg\,}\beta_i  \langle n_{A}^L|h^{-1}_2 \partial_{{\rm Arg\,}\beta_i} h_2|n_{A}^R\rangle \nonumber \\
 & {} & -
\frac{1}{2} i \oint_{\mathcal{C}_i} d{\,\rm Arg\,}\beta_i \,\partial_{{\rm Arg\,}\beta_i} \ln\prod_{n=1}^{\frac{1}{2}N}\lambda_n.
\end{eqnarray}
We exploit the identity $\sum_{n=1}^{\frac{1}{2}N} |n^L_A\rangle \langle n^R_A| = \frac{1}{2}I_{\frac{1}{2}N\times \frac{1}{2}N}$ to reduce the second term in Eq.(\ref{chiral1}) as $\frac{1}{2}i  \oint_{\mathcal{C}_i} d{\,\rm Arg\,}\beta_i \,\, \partial_{{\,\rm Arg\,}\beta_i} \ln \det h_2$. Finally, the generalized Berry phase is reduced in terms of $h_1$ and $h_2$ as follows,  
\begin{eqnarray}\label{BerryChiral}
\gamma_i
=
\frac{1}{4}i \oint_{\mathcal{C}_i} d{\,\rm Arg\,}\beta_i \, \partial_{{\,\rm Arg\,}\beta_i} \ln \det ( h_2 h_1^{-1}).
\end{eqnarray}

\begin{figure}[htb]
\includegraphics[scale=0.22]{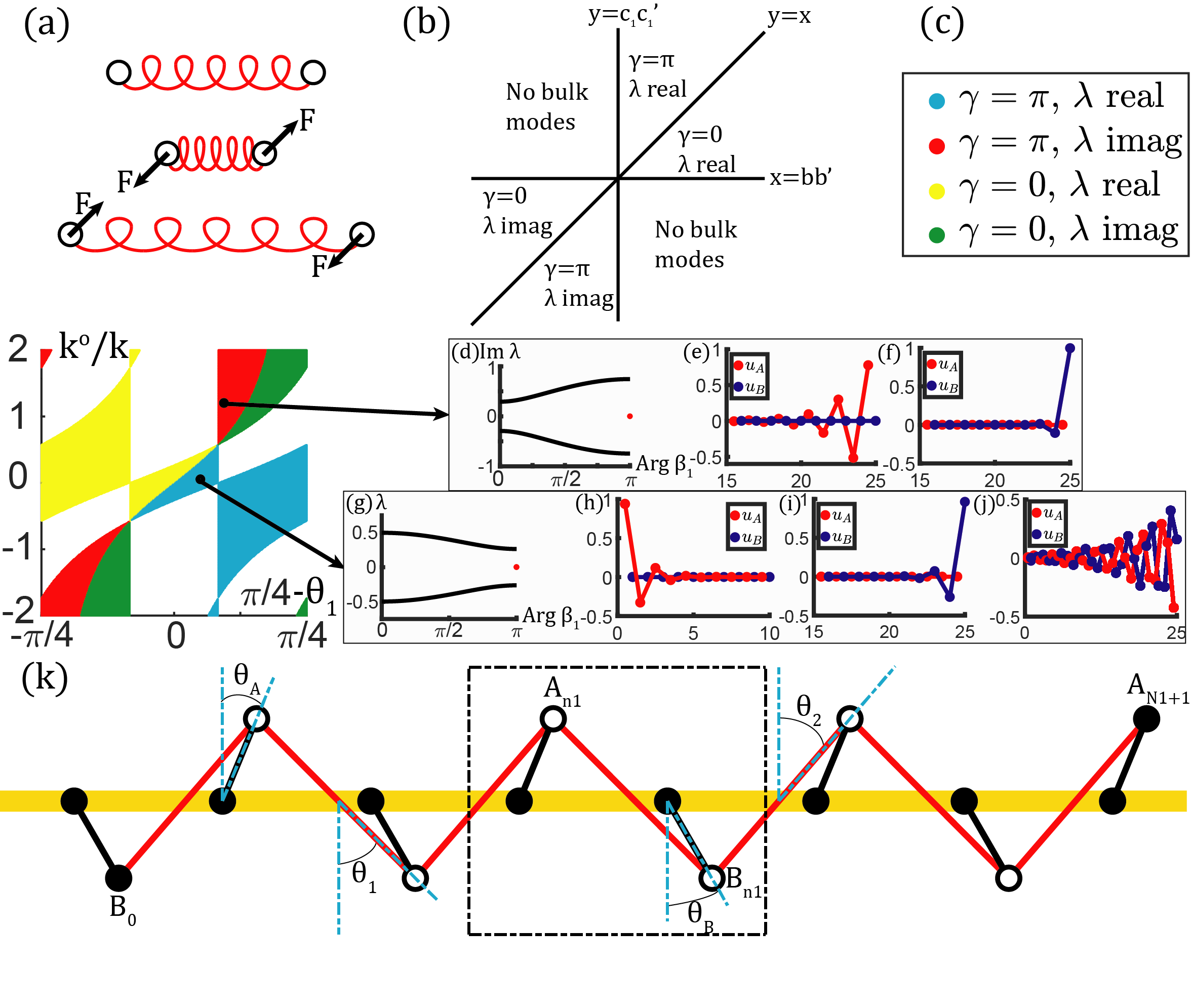}
\caption{Topological 1D rotor chain composed of 25 unit cells. (a) Basic element of the meta-beam that leads to odd elasticity. The system is imposed by a clockwise (counterclockwise) torque. (b) Topological phase diagram. (c) Given $\theta_A=\theta_B=\pi/6$ and $\theta_2=\pi/2-\theta_1$, we illustrate the topological phase diagram as the function of $\pi/4-\theta_1$ and $k^o/k$. (d) The relationship between ${\rm Im\,}\lambda$ and ${\rm Arg\,}\beta_1$ of topological configuration whose eigenvalues are complex. Topological modes are marked by the red dot. (e) and (f) Both topological modes localize on the right lattice boundary. (g) The relationship between $\lambda$ and ${\rm Arg\,}\beta_1$ of topological configuration with real eigenvalues. (h) and (i) Topological modes localize on the right and left lattice boundaries. (j) An exponentially localized bulk mode that manifests non-Hermitian skin effect. (k) 1D rotor chain subjected to fixed boundary conditions depicted by black dot rotors. 
}\label{Fig1}
\end{figure}

{It is notable that Eqs.(\ref{reflection}) and (\ref{inversion}) are the ``non-Hermitian" extensions of spatial reflection symmetry and inversion symmetry, respectively. The topological boundary modes in these systems occur at finite eigenfrequencies. This can be observed later in Figs.\ref{Fig1}(d), (g) and Fig.\ref{Fig2}(d) of rotor lattices and in Fig.\ref{Fig3}(c) of the honeycomb lattice. There is no guarantee that the band structure is symmetric with respect to the eigenfrequencies of the topological modes, which is later manifested in Fig.\ref{Fig3}(c) of the honeycomb lattice. 
On the other hand, Eq.(\ref{particlehole}) is the ``non-Hermitian" extension of the particle-hole symmetry, meaning that the symmetry operator $\Pi$ could satisfy $\Pi^2=1$ and $\{D,\Pi\}=0$, which further guarantees the eigenstates come in $\pm \lambda_n$ pairs\cite{kane2014topological}. The band structure is symmetric with respect to the zero frequency, which is in sharp contrast to the non-Hermitian extensions of spatial reflection and inversion symmetries. Thus, the band structure can be used as a key property to differentiate the behaviors of the three different symmetries. }

For non-Hermitian systems higher than 1-dimension, we can further define the generalized Berry curvature as 
\begin{eqnarray}
\Omega_{ii'}^{(n)}(\vec\beta) = \partial_{{\rm Arg\,}\beta_i}\mathcal{A}_{i'}^{(n)}-\partial_{{\rm Arg\,} \beta_{i'}}\mathcal{A}_{i}^{(n)}.
\end{eqnarray}
By exploiting the relation $\langle n'^L|\partial_{{\rm Arg\,}\beta_i} |n^R\rangle = {\langle n'^L|\partial_{{\rm Arg\,}\beta_i}D |n^R\rangle}/(\lambda_{n}-\lambda_{n'})$, it is straightforward to prove\cite{RevModPhys.82.1959} that 
\begin{eqnarray}\label{omegavanish}
\sum_{n=1}^N\Omega_{ii'}^{(n)}(\vec\beta) = 0.
\end{eqnarray}

\section{Non-Hermitian topological mechannics in rotor lattices}
\subsection{Non-Hermitian topological mechanics in the 1D rotor chain}
We first realize the basic non-Hermitian topological mechanics in the 1D rotor chain. {The chain is composed of diatomic unit cells in which the rigid rotors of length $r$ are connected to fixed pivot points separated by the distance $a$ (hence the lattice constant is $2a$). The rotors are freely rotatable about fixed pivots, and the rotation is restricted in-plane with the axis of the chain.} The other ends of the rotors are connected by particles of mass $m$, and the neighboring mass particles are connected by meta-beams that possess odd elasticity [Fig.\ref{Fig1}(k)]. The chain consists of $N_1$ unit cells with the rotors as well as the fixed pivots labeled $A_{n_1}$ and $B_{n_1}$ ($1\le n_1\le N_1$) and is subjected to fixed boundary conditions\cite{kopfler2019topologically, wakao2019higher} at rotors $B_0$ and $A_{N_1+1}$. The equilibrium configuration is that the $A$-rotors ($B$-rotors) make an angle ${\theta}_A$ (${\theta}_B$) relative to the upward (downward) normals. {Therefore, at the equilibrium state the meta-beams connecting $A_{n_1}$ and $B_{n_1}$ make an angle $\theta_1=\arctan\{[a+r(\sin\theta_B-\sin\theta_A)]/r(\cos\theta_A+\cos\theta_B)\}$ to the downward normal (the meta-beam connecting $B_{n_1}$ and $A_{n_1+1}$ make an angle $\theta_2=\arctan\{[a+r(\sin\theta_A-\sin\theta_B)]/r(\cos\theta_A+\cos\theta_B)\}$ to the upward normal).} We study the eigenvalue problem in Eq.(\ref{Newton}), where 
\begin{eqnarray}\label{OEFBC}
D & = & h_z\sigma_z
-\bigg[{\rm sgn}(b)|bb'|^{\frac{1}{2}}+\sum_{i=1}^d \beta_i^{-1} c_i |b'/b|^{\frac{1}{2}}\bigg]\sigma_+\nonumber \\
 & {} & -\bigg[{\rm sgn}(b')|bb'|^{\frac{1}{2}}+\sum_{i=1}^d \beta_i c_i'|b/b'|^{\frac{1}{2}}\bigg]\sigma_-
,
\end{eqnarray}
$d=1$ is the spatial dimension, $h_z = (a-a')/2$, $\omega$ is the eigenfrequency, $\lambda=m\omega^2-(a+a')/2$ is the eigenvalue, $u=(u_A/|b|^{\frac{1}{2}}, u_B/|b'|^{\frac{1}{2}})$ is the unit cell displacement field, ${\rm sgn}(\ldots)$ is the sign function, and $a, b, c_1, a', b', c_1'$ are constant parameters displayed in Appendix.C. Under periodic boundary conditions, we simply convert particle displacements to momentum space by replacing $\beta_i$ with $e^{iq_i}$, $q_i\in[0,2\pi)$ in Eqs.(\ref{ansatz}, \ref{OEFBC}) to obtain a non-Hermitian dynamical matrix.

Under fixed boundary conditions, we require $bb'c_ic_i'|_{i=1,\ldots,d}>0$ so that the bulk modes exist, and these bulk modes are exponentially localized near the lattice boundaries (``non-Hermitian skin effect") with the decay rate
\begin{eqnarray}\label{OEbeta}
|\beta_i| = (b'c_i/bc_i')^{1/2} \quad (bb'c_ic_i'>0) \quad i =1,2,... d.\qquad
\end{eqnarray}

The topological properties of the considered 1D rotor lattice is captured by the invariant called the generalized Berry phase. In general, the generalized Berry phase is \emph{not quantized}. However, it can be quantized by imposing the condition $h_z=0$, which is practically achieved\cite{kopfler2019topologically} by enabling the nearest neighbor springs perpendicular to each other. The generalized Berry phase is computed by Eq.(\ref{BerryChiral}). The topological phase transition occurs when $|bb'|=|c_1c_1'|$. The chain is topologically trivial when $\gamma_1=0$ and $|bb'|>|c_1c_1'|$, while the doubly degenerate topological modes\cite{PhysRevLett.121.026808, yao2018edge} emerge on lattice boundaries when $\gamma_1=\pi$ and $|bb'|<|c_1c_1'|$. They both localize on the left boundary if $|b|>|c_1|$ and $|b'|<|c_1'|$ (right boundary if $|b|<|c_1|$ and $|b'|>|c_1'|$), while they localize on different boundaries if $|b|<|c_1|$ and $|b'|<|c_1'|$. It is notable that all eigenvalues are real in the branch $bb' > 0$ and $c_1 c_1' > 0$, while they are complex in the branch $bb' < 0$ and $c_1c_1' < 0$. Although the study of topological phases apply for both branches, the mechanical responses are quite different. We leave the analysis at the end of this section due to the similar discussions of 1D and 2D rotor lattices.

\subsection{Non-Hermitian topological mechanics in the 2D rotor lattice}
\begin{figure}[htb]
\includegraphics[scale=0.15]{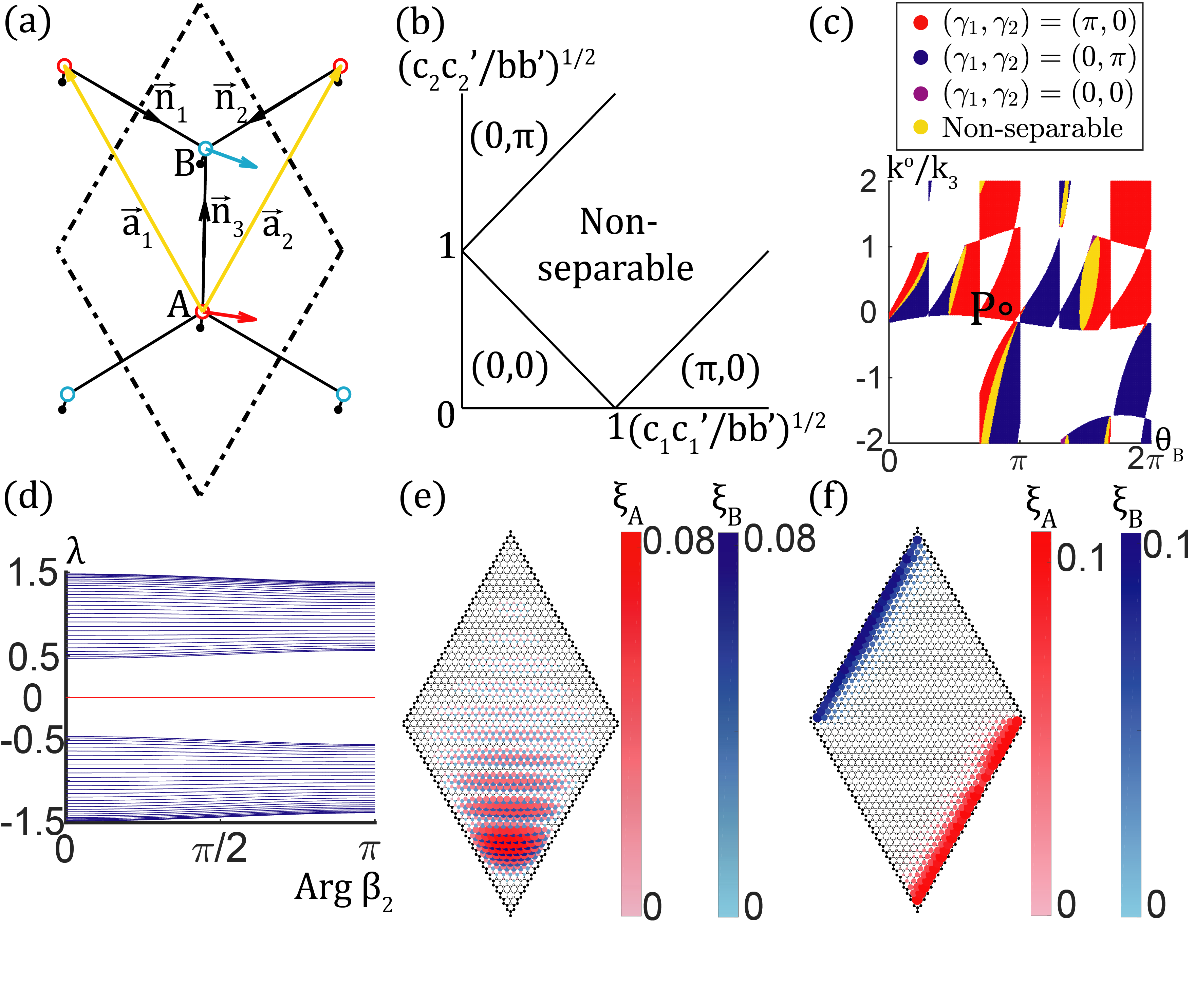}
\caption{Topological 2D rotor lattice composed of $30\times 30$ unit cells. (a) Unit cell encircled by dashed line box. Primitive vectors $\vec a_1=(-\frac{\sqrt{3}}{2},\frac{3}{2})$ and $\vec a_2=(\frac{\sqrt{3}}{2},\frac{3}{2})$ are depicted by yellow arrows. Red (blue) circle marks rotor $A$ at $\vec r_A=(0.014,0.099)$ (rotor $B$ at $\vec r_B = (0.034, 1.09)$). They are rotatable with radii $l_A=l_B=0.1$ about fixed hinges marked by black dots. Red (blue) arrow denotes the tangential unit vector which makes an angle $\theta_A=172^\circ$ ($\theta_B = 160^\circ$) to the $x$-axis. Black arrows $\hat{n}_1$, $\hat{n}_2$ and $\hat{n}_3$ mark unit vectors of three bonds. $k^o/k_3 = 0.05$. 
(b) Phase diagram of Berry phases $(\gamma_1, \gamma_2)$. (c) The configuration in (a) which leads to topological boundary modes is not unique. A finite parameter region which leads to topological configurations is illustrated here by varying $k^o/k_3$ and $\theta_B$ while we keep the rest of the parameters unchanged. 
Topological configurations whose Berry phases are $(\gamma_1, \gamma_2)=(\pi, 0)$ are represented by the red area, and the configurations whose Berry phases are $(\gamma_1, \gamma_2)=(0, \pi)$ are represented by the blue area. The unit cell configuration of (a) is marked by $P$. (d) We plot the $\lambda$ versus ${\rm Arg\,}\beta_2$ figure of the topological lattice. The Berry phases $(\gamma_1,\gamma_2) = (\pi,0)$. The red line stands for the topological edge modes localized on the top left and bottom right boundaries. (e) The displacement weight of one of the bulk modes $\xi_\alpha^{(s)}$ (defined in Eq.(\ref{weight})) on each $A$-node $\alpha=A(n_1,n_2)$ ($B$-node $\alpha=B(n_1,n_2)$) is shown using both color (see color bar on the right) and size of the disks. (f) The topological mode weight summation $\xi_\alpha$ (defined in Eq.(\ref{weightSum})) on each node. 
}\label{Fig2}
\end{figure}

Having established topological mechanics in the 1D rotor chain, we now ask if the same can be realized in the 2D rotor lattice. The unit cell is shown in Fig.\ref{Fig2}(a), 
where each rotor particle is connected to three odd-elastic meta-beams labeled $i'=1, 2$ and 3. The spring constants and the odd-elastic constants are denoted as $k_{i'}$ and $k^o_{i'}$, respectively. The lattice consists of $N_1\times N_2$ unit cells whose sites are labeled by $A_{n_1,n_2}$ and $B_{n_1,n_2}$, and is subjected to fixed boundary conditions. The equilibrium configuration is that the tangential directions of $A$-rotors and $B$-rotors make angles $\theta_A$ and $\theta_B$ to the $x$-axis, respectively, and the meta-beam labeled $i'$ makes an angle $\theta_{i'}$ to the $x$-axis. The dynamical matrix $D$ is given by Eq.(\ref{OEFBC}), where $d=2$ is the spatial dimension, and $a, b, c_1, c_2, a', b', c_1', c_2'$ are parameters detailed in Appendix.C. 

The bulk modes are exponentially localized on lattice boundaries with the decay rates $\beta_{1}$ and $\beta_{2}$ given by Eq.(\ref{OEbeta}), which manifests non-Hermitian skin effect in 2D systems. {This effect can be observed in Fig.\ref{Fig2}(e) by plotting the displacement weight 
\begin{eqnarray}\label{weight}
\xi_\alpha^{(s)}=[(u^{x(s)}_{\alpha})^2+(u^{y(s)}_{\alpha})^2]^{1/2},
\end{eqnarray} 
of one of the bulk modes, where 
$
u^{(s)}=(\vec u^{(s)}_{A,(1,1)}, \vec u^{(s)}_{B,(1,1)}, \ldots,\vec u^{(s)}_{A,(N_1,N_2)}, \vec u^{(s)}_{B,(N_1,N_2)})
$
is the $s$-th bulk mode displacement eigenvector corresponding to the eigenvalue $\lambda^{(s)}$ of the Newtonian equation of motion, and $\alpha = A(n_1, n_2)$ ($\alpha = B(n_1, n_2)$) for the mass particle $A(n_1,n_2)$ ($B(n_1, n_2)$).}

The topological properties of the 2D rotor lattice is featured by the generalized Berry phase. It is \emph{not quantized} unless $h_z=0$, which is practically achieved by letting $k_{i'}/\sin(2\theta_{i'+1}-2\theta_{i'+2})={\rm Const.}$ for $i'=1,2,3$. When the bands are separable, the topological phase is characterized by the generalized Berry phase. The lattice is free of topological edge modes when $(\gamma_1, \gamma_2)=(0,0)$ if $|bb'|^{\frac{1}{2}}>|c_1c_1'|^{\frac{1}{2}}+|c_2c_2'|^{\frac{1}{2}}$. 
The topological modes emerge on the upper-left and bottom-right lattice boundaries when $(\gamma_1, \gamma_2)=(\pi, 0)$ if $|bb'|^{\frac{1}{2}}<|c_1c_1'|^{\frac{1}{2}}-|c_2c_2'|^{\frac{1}{2}}$, while they emerge on the upper-right and bottom-left boundaries when $(\gamma_1, \gamma_2)=(0, \pi)$ if $|bb'|^{\frac{1}{2}}<|c_2c_2'|^{\frac{1}{2}}-|c_1c_1'|^{\frac{1}{2}}$. {In Fig.\ref{Fig2}(f) we illustrate the emergence of topological modes by calculating the summation of the displacement weight  
\begin{eqnarray}\label{weightSum}
\xi_\alpha=\bigg\{\frac{1}{N_{\rm topo}}\sum_{s=1}^{N_{\rm topo}}\left[(u^{x(s)}_{\alpha})^2+(u^{y(s)}_{\alpha})^2\right]\bigg\}^{1/2},
\end{eqnarray} 
where $u^{(s)}$ is the $s$-th topological mode displacement eigenvector of the Newtonian equation of motion (their eigenvalues are marked by the red dots in Fig.\ref{Fig2}(d)), $N_{\rm topo}$ is the number of topological modes, and $\alpha = A(n_1, n_2)$ ($\alpha = B(n_1, n_2)$) for the mass particle $A(n_1,n_2)$ ($B(n_1, n_2)$).} In the non-separable region when $||c_1c_1'|^{\frac{1}{2}}-|c_2c_2'|^{\frac{1}{2}}|<|bb'|^{\frac{1}{2}}<|c_1c_1'|^{\frac{1}{2}}+|c_2c_2'|^{\frac{1}{2}}$, the two bands touch at points $\vec\beta_{(w)}$ and $\vec\beta_{(w)}^*$. It is notable that in the Hermitian case when $b=b'$, $c_1=c_1'$ and $c_2=c_2'$, these two band-touching points correspond to Weyl points. 
The extension of Weyl points to non-Hermitian systems will be our future research. 

{The bulk modes of the 1D and 2D rotor lattices manifest the interesting ``non-Hermitian skin effect", in the sense that they are exponentially localized on the lattice boundaries. Consequently, the band structure and the topological properties are remarkably distinct under different boundary conditions. Here the fixed boundary condition is adopted to reflect the emergence of topological boundary modes. We note that in the following section, such ``skin effect" is absent in the honeycomb lattice and the bulk modes are extended in space, meaning that the bulk mode spectrum under fixed boundary conditions is the same as that under periodic boundary conditions. Despite the lack of the ``skin effect", the Hamiltonian (dynamical matrix) of the honeycomb lattice is still non-Hermitian.}

Both rotor lattices are stable in the branch $bb'>0$ and $c_i c_i'|_{i=1,\ldots,d}>0$. The topological modes can be excited by employing an external monochromatic shaking force. On the other hand, the lattices are unstable in the branch $bb'<0$ and $c_i c_i'|_{i=1,\ldots,d}<0$ in the dissipationless limit. Therefore, we ask all eigenmodes to decay in time by counteracting the intrinsic energy production with a damping $f= -\eta \dot{u}$, where
\begin{eqnarray}\label{eta}
\eta > \eta_c = {\rm max} \,[m\, {\rm Im}(\omega^2)/\sqrt{{\rm Re}(\omega^2)}].
\end{eqnarray} 
However, due to the real eigenvalues of topological modes, the excitations are heavily damped, which is in sharp contrast to the complex eigenvalues of the topological modes in the following honeycomb lattice.

Exceptional points emerge in both 1D and 2D rotor lattices when $bb'<0$, $c_i c_i'|_{i=1,\ldots,d}<0$, and $h_z\neq 0$. At the exceptional point $\vec\beta = \vec\beta_{\rm e}$ (here the lower index ``{\rm e}" is short for exceptional point), the eigenvalues coalesce at $\lambda_n(\vec\beta_{\rm e})=0$, and only one eigenvector $|n^R(\vec \beta_{\rm e})\rangle$ instead of two occurs at the exceptional point. In general this may not fulfill the reality condition of particle displacements. In Appendix.D we prove that, as long as the coalesce eigenvalue is real (which is true for $\lambda_n(\vec\beta_{\rm e})=0$ in rotor lattices), the two coalesce eigenvalues $\lambda(\vec\beta_{\rm e})$ and $\lambda(\vec\beta^*_{\rm e})$ must degenerate. The linear combination of $|n^R(\vec \beta_{\rm e})\rangle$ and $|n^R(\vec \beta^*_{\rm e})\rangle$ is real for all particle displacements.

\section{Non-Hermitian topological mechanics in the 2D honeycomb lattice}
\begin{figure}[htb]
\includegraphics[scale=0.16]{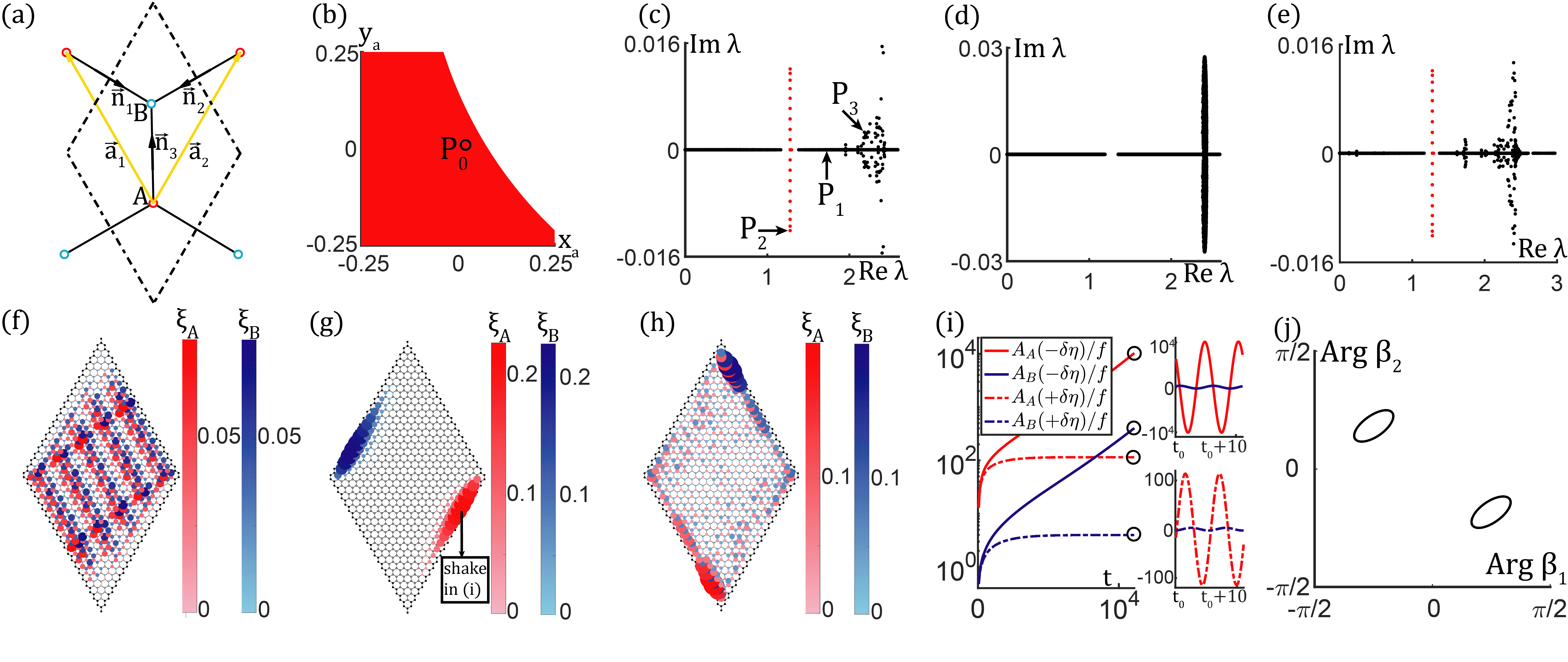}
\caption{Topological honeycomb lattice composed of $20\times 20$ unit cells. (a) The unit cell configuration. Site $A$ at $\vec r_a = (x_a, y_a) = (0.02, 0.01)$ and site $B$ at $\vec r_b = (0,1)$. Primitive vectors $\vec a_1=(-\sqrt{3}/2,3/2)$ and $\vec a_2=(\sqrt{3}/2,3/2)$. Spring constants $k_1=1$, $k_2=0.6$, and $k_3=0.9$. Odd elastic constants $k^o_1=0.300$, $k_2^o=0.0165$, and $k_3^o=-0.259$. (b) The configuration in (a) which leads to topological boundary modes is not unique. Here we illustrate a finite parameter region which leads to topological honeycomb lattices by varying $\vec r_a=(x_a, y_a)$ while the other parameters are kept unchanged. The area represents the topological configurations with Berry phases $(\gamma_1, \gamma_2)=(\pi, 0)$, and the white area is the non-seperable region in which the first and second two bands are not separated. The unit cell configuration of (a) is marked by $P_0$ at $\vec r_a = (0.02, 0.01)$. (c) The ${\rm Im\,}\lambda$ versus ${\rm Re\,}\lambda$ plot of the lattice under fixed boundary conditions. Topological and non-topological modes are marked by red and black dots, respectively. (d) The ${\rm Im\,}\lambda$ v.s. ${\rm Re\,}\lambda$ plot of the lattice under periodic boundary conditions. (e) The ${\rm Im\,}\lambda$ v.s. ${\rm Re\,}\lambda$ plot of the lattice under fixed boundary conditions but the particle mass $m_A=m_B=1$ are replaced by $m_A'=m_B'=0.7 m$ for the $n_1=1$ layer. (f) The weight of one of the bulk modes (marked by $P_1$ in (c)) $\xi_\alpha^{(s=P_1)}$ on each node on each node $\alpha$ marked by red and blue disks for $A$-sites and $B$-sites, respectively. (g) The weight of one of the topological boundary modes (marked by $P_2$ in (c)) $\xi^{(s=P_2)}_\alpha$ on each node. (h) The weight of one of the non-topological localized modes (marked by $P_3$ in (c)) $\xi^{(s=P_3)}_\alpha$ on each node. (i) Newtonian Mechanics simulation of the topological mode marked by $P_2$ in Fig.\ref{Fig3}(c): point force $\vec F_{\rm ext} = F\cos(\omega_{\rm ext}t) \hat{e}_x$ at $A_{(1, 15)}$ with $\omega_{\rm ext}=1.13$ and $F=10^{-10}$. We plot $A/F$ versus $t$ red (blue) curves for $A_{(1,15)}$ ($B_{(1,15)}$), where $A$ is the $x$-direction oscillation amplitude and $t$ varies from $0$ to $2000 \times(2\pi/\omega_{\rm ext})$. Solid curves for $\eta = \eta_c-\delta\eta$ and dashed curves for $\eta=\eta_c+\delta\eta$, where $\eta_c = 0.0107$ and $\delta\eta = 0.001\approx \eta_c/10$. Displacement-time relations are plotted from $t_0=1.10\times10^4$ to $t_0+2\times(2\pi/\omega_{\rm ext})$. (j) Two exceptional rings arise in the $({\rm Arg\,}\beta_1,{\rm Arg\,}\beta_2)$ plane when the second two bands coalesce. 
}\label{Fig3}
\end{figure}

In rotor lattices the generalized Berry phases are not quantized unless we fine-tune their parameters. Therefore, a lattice whose generalized Berry phase is quantized without fine-tuning is in need, and it is the honeycomb lattice we study as follows. Each site is connected to three odd-elastic meta-beams [Fig.\ref{Fig3}(a)]. The beam orientation, the spring constant, and the odd elastic constant are denoted as $\theta_{i'}$, $k_{i'}$ and $k_{i'}^o$, $i'=1,2,3$, respectively. The lattice is subjected to fixed boundary conditions. We study the lattice from Eq.(\ref{Newton}), where $u = (\vec u_A, \vec u_B)$ is the eigenvector, $\lambda = m\omega^2$ is the eigenvalue, $D$ is a four-band dynamical matrix
\begin{eqnarray}\label{HCD}
D=
\sigma_0\otimes h_{1,1}-
\sigma_+\otimes h_{\beta_1^{-1}, \beta_2^{-1}}-\sigma_-\otimes h_{\beta_1, \beta_2},
\end{eqnarray}
and $h_{\beta_1, \beta_2}$ is a $2\times 2$ matrix specified in Appendix.C.

\widetext

\endwidetext

The decay rates of the bulk modes, $\beta_1$ and $\beta_2$, happen to lie on a unit circle,
\begin{eqnarray}\label{3.11}
|\beta_1|=|\beta_2|=1.
\end{eqnarray}
Therefore, all bulk modes are extended in space instead of being localized on lattice boundaries. Consequently, the eigenvalues of bulk modes under fixed boundary conditions are the same as those under periodic boundary conditions. {To elaborate the absence of non-Hermitian skin effect, we select one bulk mode with the eigenvalue marked by $P_1$ in fig.\ref{Fig3}(c), to plot the displacement mode weight $\xi_\alpha^{s=P_1}$ on each node $\alpha$ in fig.\ref{Fig3}(f).}
Despite the fact that all bulk modes are extended, the dynamical matrix of the honeycomb lattice is still non-Hermitian.

To observe in-gap topological edge modes, we consider the dynamical matrix whose eigenvalues ${\rm Re\,}\lambda_{1,2}(\beta_1, \beta_2) < {\rm Re\,}\lambda_{3,4}(\beta_1', \beta_2')$ for all $(\beta_1, \beta_2)$ and $(\beta_1', \beta_2')$, meaning that the first and second two bands are separated by the band gap $\Delta$ [Fig.\ref{Fig3}(c)]. Different from the topological rotor lattices which are free of exceptional points, two exceptional rings arise in the $({\rm Arg\,}\beta_1, {\rm Arg\,}\beta_2)$ parameter space [Fig.\ref{Fig3}(j)] when the eigenvalues of the second two bands coalesce. Interestingly, the first two bands have \emph{no exceptional point}, providing a well-defined Berry phase to characterize the lattice topology. 
The dynamical matrix yields the symmetry property $\mathcal{I}D(\vec\beta)\mathcal{I}^{-1}=D(\vec\beta^*)$, where $\mathcal{I}=\sigma_x\otimes\sigma_0$. Although $\gamma_1({\rm Arg\,}\beta_2)$ is not quantized at each ${\rm Arg\,}\beta_2$, when averaged over ${\rm Arg\,}\beta_2$ the Berry phase $\gamma_1$ is quantized\cite{PhysRevB.89.155114, benalcazar2017electric}: 
\begin{eqnarray}\label{BerryInversion}
\gamma_1 = \frac{1}{2\pi}\int_{-\pi}^\pi \gamma_1({\rm Arg\,}\beta_2) d{\rm Arg\,}\beta_2  =0 \,\,\,{\rm or}\,\,\, \pi.
\end{eqnarray}
Similarly, $\gamma_2$ is also quantized to $0$ or $\pi$ when $\gamma_2({\rm Arg\,}\beta_1)$ is averaged over ${\rm Arg\,}\beta_1$. {The topological modes arise on the upper-left and bottom-right (upper-right and bottom-left) lattice boundaries when the Berry phases $(\gamma_1, \gamma_2) = (\pi, 0)$ ($(\gamma_1, \gamma_2) = (0, \pi)$). Fig.\ref{Fig3} depicts a topological honeycomb lattice with the generalized Berry phase $(\gamma_1, \gamma_2) = (\pi, 0)$. In contrast to the degenerate topological modes in the 2D rotor lattice, the eigenfrequencies of the topological modes in the honeycomb lattice is distinguishable [red dots in Fig.\ref{Fig3}(c)]. It is straightforward to choose one topological boundary mode marked by $P_2$ in Fig.\ref{Fig3}(c), and plot the displacement weight $\xi_\alpha^{(s=P_2)}$ on each node in Fig.\ref{Fig3}(g).}





This model presents {an interesting} feature not observed in the Hermitian counterpart. The eigenvalues of topological modes are complex due to the breaking of $\mathcal{PT}$-symmetry, meaning that in the dissipationless limit, the eigenmode excitations exponentially grow in time. 
An external damping can neutralize the internal energy production. However due to the large damping, both topological modes and bulk modes are excited when the lattice is shaken by an external force with frequency ${\rm Re\,}(\omega_{\rm topo}^{(s)})$, where  $\omega_{\rm topo}^{(s)}$ ($1\le s\le N_{\rm topo}$) is a topological mode eigenfrequency.
Hence, we ask the excitations of bulk modes to be small by imposing a lower bound of the bandgap $\Delta\gg \eta \,{\rm Re\,}(\omega_{\rm topo})$. The unique consequence of complex eigenvalues is that the excitations of topological modes exponentially increase in time if $\eta\lesssim\eta_c$ ($\eta_c$ given by Eq.(\ref{eta})), and they saturate if $\eta\gtrsim\eta_c$ [Fig.\ref{Fig3}(i)].

Besides the bulk modes which are extended in space, we observe additional modes whose eigenvalues are not separated from the bulk band, and they are localized on the lattice boundaries. Compared to topological modes, these localized modes are not topologically protected and are easily affected by perturbations [Fig.\ref{Fig3}(e)]. {For instance, the displacement weight $\xi_\alpha^{(s=P_3)}$ of one of such localized modes with the eigenvalue marked by $P_3$ in Fig.\ref{Fig3}(c) is plotted in Fig.\ref{Fig3}(h) on each node $\alpha$}. It is therefore interesting to seek exotic edge mode responses driven by external force with bulk mode frequencies.

\section{Summary and Perspectives}
{In this paper, we extend the notion of ``non-Hermitian topological theory" to mechanical systems. We study the 1D, 2D rotor lattices and the honeycomb lattice to illustrate our idea. The rotor lattices and the honeycomb lattice feature unique properties that cannot be replaced by one another. First, unlike Hermitian systems, non-Hermitian structures enjoy the additional attributes that the bulk modes are localized on the lattice boundaries and the subsequent sensitivity of band structure and lattice topology to the boundary conditions. This unique property, namely ``non-Hermitian skin effect", is reproduced within our rotor lattices. While this effect is missing in the honeycomb lattice, the eigenfrequencies of its topological modes are complex. Due to the large damping which counteracts the intrinsic energy production to stabilize the lattice, the excitations of the topological modes in rotor lattices are heavily damped. However, the topological mode excitations of the honeycomb lattice are still strong even in the heavily damped regime. }

Our study is based on but not limited to odd elasticity. Any active matter that injects energy could be the element of non-Hermitian metamaterials. Now we discuss potential applications in biological structures. 

First, the dynamics of biological structures are challenged by the ubiquitous heavy dissipation\cite{RevModPhys.85.1143, toner2005hydrodynamics, geyer2018sounds}. It is worth applying non-Hermitian mechanics to active biological systems in which the heavy damping is counteracted by energy production, such as motor activities in fiber networks and motility of cell sheets. So far the study of non-Hermitian systems is limited to periodic lattice structures. Thus, the extension of non-Hermitian mechanics to amorphous structures is useful to explore energetic biological systems.

Second, in addition to topological modes, there also exist non-topological localized modes in non-Hermitian systems. It is thus intriguing to ask if these modes are associated with the in-band boundary softness or exotic mechanical responses of active biological systems\cite{petrie2012leading, petrie2016multiple}.

\begin{acknowledgements}
D. Z. would like to thank insightful discussions with Vincenzo Vitelli, Po-Yao Chang, Anton Souslov and Colin Scheibner. D. Z. is supported in part by the National Science Foundation (Grant No. NSF-EFRI-1741618). 

\emph{Note added.---After this work was completed, we became aware of an independent work by C. Scheibner, W. Irvine, and V. Vitelli\cite{scheibner2020non}. We thank them for bringing their research to our attention.}
\end{acknowledgements}

\appendix

\section{Biorthogonality and completeness of generalized eigenvectors in non-Hermitian matrix}
We start by considering a non-Hermitian $N\times N$ matrix $D$ with eigenvalues denoted as $\lambda_n$, $1\le n \le N$. The corresponding generalized right eigenvector and left eigenvector of rank $l$ are denoted as $|n_{(l)}^R\rangle$ and $\langle n_{(l)}^L|$, respectively. They satisfy 
\begin{eqnarray}\label{A1}
 & {} & (D-\lambda_n I)^l |n^R_{(l)}\rangle = 0, \nonumber \\
 & {\rm but} \qquad &  (D-\lambda_n I)^{l'} |n^R_{(l)}\rangle \neq 0, \qquad \forall l'<l
\end{eqnarray}
and 
\begin{eqnarray}\label{A2}
 & {} & \langle n^L_{(l)}|(D-\lambda_n I)^l = 0, \nonumber \\
 & {\rm but} \qquad & \langle n^L_{(l)}|(D-\lambda_n I)^{l'} \neq 0,  \qquad \forall l'<l.
\end{eqnarray}
The generalized eigenvectors is a complete basis. In particular, ordinary eigenvectors (denoted as $|n^R\rangle$ and $\langle n^L|$) are rank 1 generalized eigenvectors.


It is straightforward to prove that acting $D-\lambda_n I$ on the generalized right (left) eigenvector $|n_{(l)}^R\rangle$ ($\langle n_{(l)}^L|$) gives the linear superposition of lower rank generalized right (left) eigenvectors:
\begin{eqnarray}\label{A3}
 & {} & (D-\lambda_n I)|n^R_{(l)}\rangle = \sum_{k=1}^{l-1} R_{n,(k,l)} |n^R_{(k)}\rangle\nonumber \\
 & {\rm and\quad} & \langle n^L_{(l)}|(D-\lambda_n I) = \sum_{k=1}^{l-1} L_{n,(k,l)} \langle n^L_{(k)}|,
\end{eqnarray}
where $R_{n,(k,l)}$ ($L_{n,(k,l)}$) is the coefficient of the rank $k$ generalized right (left) eigenvector of the eigenvalue $\lambda_n$. It is obvious that $R_{n,(l-1,l)}\neq 0$ ($L_{n,(l-1,l)}\neq 0$).

We denote the highest rank of left (right) generalized eigenvector of the eigenvalue $\lambda_n$ as $l_n$ ($r_n$). It is easy to prove that $l_n=r_n$. To this end, without loss of generality we assume $r_n> l_n$ (i.e., $r_n\ge l_n+1$). We calculate $0=\langle n^L_{(l_n-l)}|(H-\lambda_n I)^{r_n-1} |n^R_{(r_n)}\rangle$ for all $l=0,1,\ldots, l_n-1$ to find $|n^R\rangle=0$ which is not true. Therefore, in what follows we use $l_n$ to represent the highest rank of the left and right generalized eigenvectors. From now on we define $R_{n,(l',l)}|_{l'\ge l}=L_{n,(l',l)}|_{l'\ge l}=0$ and then Eq.(\ref{A3}) can be re-written as 
\begin{eqnarray}\label{A10}
 & {} & (D-\lambda_n I)|n^R_{(l)}\rangle = \sum_{k=1}^{l_n} R_{n,(k,l)} |n^R_{(k)}\rangle\nonumber \\
 & {\rm and\quad} & \langle n^L_{(l)}|(D-\lambda_n I) = \sum_{k=1}^{l_n} L_{n,(k,l)} \langle n^L_{(k)}|,
\end{eqnarray}

For different eigenvalues $\lambda_n\neq\lambda_{n'}$, the left and right generalized eigenvectors obey the biorthogonality
\begin{eqnarray}\label{A4}
\langle n'^L_{(l')}|n^R_{(l)}\rangle = \langle n^R_{(l)}|n'^L_{(l')}\rangle = 0\quad \forall l\le l_n, \quad \forall l'\le l_{n'}.\qquad
\end{eqnarray}
This can be proved by substituting Eq.(\ref{A3}) into $\langle n'^L_{(l')}|(D-\lambda_n I)|n^R_{(l)}\rangle$ repeatly in every step of the second principle of mathematical induction. 
Next, for the left and right generalized eigenvectors of the same eigenvalue $\lambda_n$, the biorthogonality is presented as follows, 
\begin{eqnarray}\label{A5}
 & {} & \langle n^L_{(l')}|n^R_{(l)}\rangle,\,\,  \langle n^R_{(l)}|n^L_{(l')}\rangle \neq 0 \qquad l+l'=l_n+1\nonumber \\
 & {} & \langle n^L_{(l')}|n^R_{(l)}\rangle = \langle n^R_{(l)}|n^L_{(l')}\rangle = 0 \qquad{\rm otherwise}.
\end{eqnarray}
This is proved by using the mathamatical induction and computing $\langle n^L_{(l')}|(D-\lambda_n I)|n^R_{(l)}\rangle$ in every step. Rescaling the right eigenvector $|n^R_{(l)}\rangle \to |n^R_{(l)}\rangle/\langle n^L_{(l')}|n^R_{(l)}\rangle$, and summarizing Eqs.(\ref{A4}, \ref{A5}), we obtain the normalized biorthogonality of the generalized eigenvectors,
\begin{eqnarray}\label{A6}
 \langle n'^L_{(l')}|n^R_{(l)}\rangle =\langle n^R_{(l)}|n'^L_{(l')}\rangle =\delta_{n,n'}\delta_{l+l',l_n+1}.
\end{eqnarray}
Based on Eq.(\ref{A6}) it is easy to prove the completeness of the generalized eigenvectors, 
\begin{eqnarray}\label{A7}
I_{N\times N} & = & \sum_{n}\sum_{l=1}^{l_n} |n^R_{(l)}\rangle \langle n^L_{(l_n+1-l)}| \nonumber \\
 & = & \sum_{n}\sum_{l=1}^{l_n}  | n^L_{(l_n+1-l)}\rangle \langle n^R_{(l)}|.
\end{eqnarray}
In particular, if all highest ranks of generalized eigenvectors are $l_n=1$, the generalized eigenbasis is the same as the ordinary eigenbasis, and there is no exceptional point in the non-Hermitian matrix $D$. Eqs.(\ref{A6}, \ref{A7}) are reduced to 
\begin{eqnarray}\label{A8}
 \langle n'^L|n^R\rangle=\langle n^R|n'^L\rangle =\delta_{n,n'},
\end{eqnarray}
and 
\begin{eqnarray}\label{A9}
I_{N\times N} = \sum_{n}|n^R\rangle \langle n^L| = \sum_{n}|n^L\rangle \langle n^R|.
\end{eqnarray}
Finally, based on Eq.(\ref{A6}) we further show the coefficients $R_{n,(k,l)}$ and $L_{n,(k,l)}$ in Eq.(\ref{A10}) are related. To this end, we calculate $\langle n^L_{(l)} |H-\lambda_n I|n^R_{(r)}\rangle$ to find 
\begin{eqnarray}\label{A11}
R_{n,(l_n+1-l,l_n+1-l')} = L_{n,(l',l)},
\end{eqnarray}
which is valid for all $l_n\ge l, l'\ge 1$.


\section{Generalized Berry phase in non-Hermitian classical systems}
Newtonian equation of motion is the second-order derivative in time, while Schrodinger's equation is the first-order derivative. In order to define the generalized Berry phase in the Newtonian equation of motion, we consider the simple case that for the $N\times N$ non-Hermitian dynamical matrix $D$, all bands are separated from each other\cite{shen2018topological}. Hence the $D$-matrix is free of exceptional points, and the left and right eigenbasis are complete. We use the eigenvalue equation,
\begin{eqnarray}
D|n^R\rangle = \lambda_n |n^R\rangle,
\end{eqnarray}
where $|n^R\rangle$ is the right eigenvector corresponding to eigenvalue $\lambda_n$. We then define an auxiliary wave function 
\begin{eqnarray}
|u_n^R(t')\rangle = |n^R\rangle e^{-i\lambda_n t'},
\end{eqnarray}
which evolves as the auxiliary parameter $t'$ advances. It also satisfies the eigenvalue equation $D|u^R_n(t')\rangle = \lambda_n |u^R_n(t')\rangle$. Since any $N$-dimensional wave function $|\psi^R(t')\rangle$ can be expressed as the linear superposition of the complete basis $\{|u^R_n(t')\rangle\}$, we find that $|\psi^R(t')\rangle$ is subjected to the Schrodinger-like equation of motion
\begin{eqnarray}\label{Slike}
D|\psi^R(t')\rangle = i\partial_{t'}|\psi^R(t')\rangle.
\end{eqnarray}
Starting from Eq.(\ref{Slike}), we derive the generalized Berry phase of non-Hermitian mechanical systems.

Based on Eq.(\ref{ansatz}) of main text, we use the ansartz $(\vec u_{A,(n_1,\ldots,n_d)}, \vec u_{B,(n_1,\ldots,n_d)}) = \beta_1^{n_1}\ldots \beta_d^{n_d}(\vec u_{A}, \vec u_{B})$ to reduce the lattice equation of motion as $D(\vec\beta)u=\lambda u$, where the dynamical matrix $D(\vec\beta)$ is the function of $\vec\beta$. We adiabatically evolve $\vec \beta(t')$ as the auxiliary parameter $t'$ advances. In sections.\uppercase\expandafter{\romannumeral4} and \uppercase\expandafter{\romannumeral5} we prove that, for all three models we study in this paper, the modulus $|\beta_i|$ keeps constant while the complex phase ${\rm Arg\,}\beta_i$ varies from 0 to $2\pi$. 
The starting eigenvector is denoted as $|n^R(\vec\beta(t'=0))\rangle$. 
According to the quantum adiabatic theorem\cite{kato1950adiabatic, messiah1962quantum, xiao2010berry}, 
a system initially in one of its eigenstate $|n^R(\vec\beta(0))\rangle$ will stay as an instantaneous eigenstate of the Hamiltonian $D(\vec\beta(t'))$ throughout the process. The only degree of freedom is the phase of the state. Write the state at $t'$ as 
\begin{eqnarray}
|\psi_n^R(t')\rangle = e^{i\gamma_i^{(n)}(t')} e^{-i\int_0^{t'}dt'' \,\lambda_n(\vec\beta(t''))}|n^R(\vec\beta(t'))\rangle,\quad
\end{eqnarray}
and insert it into Eq.(\ref{Slike}), we obtain
\begin{eqnarray}
(d\gamma_i^{(n)}/dt') |n^R(\vec\beta(t'))\rangle = i \partial_{t'} |n^R(\vec\beta(t'))\rangle.
\end{eqnarray}
Next, according to the normalized biorthogonality and completeness of the eigenbasis, we multiply the above equation on the left hand side with the left eigenvector $\langle n^L(\vec\beta(t'))|$, to obtain 
\begin{eqnarray}
\gamma_i^{(n)}
= \oint_{\mathcal{C}} \mathcal{A}^{(n)}_i(\vec\beta)\,\, d{\,\rm Arg\,}\beta_i,
\end{eqnarray}
where $\mathcal{C}$ is the closed path such that the variable $\vec\beta(0) = \vec\beta(T)$ goes back to itself, and 
\begin{eqnarray}
\mathcal{A}^{(n)}_i(\vec\beta)
= i \langle n^L| \partial_{{\rm Arg\,}\beta_i} |n^R\rangle 
\end{eqnarray}
is the generalized Berry connection. 

\section{Non-Hermitian skin effect in the 1D, 2D rotor lattices, and the 2D honeycomb lattice}
\subsection{Non-Hermitian skin effect in the 1D rotor chain}
Here we show calculation details of non-Hermitian skin effect in 1D rotor chain. Let us first denote $\Theta_A = \theta_A+\theta_1$, $\Theta_A' = \theta_2-\theta_A$, $\Theta_B =\theta_1-\theta_B$ and $\Theta_B' = \theta_B+\theta_2$. The parameters are displayed as follows,
\begin{eqnarray}
 & {} & a = (k+k^o\cot\Theta_A)\sin^2\Theta_A+(k-k^o\cot\Theta_A')\sin^2\Theta_A'\nonumber \\
 & {} & b = (k+k^o\cot\Theta_A)\sin\Theta_A\sin\Theta_B\nonumber \\
 & {} & c_1 = (k-k^o\cot\Theta_A')\sin\Theta_A'\sin\Theta_B'\nonumber \\
 & {} & a' = (k+k^o\cot\Theta_B)\sin^2\Theta_B+(k-k^o\cot\Theta_B')\sin^2\Theta_B'\nonumber \\
 & {} & b' = (k+k^o\cot\Theta_B)\sin\Theta_A\sin\Theta_B\nonumber \\
 & {} & c_1' = (k-k^o\cot\Theta_B')\sin\Theta_A'\sin\Theta_B'.
\end{eqnarray}
The displacement field of the chain is denoted as $\{u_{An}, u_{Bn}\}$, which stretches the meta-beams connecting the rotor particles. Newton's equation of motion is 
\begin{eqnarray}
 & {} & m \ddot{u}_{An} = (b u_{Bn}+c_1 u_{B,n-1})-a u_{An}\nonumber \\
 & {} & m \ddot{u}_{Bn} = (b' u_{An}+c_1' u_{A,n+1})-a' u_{Bn},
\end{eqnarray}
subjected to fixed boundary conditions 
\begin{eqnarray}
u_{B0}=u_{A,N+1}= 0. 
\end{eqnarray}
We use the ansartz $(u_{An},u_{Bn}) = \beta_1^n(u_A, u_B)$ to simplify the equation of motion as ${D} u = \lambda u$, with $u=(u_A/|b|^{\frac{1}{2}}, u_B/|b'|^{\frac{1}{2}})$ and 
\begin{eqnarray}
D & = & h_z\sigma_z
-\left[{\rm sgn}(b)|bb'|^{\frac{1}{2}}+ \beta_1^{-1} c_1 |b'/b|^{\frac{1}{2}}\right]\sigma_+\nonumber \\
 & {} & -\left[{\rm sgn}(b')|bb'|^{\frac{1}{2}}+ \beta_1 c_1'|b/b'|^{\frac{1}{2}}\right]\sigma_-
,
\end{eqnarray}
where $h_z = (a-a')/2$ and $\lambda = m\omega^2-(a+a')/2$. Given the eigenvalue $\lambda$, we find that $\det (D-\lambda I)=0$ is the second order equation of $\beta_1$. Solving this equation gives us two solutions $\beta_1^{(1)}$ and $\beta_1^{(2)}$ which satisfy 
\begin{eqnarray}\label{delta}
\beta_1^{(1,2)} = -(1\mp\sqrt{1-{4AC}/{B^2}}) (B/2A), 
\end{eqnarray}
and
\begin{eqnarray}
\beta^{(1)}_1\beta^{(2)}_1 = b'c_1/bc'_1,
\end{eqnarray}
where $A = bc_1'$, $B = bb'+c_1c_1'-\lambda^2+h_z^2$, and  $C = b'c_1$. The eigenvector correspoding to $\beta_1^{(i)}$ is denoted as $u^{(i)}$. The general wave function is given by 
\begin{eqnarray}
u_n=\beta_1^{(1)n}u^{(1)}+\beta_1^{(2)n}u^{(2)},
\end{eqnarray}
which is subjected to the fixed boundary conditions. Eliminating the eigenvectors gives us
\begin{eqnarray}\label{1.1}
(\beta_1^{(1)})^{N+1}  (b'+c'\beta_1^{(2)})
=
(\beta_1^{(2)})^{N+1}  (b'+c'\beta_1^{(1)}).\qquad
\end{eqnarray}

We are concerned with the bulk modes for a long chain in the limit $N\to\infty$, which requires $|\beta_1^{(1)}| = |\beta_1^{(2)}|$ for the bulk modes\cite{yao2018edge}. If the bulk mode condition is not fulfilled, without loss of generality we let $|\beta_1^{(1)}|>|\beta_1^{(2)}|$. By taking the $\lim_{N\to\infty}(\beta_1^{(2)}/\beta_1^{(1)})^N\to 0$, Eq.(\ref{1.1}) is simplified as $b'+c'\beta_1^{(2)}=0$, which gives a single $\beta_1$ solution instead of two, and this solution is independent of the chain length $N$. In order to have the bulk mode condition $|\beta_1^{(1)}| = |\beta_1^{(2)}|$, additional constraints are imposed on Eq.(\ref{delta}):
\begin{eqnarray}
1-4AC/B^2<0 \qquad {\rm and}\qquad {\rm Im\,}(AC/B^2)=0,\qquad
\end{eqnarray}
which in turn gives the constraint $bb'c_1c_1'>0$. In other words, there is no bulk mode solution if $bb'c_1c_1'<0$. Consequently, all eigenmodes are localized modes near the lattice boundaries if $bb'c_1c_1'<0$.

Based on these results, the bulk mode condition is mathematically formulated as 
\begin{eqnarray}
|\beta_1^{(1,2)}| = |\beta_1|= \sqrt{b'c_1/bc'_1} \qquad (bb'c_1c_1'>0).
\end{eqnarray}
In general, the decay rate $|\beta_1|$ of the \emph{bulk modes} is not 1, a unique feature of non-Hermitian systems dubbed the ``skin effect".

\subsection{Non-Hermitian skin effect in the 2D rotor lattice}
To study the non-Hermitian skin effect in the 2D honeycomb rotor lattice, we first denote a set of parameters which will be used later: $\Theta_{Ai} = \theta_A-\theta_i$, $\Theta_{Bi} = \theta_B-\theta_i$ for $i=1,2,3$, and 
\begin{eqnarray}
a  & = &  \sum_{i=1}^3 (k_i+k^o_i \tan\Theta_{Ai}) \cos^2\Theta_{Ai}\nonumber \\
b  & = &  (k_3+k^o_3\tan\Theta_{A3})\cos\Theta_{A3}\cos\Theta_{B3} \nonumber \\
c_1  & = & (k_1+k^o_1 \tan\Theta_{A1})\cos\Theta_{A1}\cos\Theta_{B1} \nonumber \\
c_2  & = & (k_2+k^o_2\tan\Theta_{A2})\cos\Theta_{A2}\cos\Theta_{B2} \nonumber \\
a'  & = &  \sum_{i=1}^3 (k_i+k^o_i \tan\Theta_{Bi})\cos^2\Theta_{Bi} \nonumber \\
b'  & = &  (k_3+k^o_3\tan\Theta_{B3})\cos\Theta_{A3}\cos\Theta_{B3} \nonumber \\
c_1'  & = &  (k_1+k^o_1 \tan\Theta_{B1})\cos\Theta_{B1}\cos\Theta_{A1}\nonumber \\
c_2'  & = &  (k_2+k^o_2\tan\Theta_{B2})\cos\Theta_{B2}\cos\Theta_{A2}.
\end{eqnarray}
The Newtonian equation of motion reads
\begin{eqnarray}
 & {} & -a u_{An_1n_2}+bu_{Bn_1n_2}+c_1 u_{Bn_1-1,n_2}+c_2 u_{Bn_1,n_2-1} \nonumber \\
 & {} & = m\ddot{u}_{An_1n_2}\nonumber \\
 & {} & -a' u_{Bn_1n_2}+b'u_{An_1n_2}+c_1'u_{An_1+1,n_2}+c_2' u_{An_1,n_2+1}\nonumber \\
 & {} &  = m\ddot{u}_{Bn_1n_2},
\end{eqnarray}
subjected to fixed boundary conditions,
\begin{eqnarray}
u_A(N_1+1,n_2) = u_A(n_1,N_2+1) \nonumber \\
= u_B(0,n_2) = u_B(n_1,0) = 0  
\end{eqnarray}
for all $ n_1=1,2,...,N_1$ and $ n_2 =1,2,...,N_2$. By applying the ansartz $({u}_{An_1,n_2}, {u}_{Bn_1,n_2}) = \beta_1^{n_1}\beta_2^{n_2}({u}_{A}, {u}_{B})$, the Newton's equation of motion is simpified as $D u = \lambda u$, where 
\begin{eqnarray}
D & = & h_z\sigma_z
-\bigg[{\rm sgn}(b)|bb'|^{\frac{1}{2}}+\sum_{i=1}^2 \beta_i^{-1} c_i |b'/b|^{\frac{1}{2}}\bigg]\sigma_+\nonumber \\
 & {} & -\bigg[{\rm sgn}(b')|bb'|^{\frac{1}{2}}+\sum_{i=1}^2 \beta_i c_i'|b/b'|^{\frac{1}{2}}\bigg]\sigma_-
,
\end{eqnarray}
and $u=(u_A/|b|^{\frac{1}{2}}, u_B/|b'|^{\frac{1}{2}})$. Similar as the 1D rotor chain, given the eigenvalue $\lambda$ and decay rate $\beta_2$ ($\beta_1$), we find $\det(D -\lambda I) = 0$ is the second order equation of $\beta_1$ ($\beta_2$). Solving this equation gives us two solutions $\beta_1^{(1)}$ and $\beta_1^{(2)}$ which satisfy
\begin{eqnarray}\label{delta2}
\beta_1^{(1,2)} = -(1\mp\sqrt{1-{4AC}/{B^2}}) (B/2A), 
\end{eqnarray}
and
\begin{eqnarray}\label{2.3}
\beta_1^{(1)}\beta_1^{(2)} =c_1 (b'+c_2'\beta_2) /c_1' (b+c_2\beta_2^{-1}),
\end{eqnarray}
where $A = c_1'(b+c_2/\beta_2)$, $B = bb'+c_1c_1'+c_2c_2'+c_2'b\beta_2+c_2b'/\beta_2-\lambda^2+h_z^2$, and $C = c_1(b'+c_2'\beta_2)$. By denoting the eigenvector as $u^{(i)}$ which corresponds to $\beta_1^{(i)}$ , we then express the general wave function as the linear superposition of the eigenvectors $u^{(1)}$ and $u^{(2)}$:
\begin{eqnarray}
u_{n_1n_2}
=
\beta_2^{n_2}(\beta_1^{(1)n_1}u^{(1)}+\beta_1^{(2)n_1}u^{(2)}),
\end{eqnarray}
subjected to fixed boundary conditions. Eliminating $u^{(i)}$ gives us the following relationship, 
\begin{eqnarray}\label{2.1}
 & {} & 
(b+c_1/\beta_1^{(1)}+c_2/\beta_2)(\beta_1^{(1)})^{N_1+1}\nonumber \\
 & = & 
(b+c_1/\beta_1^{(2)}+c_2/\beta_2)(\beta_1^{(2)})^{N_1+1}.
\end{eqnarray}

The bulk mode condition\cite{yao2018edge} requires $|\beta_1^{(1)}| = |\beta_1^{(2)}|$, which imposes additional conditions on Eq.(\ref{delta2}), 
\begin{eqnarray}\label{2.2}
1-4AC/B^2<0 \qquad {\rm and}\qquad {\rm Im\,}(AC/B^2)=0.\qquad
\end{eqnarray}
Solving Eq.(\ref{2.2}) gives us the decay rate of $\beta_2$:
\begin{eqnarray}\label{2.10}
|\beta_2| = \sqrt{b'c_2/bc_2'}, \qquad (bb'c_2c_2'>0).
\end{eqnarray}
By substituting $|\beta_2|$ into Eq.(\ref{2.3}) we further obtain the decay rate $\beta_1$,
\begin{eqnarray}
|\beta_1| = \sqrt{b'c_1/bc_1'}\qquad (bb'c_1c_1'>0).
\end{eqnarray}
In general, these bulk mode decay rates are not 1, which are the manifestation of the non-Hermitian skin effect in 2D systems.

\subsection{Non-Hermitian skin effect in the 2D honeycomb lattice}
To study the honeycomb lattice, the ansartz we adopt is $\vec u_{A,B(n_1,n_2)} = \beta_1^{n_1}\beta_2^{n_2}\vec u_{A,B}$, where $\beta_{i=1,2}$ are the decay rates of the bulk modes along the lattice directions. The Newton's equation of motion is simplified as $D u = \lambda u$, where $\lambda = m\omega^2$, $u=(u_A^x, u_A^y, u_B^x, u_B^y)$, and the dynamical matrix 
\begin{eqnarray}\label{DMHC}
D=
\sigma_0\otimes h_{1,1}-
\sigma_+\otimes h_{\beta_1^{-1}, \beta_2^{-1}}-\sigma_-\otimes h_{\beta_1, \beta_2}.\qquad
\end{eqnarray}
$h_{\beta_1, \beta_2}$ is a $2\times 2$ matrix, with the matrix element specified as follows, 
\widetext
\begin{eqnarray}\label{2by2}
 & {} & h_{\beta_1, \beta_2}(1,1) = (k_3\cos\theta_3-k^o_3\sin\theta_3)\cos\theta_3+\beta_1 (k_1\cos\theta_1-k^o_1\sin\theta_1)\cos\theta_1+\beta_2 (k_2\cos\theta_2-k^o_2\sin\theta_2)\cos\theta_2\nonumber \\
 & {} & h_{\beta_1, \beta_2}(1,2) = (k_3\cos\theta_3-k^o_3\sin\theta_3)\sin\theta_3+\beta_1 (k_1\cos\theta_1-k^o_1\sin\theta_1)\sin\theta_1+\beta_2 (k_2\cos\theta_2-k^o_2\sin\theta_2)\sin\theta_2\nonumber \\
 & {} & h_{\beta_1, \beta_2}(2,1) = (k_3\sin\theta_3+k^o_3\cos\theta_3)\cos\theta_3+\beta_1 (k_1\sin\theta_1+k^o_1\cos\theta_1)\cos\theta_1+\beta_2 (k_2\sin\theta_2+k^o_2\cos\theta_2)\cos\theta_2\nonumber \\
 & {} & h_{\beta_1, \beta_2}(2,2) = (k_3\sin\theta_3+k^o_3\cos\theta_3)\sin\theta_3+\beta_1 (k_1\sin\theta_1+k^o_1\cos\theta_1)\sin\theta_1+\beta_2 (k_2\sin\theta_2+k^o_2\cos\theta_2)\sin\theta_2.
\end{eqnarray}
\endwidetext
\noindent The fixed boundary conditions are given by 
\begin{eqnarray}
u_A^{x,y}(N_1+1,n_2)=u_A^{x,y}(n_1,N_2+1)\nonumber \\
=u_B^{x,y}(0,n_2)=u_B^{x,y}(n_1,0)= 0
\end{eqnarray}
for all $ n_1=1,2,...,N_1$ and $ n_2=1,2,...,N_2$.

To study the non-Hermitian skin effect of bulk modes, we have to introduce the following three properties of the equation $\det(D-\lambda I)=0$. (1) For given $\lambda$ and $\beta_2$ ($\beta_1$), $\det (D-\lambda I)=0$ is the second-order equation of $\beta_1$ ($\beta_2$). Thus, one can write this equation as the following polynomial form of $\beta_1$ and $\beta_2$,
\begin{eqnarray}
\det(D-\lambda I) = a_1\beta_1+a_1'/\beta_1+a_2\beta_2+a_2'/\beta_2\nonumber \\
+a_3\beta_1/\beta_2+a_3'\beta_2/\beta_1+a_4=0,\qquad
\end{eqnarray}
where $a_{1,2,3}$, $a_{1,2,3}'$ and $a_4$ are constants given by $D$ and $\lambda$. (2) If $\lambda$ is real, we can prove that $a_{1,2,3}$, $a'_{1,2,3}$ and $a_4$ are real, and they satisfy $a_1=a_1'$, $a_2=a_2'$, and $a_3=a_3'$. (3) If $\det(D(\beta_1, \beta_2)-\lambda I)=0$, then $\det(D(\beta_1^{-1}, \beta_2^{-1})-\lambda I)=0$ as well.

Equipped with the above three properties, we are now ready to study the non-Hermitian skin effect. Given $\lambda$ and $\beta_2$, solving the equation $\det(D(\beta_1, \beta_2)-\lambda I)=0$ gives us two solutions $\beta_1^{(1)}$ and $\beta_1^{(2)}$ which satisfy
\begin{eqnarray}\label{3.2}
\beta_1^{(1,2)} = -(1\mp \sqrt{1-{4AC}/{B^2}})(B/2A),
\end{eqnarray}
and
\begin{eqnarray}\label{3.4}
\beta_1^{(1)}\beta_1^{(2)} = C/A, 
\end{eqnarray}
where $A = a_1+{a_3}/{\beta_2}$, $B = a_2\beta_2+{a_2'}/{\beta_2}+a_4$ and $C = a_1'+a_3'\beta_2$. The eigenvector which corresponds to $\beta_1^{(i)}$ is denoted as $u^{(i)}$. Thus, the wave function of the general form is 
\begin{eqnarray}
u_{n_1n_2} = \beta_2^{n_2} (\beta_1^{(1)n_1}u^{(1)}+\beta_1^{(2)n_1}u^{(2)}).
\end{eqnarray}
Employing fixed boundary conditions and eliminating $u^{(i)}$ gives us the simplified relation between $\beta_1^{(1)}$ and $\beta_1^{(2)}$,
\begin{eqnarray}
 \det \left[
 \frac{h(\beta_1^{(1)-1},\beta_2^{-1})}{(\beta_1^{(1)})^{N_1+1}}
-
 \frac{h(\beta_1^{(2)-1},\beta_2^{-1})}{(\beta_1^{(2)})^{N_1+1}}
\right]=0.\qquad
\end{eqnarray}
Similar as the rotor lattices, the bulk mode condition requires $|\beta_1^{(1)}|=|\beta_1^{(2)}|$, which in turn demands
\begin{eqnarray}\label{3.10}
1-4AC/B^2<0 \qquad {\rm and}\qquad {\rm Im\,}(AC/B^2)=0\qquad
\end{eqnarray}
in Eq.(\ref{3.2}). We find that if $|\beta_2|=1$ and $\lambda$ is real, the conditions in Eq.(\ref{3.10}) are validated. Substituting $|\beta_2|=1$ into Eq.(\ref{3.4}) gives us $|\beta_1^{(1)}|=|\beta_1^{(2)}|=1$. In summary, we find 
\begin{eqnarray}\label{3.11}
|\beta_1|=|\beta_2|=1.
\end{eqnarray}
Eq.(\ref{3.11}) is obtained by solving $\det(D(\beta_1, \beta_2)-\lambda I)=0$ in terms of $\beta_1$, but solving this determinant in terms of $\beta_2$ offers us the same result. It means that instead of localizing on the lattice boundaries, all bulk modes are extended. The generalized Brillouin zone happens to be a unit circle. An important consequence of Eq.(\ref{3.11}) is that $\beta_i^* = \beta_i^{-1}$, which is essential for the symmetry property $\mathcal{I}D(\vec\beta)\mathcal{I}^{-1}=D(\vec\beta^*)$ which quantizes the generalized Berry phase. Despite the fact that all bulk modes are extended, the dynamical matrix of the honeycomb lattice is still non-Hermitian.

\section{Reality condition of mechanical waves at exceptional points}
At the exceptional point $\vec\beta = \vec\beta_{\rm e}$, the eigenvalue $\lambda_n(\vec\beta_{\rm e})$ as well as the eigenvector $|n^R(\vec\beta_{\rm e})\rangle$ coalesce. The Newton's equation of motion for the exceptional point reads 
\begin{eqnarray}\label{EP1}
D(\vec\beta_{\rm e}) |n^R(\vec\beta_{\rm e})\rangle=\lambda_{n}(\vec\beta_{\rm e})|n^R(\vec\beta_{\rm e})\rangle.
\end{eqnarray}
The complex conjugation of Eq.(\ref{EP1}) gives another equation of motion, 
\begin{eqnarray}\label{EP2}
D(\vec\beta_{\rm e}^*) |n^R(\vec\beta_{\rm e})\rangle^*=\lambda_{n}^*(\vec\beta_{\rm e})|n^R(\vec\beta_{\rm e})\rangle^*,
\end{eqnarray}
where we have employed the property $D^*(\vec\beta_{\rm e})=D(\vec\beta_{\rm e}^*)$ in real space $D^*=D$. From Eq.(\ref{EP2}) it is straightforward to prove that $\vec\beta_{\rm e}^*$ is also an exceptional point with the corresponding eigenvalue $\lambda_n(\vec\beta_{\rm e}^*) = \lambda^*_n(\vec\beta_{\rm e})$ and the eigenvector $|n^R(\vec\beta_{\rm e}^*)\rangle = |n^R(\vec\beta_{\rm e})\rangle^*$. 

If $\lambda_n(\vec\beta_{\rm e})$ is real, the eigenvalues of the exceptional points $\vec\beta_{\rm e}$ and $\vec\beta_{\rm e}^*$ must degenerate: $\lambda_n(\vec\beta_{\rm e})=\lambda_n^*(\vec\beta_{\rm e})$. Hence, the linear combination of the eigenmodes $|n^R(\vec \beta_{\rm e})\rangle$ and $|n^R(\vec \beta^*_{\rm e})\rangle$ indicates that all particle displacements are real.

\section{The configurations of Honeycomb lattice with complex eigenfrequencies of topological modes}
In order to find a honeycomb lattice configuration that possesses complex eigenfrequencies of topological boundary modes, this four-band non-Hermitian dynamical matrix has to fulfill five criteria. (1) The first and the second two bands are separated by the bandgap $\Delta$. (2) The generalized Berry phase of the first two bands $(\gamma_1, \gamma_2) = (\pi,0)$ or $(\gamma_1, \gamma_2) = (0,\pi)$. (3) The eigenfrequencies of topological edge modes are complex. (4) The critical damping coefficient $\eta_c$, the topological mode eigenfrequencies $\omega_{\rm topo}$, and bulk mode eigenfrequencies $\omega_{\rm bulk}$ have to satisfy the following relationship
\begin{eqnarray}\label{eta5}
\eta_c & = & {\rm max} \,\{m\, {\rm Im}(\omega_{\rm topo}^2)/[{\rm Re}(\omega_{\rm topo}^2)]^{1/2}\}\nonumber \\
 & > & {\rm max} \,\{m\, {\rm Im}(\omega_{\rm bulk}^2)/[{\rm Re}(\omega_{\rm bulk}^2)]^{1/2}\}.
\end{eqnarray}
Eq.(\ref{eta5}) implies that the excitation of topological boundary modes is marginal. The amplitude exponentially grows in time when $\eta \lesssim \eta_c$, while it yields a constant if $\eta\gtrsim\eta_c$. The excitations of bulk modes are bounded in both cases. (5) Driven an external shaking with the frequency $\omega_{\rm ext} = {\rm Re\,}(\omega_{\rm topo})$, both of the topological modes and the bulk modes are excited due to the damping. We impose a lower bound to the bandgap $\Delta \gg \eta \,{\rm Re\,}(\omega_{\rm topo})$ (here we let $\Delta \gtrapprox 5 \eta \,{\rm Re\,}(\omega_{\rm topo})$ in the main text), and then the excitations of bulk modes are weak compared to the topological modes.

Most of the topological lattice configurations which fulfil criteria (1), (2) and (3) do not accomplish the rest two criteria (4) and (5). To seek the parameter region which satisfies all five criteria presented above, we start from a special configuration in which all eigenvalues of this non-Hermitian dynamical matrix are real and positive. Small parameter deviations from this configuration render eigenfrequencies with small imaginary parts, which means a small damping $\eta$ is strong enough to counteract the energy injection, and then criterion (5) is assured. 
Finally, we go through the neighborhood of this special configuration to find the parameter region which satisfies criterion (4). One of such a set of parameters which satisfy all five criteria is depicted by $P_0$ in Fig.\ref{Fig3}(b) of the main text.

We now discuss how to find this special configuration, in which all eigenvalues $\lambda=m\omega^2$ are real and positive. This configuration can be achieved by enabling $h(1,1)$ of Eq.(\ref{HCD}) to be an identity, which in turn demands 
\begin{eqnarray}\label{special}
 & {} & \sum_{i=1}^3 (k_i\cos 2\theta_i-k_i^o \sin 2\theta_i) =\nonumber \\
 & {} & \sum_{i=1}^3 (k_i\sin 2\theta_i+k_i^o \cos 2\theta_i) = \sum_{i=1}^3 k_i^o = 0.
\end{eqnarray}
Given the normal spring constants $k_1$, $k_2$ and $k_3$ and bond orientations $\theta_1$, $\theta_2$ and $\theta_3$, one can determine the odd elastic constants $k_1^o$, $k_2^o$ and $k_3^o$ accordingly. Through analytic calculations, we prove that all bulk mode eigenvalues are real and positive. We then numerically solve the topological boundary modes and find that their eigenvalues are real and positive, too. In summary, all eigenvalues are real and positive numbers in this special configuration.

Deviations of the parameters $k_1$, $k_2$, $k_3$, $\theta_1$, $\theta_2$, $\theta_3$, $k_1^o$, $k_2^o$, and $k_3^o$ lead to a wide range of lattice configurations with complex topological mode eigenfrequencies, which is depicted in Fig.\ref{Fig3}(a).


\begin{thebibliography}{118}%
\makeatletter
\providecommand \@ifxundefined [1]{%
 \@ifx{#1\undefined}
}%
\providecommand \@ifnum [1]{%
 \ifnum #1\expandafter \@firstoftwo
 \else \expandafter \@secondoftwo
 \fi
}%
\providecommand \@ifx [1]{%
 \ifx #1\expandafter \@firstoftwo
 \else \expandafter \@secondoftwo
 \fi
}%
\providecommand \natexlab [1]{#1}%
\providecommand \enquote  [1]{``#1''}%
\providecommand \bibnamefont  [1]{#1}%
\providecommand \bibfnamefont [1]{#1}%
\providecommand \citenamefont [1]{#1}%
\providecommand \href@noop [0]{\@secondoftwo}%
\providecommand \href [0]{\begingroup \@sanitize@url \@href}%
\providecommand \@href[1]{\@@startlink{#1}\@@href}%
\providecommand \@@href[1]{\endgroup#1\@@endlink}%
\providecommand \@sanitize@url [0]{\catcode `\\12\catcode `\$12\catcode
  `\&12\catcode `\#12\catcode `\^12\catcode `\_12\catcode `\%12\relax}%
\providecommand \@@startlink[1]{}%
\providecommand \@@endlink[0]{}%
\providecommand \url  [0]{\begingroup\@sanitize@url \@url }%
\providecommand \@url [1]{\endgroup\@href {#1}{\urlprefix }}%
\providecommand \urlprefix  [0]{URL }%
\providecommand \Eprint [0]{\href }%
\providecommand \doibase [0]{http://dx.doi.org/}%
\providecommand \selectlanguage [0]{\@gobble}%
\providecommand \bibinfo  [0]{\@secondoftwo}%
\providecommand \bibfield  [0]{\@secondoftwo}%
\providecommand \translation [1]{[#1]}%
\providecommand \BibitemOpen [0]{}%
\providecommand \bibitemStop [0]{}%
\providecommand \bibitemNoStop [0]{.\EOS\space}%
\providecommand \EOS [0]{\spacefactor3000\relax}%
\providecommand \BibitemShut  [1]{\csname bibitem#1\endcsname}%
\let\auto@bib@innerbib\@empty
\bibitem [{\citenamefont {Huber}(2016)}]{huber2016topological}%
  \BibitemOpen
  \bibfield  {author} {\bibinfo {author} {\bibfnamefont {S.~D.}\ \bibnamefont
  {Huber}},\ }\href@noop {} {\bibfield  {journal} {\bibinfo  {journal} {Nature
  Physics}\ }\textbf {\bibinfo {volume} {12}},\ \bibinfo {pages} {621}
  (\bibinfo {year} {2016})}\BibitemShut {NoStop}%
\bibitem [{\citenamefont {S{\"u}sstrunk}\ and\ \citenamefont
  {Huber}(2015)}]{susstrunk2015observation}%
  \BibitemOpen
  \bibfield  {author} {\bibinfo {author} {\bibfnamefont {R.}~\bibnamefont
  {S{\"u}sstrunk}}\ and\ \bibinfo {author} {\bibfnamefont {S.~D.}\ \bibnamefont
  {Huber}},\ }\href@noop {} {\bibfield  {journal} {\bibinfo  {journal}
  {Science}\ }\textbf {\bibinfo {volume} {349}},\ \bibinfo {pages} {47}
  (\bibinfo {year} {2015})}\BibitemShut {NoStop}%
\bibitem [{\citenamefont {Kane}\ and\ \citenamefont
  {Lubensky}(2014)}]{kane2014topological}%
  \BibitemOpen
  \bibfield  {author} {\bibinfo {author} {\bibfnamefont {C.}~\bibnamefont
  {Kane}}\ and\ \bibinfo {author} {\bibfnamefont {T.}~\bibnamefont
  {Lubensky}},\ }\href@noop {} {\bibfield  {journal} {\bibinfo  {journal}
  {Nature Physics}\ }\textbf {\bibinfo {volume} {10}},\ \bibinfo {pages} {39}
  (\bibinfo {year} {2014})}\BibitemShut {NoStop}%
\bibitem [{\citenamefont {Yang}\ \emph {et~al.}(2015)\citenamefont {Yang},
  \citenamefont {Gao}, \citenamefont {Shi}, \citenamefont {Lin}, \citenamefont
  {Gao}, \citenamefont {Chong},\ and\ \citenamefont
  {Zhang}}]{yang2015topological}%
  \BibitemOpen
  \bibfield  {author} {\bibinfo {author} {\bibfnamefont {Z.}~\bibnamefont
  {Yang}}, \bibinfo {author} {\bibfnamefont {F.}~\bibnamefont {Gao}}, \bibinfo
  {author} {\bibfnamefont {X.}~\bibnamefont {Shi}}, \bibinfo {author}
  {\bibfnamefont {X.}~\bibnamefont {Lin}}, \bibinfo {author} {\bibfnamefont
  {Z.}~\bibnamefont {Gao}}, \bibinfo {author} {\bibfnamefont {Y.}~\bibnamefont
  {Chong}}, \ and\ \bibinfo {author} {\bibfnamefont {B.}~\bibnamefont
  {Zhang}},\ }\href@noop {} {\bibfield  {journal} {\bibinfo  {journal}
  {Physical review letters}\ }\textbf {\bibinfo {volume} {114}},\ \bibinfo
  {pages} {114301} (\bibinfo {year} {2015})}\BibitemShut {NoStop}%
\bibitem [{\citenamefont {Mao}\ and\ \citenamefont
  {Lubensky}(2018)}]{mao2018maxwell}%
  \BibitemOpen
  \bibfield  {author} {\bibinfo {author} {\bibfnamefont {X.}~\bibnamefont
  {Mao}}\ and\ \bibinfo {author} {\bibfnamefont {T.~C.}\ \bibnamefont
  {Lubensky}},\ }\href@noop {} {\bibfield  {journal} {\bibinfo  {journal}
  {Annual Review of Condensed Matter Physics}\ }\textbf {\bibinfo {volume}
  {9}},\ \bibinfo {pages} {413} (\bibinfo {year} {2018})}\BibitemShut {NoStop}%
\bibitem [{\citenamefont {Barlas}\ and\ \citenamefont
  {Prodan}(2018)}]{barlas2018topological}%
  \BibitemOpen
  \bibfield  {author} {\bibinfo {author} {\bibfnamefont {Y.}~\bibnamefont
  {Barlas}}\ and\ \bibinfo {author} {\bibfnamefont {E.}~\bibnamefont
  {Prodan}},\ }\href@noop {} {\bibfield  {journal} {\bibinfo  {journal}
  {Physical Review B}\ }\textbf {\bibinfo {volume} {98}},\ \bibinfo {pages}
  {094310} (\bibinfo {year} {2018})}\BibitemShut {NoStop}%
\bibitem [{\citenamefont {Nash}\ \emph {et~al.}(2015)\citenamefont {Nash},
  \citenamefont {Kleckner}, \citenamefont {Read}, \citenamefont {Vitelli},
  \citenamefont {Turner},\ and\ \citenamefont {Irvine}}]{nash2015topological}%
  \BibitemOpen
  \bibfield  {author} {\bibinfo {author} {\bibfnamefont {L.~M.}\ \bibnamefont
  {Nash}}, \bibinfo {author} {\bibfnamefont {D.}~\bibnamefont {Kleckner}},
  \bibinfo {author} {\bibfnamefont {A.}~\bibnamefont {Read}}, \bibinfo {author}
  {\bibfnamefont {V.}~\bibnamefont {Vitelli}}, \bibinfo {author} {\bibfnamefont
  {A.~M.}\ \bibnamefont {Turner}}, \ and\ \bibinfo {author} {\bibfnamefont
  {W.~T.}\ \bibnamefont {Irvine}},\ }\href@noop {} {\bibfield  {journal}
  {\bibinfo  {journal} {Proceedings of the National Academy of Sciences}\
  }\textbf {\bibinfo {volume} {112}},\ \bibinfo {pages} {14495} (\bibinfo
  {year} {2015})}\BibitemShut {NoStop}%
\bibitem [{\citenamefont {Zhou}\ \emph {et~al.}(2020)\citenamefont {Zhou},
  \citenamefont {Ma}, \citenamefont {Sun}, \citenamefont {Gonella},\ and\
  \citenamefont {Mao}}]{PhysRevB.101.104106}%
  \BibitemOpen
  \bibfield  {author} {\bibinfo {author} {\bibfnamefont {D.}~\bibnamefont
  {Zhou}}, \bibinfo {author} {\bibfnamefont {J.}~\bibnamefont {Ma}}, \bibinfo
  {author} {\bibfnamefont {K.}~\bibnamefont {Sun}}, \bibinfo {author}
  {\bibfnamefont {S.}~\bibnamefont {Gonella}}, \ and\ \bibinfo {author}
  {\bibfnamefont {X.}~\bibnamefont {Mao}},\ }\href {\doibase
  10.1103/PhysRevB.101.104106} {\bibfield  {journal} {\bibinfo  {journal}
  {Phys. Rev. B}\ }\textbf {\bibinfo {volume} {101}},\ \bibinfo {pages}
  {104106} (\bibinfo {year} {2020})}\BibitemShut {NoStop}%
\bibitem [{\citenamefont {Rocklin}\ \emph {et~al.}(2017)\citenamefont
  {Rocklin}, \citenamefont {Zhou}, \citenamefont {Sun},\ and\ \citenamefont
  {Mao}}]{rocklin2017transformable}%
  \BibitemOpen
  \bibfield  {author} {\bibinfo {author} {\bibfnamefont {D.~Z.}\ \bibnamefont
  {Rocklin}}, \bibinfo {author} {\bibfnamefont {S.}~\bibnamefont {Zhou}},
  \bibinfo {author} {\bibfnamefont {K.}~\bibnamefont {Sun}}, \ and\ \bibinfo
  {author} {\bibfnamefont {X.}~\bibnamefont {Mao}},\ }\href@noop {} {\bibfield
  {journal} {\bibinfo  {journal} {Nature communications}\ }\textbf {\bibinfo
  {volume} {8}},\ \bibinfo {pages} {14201} (\bibinfo {year}
  {2017})}\BibitemShut {NoStop}%
\bibitem [{\citenamefont {Meeussen}\ \emph {et~al.}(2016)\citenamefont
  {Meeussen}, \citenamefont {Paulose},\ and\ \citenamefont
  {Vitelli}}]{meeussen2016geared}%
  \BibitemOpen
  \bibfield  {author} {\bibinfo {author} {\bibfnamefont {A.~S.}\ \bibnamefont
  {Meeussen}}, \bibinfo {author} {\bibfnamefont {J.}~\bibnamefont {Paulose}}, \
  and\ \bibinfo {author} {\bibfnamefont {V.}~\bibnamefont {Vitelli}},\
  }\href@noop {} {\bibfield  {journal} {\bibinfo  {journal} {Physical Review
  X}\ }\textbf {\bibinfo {volume} {6}},\ \bibinfo {pages} {041029} (\bibinfo
  {year} {2016})}\BibitemShut {NoStop}%
\bibitem [{\citenamefont {Paulose}\ \emph {et~al.}(2015)\citenamefont
  {Paulose}, \citenamefont {Chen},\ and\ \citenamefont
  {Vitelli}}]{paulose2015topological}%
  \BibitemOpen
  \bibfield  {author} {\bibinfo {author} {\bibfnamefont {J.}~\bibnamefont
  {Paulose}}, \bibinfo {author} {\bibfnamefont {B.~G.-g.}\ \bibnamefont
  {Chen}}, \ and\ \bibinfo {author} {\bibfnamefont {V.}~\bibnamefont
  {Vitelli}},\ }\href@noop {} {\bibfield  {journal} {\bibinfo  {journal}
  {Nature Physics}\ }\textbf {\bibinfo {volume} {11}},\ \bibinfo {pages} {153}
  (\bibinfo {year} {2015})}\BibitemShut {NoStop}%
\bibitem [{\citenamefont {Zhou}\ \emph
  {et~al.}(2018{\natexlab{a}})\citenamefont {Zhou}, \citenamefont {Zhang},\
  and\ \citenamefont {Mao}}]{zhou2018topological}%
  \BibitemOpen
  \bibfield  {author} {\bibinfo {author} {\bibfnamefont {D.}~\bibnamefont
  {Zhou}}, \bibinfo {author} {\bibfnamefont {L.}~\bibnamefont {Zhang}}, \ and\
  \bibinfo {author} {\bibfnamefont {X.}~\bibnamefont {Mao}},\ }\href@noop {}
  {\bibfield  {journal} {\bibinfo  {journal} {Physical review letters}\
  }\textbf {\bibinfo {volume} {120}},\ \bibinfo {pages} {068003} (\bibinfo
  {year} {2018}{\natexlab{a}})}\BibitemShut {NoStop}%
\bibitem [{\citenamefont {Zhou}\ \emph {et~al.}(2019)\citenamefont {Zhou},
  \citenamefont {Zhang},\ and\ \citenamefont {Mao}}]{zhou2019topological}%
  \BibitemOpen
  \bibfield  {author} {\bibinfo {author} {\bibfnamefont {D.}~\bibnamefont
  {Zhou}}, \bibinfo {author} {\bibfnamefont {L.}~\bibnamefont {Zhang}}, \ and\
  \bibinfo {author} {\bibfnamefont {X.}~\bibnamefont {Mao}},\ }\href@noop {}
  {\bibfield  {journal} {\bibinfo  {journal} {Physical Review X}\ }\textbf
  {\bibinfo {volume} {9}},\ \bibinfo {pages} {021054} (\bibinfo {year}
  {2019})}\BibitemShut {NoStop}%
\bibitem [{\citenamefont {Bandres}\ \emph {et~al.}(2016)\citenamefont
  {Bandres}, \citenamefont {Rechtsman},\ and\ \citenamefont
  {Segev}}]{bandres2016topological}%
  \BibitemOpen
  \bibfield  {author} {\bibinfo {author} {\bibfnamefont {M.~A.}\ \bibnamefont
  {Bandres}}, \bibinfo {author} {\bibfnamefont {M.~C.}\ \bibnamefont
  {Rechtsman}}, \ and\ \bibinfo {author} {\bibfnamefont {M.}~\bibnamefont
  {Segev}},\ }\href@noop {} {\bibfield  {journal} {\bibinfo  {journal}
  {Physical Review X}\ }\textbf {\bibinfo {volume} {6}},\ \bibinfo {pages}
  {011016} (\bibinfo {year} {2016})}\BibitemShut {NoStop}%
\bibitem [{\citenamefont {Tran}\ \emph {et~al.}(2015)\citenamefont {Tran},
  \citenamefont {Dauphin}, \citenamefont {Goldman},\ and\ \citenamefont
  {Gaspard}}]{tran2015topological}%
  \BibitemOpen
  \bibfield  {author} {\bibinfo {author} {\bibfnamefont {D.-T.}\ \bibnamefont
  {Tran}}, \bibinfo {author} {\bibfnamefont {A.}~\bibnamefont {Dauphin}},
  \bibinfo {author} {\bibfnamefont {N.}~\bibnamefont {Goldman}}, \ and\
  \bibinfo {author} {\bibfnamefont {P.}~\bibnamefont {Gaspard}},\ }\href@noop
  {} {\bibfield  {journal} {\bibinfo  {journal} {Physical Review B}\ }\textbf
  {\bibinfo {volume} {91}},\ \bibinfo {pages} {085125} (\bibinfo {year}
  {2015})}\BibitemShut {NoStop}%
\bibitem [{\citenamefont {Varjas}\ \emph {et~al.}(2019)\citenamefont {Varjas},
  \citenamefont {Lau}, \citenamefont {P{\"o}yh{\"o}nen}, \citenamefont
  {Akhmerov}, \citenamefont {Pikulin},\ and\ \citenamefont
  {Fulga}}]{varjas2019topological}%
  \BibitemOpen
  \bibfield  {author} {\bibinfo {author} {\bibfnamefont {D.}~\bibnamefont
  {Varjas}}, \bibinfo {author} {\bibfnamefont {A.}~\bibnamefont {Lau}},
  \bibinfo {author} {\bibfnamefont {K.}~\bibnamefont {P{\"o}yh{\"o}nen}},
  \bibinfo {author} {\bibfnamefont {A.~R.}\ \bibnamefont {Akhmerov}}, \bibinfo
  {author} {\bibfnamefont {D.~I.}\ \bibnamefont {Pikulin}}, \ and\ \bibinfo
  {author} {\bibfnamefont {I.~C.}\ \bibnamefont {Fulga}},\ }\href@noop {}
  {\bibfield  {journal} {\bibinfo  {journal} {arXiv preprint arXiv:1904.07242}\
  } (\bibinfo {year} {2019})}\BibitemShut {NoStop}%
\bibitem [{\citenamefont {Mitchell}\ \emph
  {et~al.}(2018{\natexlab{a}})\citenamefont {Mitchell}, \citenamefont {Nash},
  \citenamefont {Hexner}, \citenamefont {Turner},\ and\ \citenamefont
  {Irvine}}]{mitchell2018amorphous}%
  \BibitemOpen
  \bibfield  {author} {\bibinfo {author} {\bibfnamefont {N.~P.}\ \bibnamefont
  {Mitchell}}, \bibinfo {author} {\bibfnamefont {L.~M.}\ \bibnamefont {Nash}},
  \bibinfo {author} {\bibfnamefont {D.}~\bibnamefont {Hexner}}, \bibinfo
  {author} {\bibfnamefont {A.~M.}\ \bibnamefont {Turner}}, \ and\ \bibinfo
  {author} {\bibfnamefont {W.~T.}\ \bibnamefont {Irvine}},\ }\href@noop {}
  {\bibfield  {journal} {\bibinfo  {journal} {Nature Physics}\ }\textbf
  {\bibinfo {volume} {14}},\ \bibinfo {pages} {380} (\bibinfo {year}
  {2018}{\natexlab{a}})}\BibitemShut {NoStop}%
\bibitem [{\citenamefont {Bender}\ and\ \citenamefont
  {Boettcher}(1998)}]{PhysRevLett.80.5243}%
  \BibitemOpen
  \bibfield  {author} {\bibinfo {author} {\bibfnamefont {C.~M.}\ \bibnamefont
  {Bender}}\ and\ \bibinfo {author} {\bibfnamefont {S.}~\bibnamefont
  {Boettcher}},\ }\href {\doibase 10.1103/PhysRevLett.80.5243} {\bibfield
  {journal} {\bibinfo  {journal} {Phys. Rev. Lett.}\ }\textbf {\bibinfo
  {volume} {80}},\ \bibinfo {pages} {5243} (\bibinfo {year}
  {1998})}\BibitemShut {NoStop}%
\bibitem [{\citenamefont {Heiss}(2012)}]{heiss2012physics}%
  \BibitemOpen
  \bibfield  {author} {\bibinfo {author} {\bibfnamefont {W.}~\bibnamefont
  {Heiss}},\ }\href@noop {} {\bibfield  {journal} {\bibinfo  {journal} {Journal
  of Physics A: Mathematical and Theoretical}\ }\textbf {\bibinfo {volume}
  {45}},\ \bibinfo {pages} {444016} (\bibinfo {year} {2012})}\BibitemShut
  {NoStop}%
\bibitem [{\citenamefont {Lee}(2016)}]{lee2016anomalous}%
  \BibitemOpen
  \bibfield  {author} {\bibinfo {author} {\bibfnamefont {T.~E.}\ \bibnamefont
  {Lee}},\ }\href@noop {} {\bibfield  {journal} {\bibinfo  {journal} {Physical
  review letters}\ }\textbf {\bibinfo {volume} {116}},\ \bibinfo {pages}
  {133903} (\bibinfo {year} {2016})}\BibitemShut {NoStop}%
\bibitem [{\citenamefont {Xu}\ \emph {et~al.}(2017)\citenamefont {Xu},
  \citenamefont {Wang},\ and\ \citenamefont {Duan}}]{xu2017weyl}%
  \BibitemOpen
  \bibfield  {author} {\bibinfo {author} {\bibfnamefont {Y.}~\bibnamefont
  {Xu}}, \bibinfo {author} {\bibfnamefont {S.-T.}\ \bibnamefont {Wang}}, \ and\
  \bibinfo {author} {\bibfnamefont {L.-M.}\ \bibnamefont {Duan}},\ }\href@noop
  {} {\bibfield  {journal} {\bibinfo  {journal} {Physical review letters}\
  }\textbf {\bibinfo {volume} {118}},\ \bibinfo {pages} {045701} (\bibinfo
  {year} {2017})}\BibitemShut {NoStop}%
\bibitem [{\citenamefont {Esaki}\ \emph {et~al.}(2011)\citenamefont {Esaki},
  \citenamefont {Sato}, \citenamefont {Hasebe},\ and\ \citenamefont
  {Kohmoto}}]{PhysRevB.84.205128}%
  \BibitemOpen
  \bibfield  {author} {\bibinfo {author} {\bibfnamefont {K.}~\bibnamefont
  {Esaki}}, \bibinfo {author} {\bibfnamefont {M.}~\bibnamefont {Sato}},
  \bibinfo {author} {\bibfnamefont {K.}~\bibnamefont {Hasebe}}, \ and\ \bibinfo
  {author} {\bibfnamefont {M.}~\bibnamefont {Kohmoto}},\ }\href {\doibase
  10.1103/PhysRevB.84.205128} {\bibfield  {journal} {\bibinfo  {journal} {Phys.
  Rev. B}\ }\textbf {\bibinfo {volume} {84}},\ \bibinfo {pages} {205128}
  (\bibinfo {year} {2011})}\BibitemShut {NoStop}%
\bibitem [{\citenamefont {Leykam}\ \emph {et~al.}(2017)\citenamefont {Leykam},
  \citenamefont {Bliokh}, \citenamefont {Huang}, \citenamefont {Chong},\ and\
  \citenamefont {Nori}}]{PhysRevLett.118.040401}%
  \BibitemOpen
  \bibfield  {author} {\bibinfo {author} {\bibfnamefont {D.}~\bibnamefont
  {Leykam}}, \bibinfo {author} {\bibfnamefont {K.~Y.}\ \bibnamefont {Bliokh}},
  \bibinfo {author} {\bibfnamefont {C.}~\bibnamefont {Huang}}, \bibinfo
  {author} {\bibfnamefont {Y.~D.}\ \bibnamefont {Chong}}, \ and\ \bibinfo
  {author} {\bibfnamefont {F.}~\bibnamefont {Nori}},\ }\href {\doibase
  10.1103/PhysRevLett.118.040401} {\bibfield  {journal} {\bibinfo  {journal}
  {Phys. Rev. Lett.}\ }\textbf {\bibinfo {volume} {118}},\ \bibinfo {pages}
  {040401} (\bibinfo {year} {2017})}\BibitemShut {NoStop}%
\bibitem [{\citenamefont {Lieu}(2018)}]{PhysRevB.97.045106}%
  \BibitemOpen
  \bibfield  {author} {\bibinfo {author} {\bibfnamefont {S.}~\bibnamefont
  {Lieu}},\ }\href {\doibase 10.1103/PhysRevB.97.045106} {\bibfield  {journal}
  {\bibinfo  {journal} {Phys. Rev. B}\ }\textbf {\bibinfo {volume} {97}},\
  \bibinfo {pages} {045106} (\bibinfo {year} {2018})}\BibitemShut {NoStop}%
\bibitem [{\citenamefont {Kunst}\ \emph {et~al.}(2018)\citenamefont {Kunst},
  \citenamefont {Edvardsson}, \citenamefont {Budich},\ and\ \citenamefont
  {Bergholtz}}]{PhysRevLett.121.026808}%
  \BibitemOpen
  \bibfield  {author} {\bibinfo {author} {\bibfnamefont {F.~K.}\ \bibnamefont
  {Kunst}}, \bibinfo {author} {\bibfnamefont {E.}~\bibnamefont {Edvardsson}},
  \bibinfo {author} {\bibfnamefont {J.~C.}\ \bibnamefont {Budich}}, \ and\
  \bibinfo {author} {\bibfnamefont {E.~J.}\ \bibnamefont {Bergholtz}},\ }\href
  {\doibase 10.1103/PhysRevLett.121.026808} {\bibfield  {journal} {\bibinfo
  {journal} {Phys. Rev. Lett.}\ }\textbf {\bibinfo {volume} {121}},\ \bibinfo
  {pages} {026808} (\bibinfo {year} {2018})}\BibitemShut {NoStop}%
\bibitem [{\citenamefont {Yao}\ and\ \citenamefont {Wang}(2018)}]{yao2018edge}%
  \BibitemOpen
  \bibfield  {author} {\bibinfo {author} {\bibfnamefont {S.}~\bibnamefont
  {Yao}}\ and\ \bibinfo {author} {\bibfnamefont {Z.}~\bibnamefont {Wang}},\
  }\href@noop {} {\bibfield  {journal} {\bibinfo  {journal} {Physical review
  letters}\ }\textbf {\bibinfo {volume} {121}},\ \bibinfo {pages} {086803}
  (\bibinfo {year} {2018})}\BibitemShut {NoStop}%
\bibitem [{\citenamefont {Yao}\ \emph {et~al.}(2018)\citenamefont {Yao},
  \citenamefont {Song},\ and\ \citenamefont {Wang}}]{yao2018non}%
  \BibitemOpen
  \bibfield  {author} {\bibinfo {author} {\bibfnamefont {S.}~\bibnamefont
  {Yao}}, \bibinfo {author} {\bibfnamefont {F.}~\bibnamefont {Song}}, \ and\
  \bibinfo {author} {\bibfnamefont {Z.}~\bibnamefont {Wang}},\ }\href@noop {}
  {\bibfield  {journal} {\bibinfo  {journal} {Physical review letters}\
  }\textbf {\bibinfo {volume} {121}},\ \bibinfo {pages} {136802} (\bibinfo
  {year} {2018})}\BibitemShut {NoStop}%
\bibitem [{\citenamefont {Song}\ \emph {et~al.}(2019)\citenamefont {Song},
  \citenamefont {Yao},\ and\ \citenamefont {Wang}}]{song2019non}%
  \BibitemOpen
  \bibfield  {author} {\bibinfo {author} {\bibfnamefont {F.}~\bibnamefont
  {Song}}, \bibinfo {author} {\bibfnamefont {S.}~\bibnamefont {Yao}}, \ and\
  \bibinfo {author} {\bibfnamefont {Z.}~\bibnamefont {Wang}},\ }\href@noop {}
  {\bibfield  {journal} {\bibinfo  {journal} {arXiv preprint arXiv:1905.02211}\
  } (\bibinfo {year} {2019})}\BibitemShut {NoStop}%
\bibitem [{\citenamefont {Shen}\ \emph {et~al.}(2018)\citenamefont {Shen},
  \citenamefont {Zhen},\ and\ \citenamefont {Fu}}]{shen2018topological}%
  \BibitemOpen
  \bibfield  {author} {\bibinfo {author} {\bibfnamefont {H.}~\bibnamefont
  {Shen}}, \bibinfo {author} {\bibfnamefont {B.}~\bibnamefont {Zhen}}, \ and\
  \bibinfo {author} {\bibfnamefont {L.}~\bibnamefont {Fu}},\ }\href@noop {}
  {\bibfield  {journal} {\bibinfo  {journal} {Physical review letters}\
  }\textbf {\bibinfo {volume} {120}},\ \bibinfo {pages} {146402} (\bibinfo
  {year} {2018})}\BibitemShut {NoStop}%
\bibitem [{\citenamefont {Yuce}(2018)}]{yuce2018edge}%
  \BibitemOpen
  \bibfield  {author} {\bibinfo {author} {\bibfnamefont {C.}~\bibnamefont
  {Yuce}},\ }\href@noop {} {\bibfield  {journal} {\bibinfo  {journal} {Physical
  Review A}\ }\textbf {\bibinfo {volume} {97}},\ \bibinfo {pages} {042118}
  (\bibinfo {year} {2018})}\BibitemShut {NoStop}%
\bibitem [{\citenamefont {Lee}\ and\ \citenamefont
  {Thomale}(2019)}]{PhysRevB.99.201103}%
  \BibitemOpen
  \bibfield  {author} {\bibinfo {author} {\bibfnamefont {C.~H.}\ \bibnamefont
  {Lee}}\ and\ \bibinfo {author} {\bibfnamefont {R.}~\bibnamefont {Thomale}},\
  }\href {\doibase 10.1103/PhysRevB.99.201103} {\bibfield  {journal} {\bibinfo
  {journal} {Phys. Rev. B}\ }\textbf {\bibinfo {volume} {99}},\ \bibinfo
  {pages} {201103} (\bibinfo {year} {2019})}\BibitemShut {NoStop}%
\bibitem [{\citenamefont {Zhou}\ \emph
  {et~al.}(2018{\natexlab{b}})\citenamefont {Zhou}, \citenamefont {Peng},
  \citenamefont {Yoon}, \citenamefont {Hsu}, \citenamefont {Nelson},
  \citenamefont {Fu}, \citenamefont {Joannopoulos}, \citenamefont
  {Solja{\v{c}}i{\'c}},\ and\ \citenamefont {Zhen}}]{zhou2018observation}%
  \BibitemOpen
  \bibfield  {author} {\bibinfo {author} {\bibfnamefont {H.}~\bibnamefont
  {Zhou}}, \bibinfo {author} {\bibfnamefont {C.}~\bibnamefont {Peng}}, \bibinfo
  {author} {\bibfnamefont {Y.}~\bibnamefont {Yoon}}, \bibinfo {author}
  {\bibfnamefont {C.~W.}\ \bibnamefont {Hsu}}, \bibinfo {author} {\bibfnamefont
  {K.~A.}\ \bibnamefont {Nelson}}, \bibinfo {author} {\bibfnamefont
  {L.}~\bibnamefont {Fu}}, \bibinfo {author} {\bibfnamefont {J.~D.}\
  \bibnamefont {Joannopoulos}}, \bibinfo {author} {\bibfnamefont
  {M.}~\bibnamefont {Solja{\v{c}}i{\'c}}}, \ and\ \bibinfo {author}
  {\bibfnamefont {B.}~\bibnamefont {Zhen}},\ }\href@noop {} {\bibfield
  {journal} {\bibinfo  {journal} {Science}\ }\textbf {\bibinfo {volume}
  {359}},\ \bibinfo {pages} {1009} (\bibinfo {year}
  {2018}{\natexlab{b}})}\BibitemShut {NoStop}%
\bibitem [{\citenamefont {Xiong}(2018)}]{xiong2018does}%
  \BibitemOpen
  \bibfield  {author} {\bibinfo {author} {\bibfnamefont {Y.}~\bibnamefont
  {Xiong}},\ }\href@noop {} {\bibfield  {journal} {\bibinfo  {journal} {Journal
  of Physics Communications}\ }\textbf {\bibinfo {volume} {2}},\ \bibinfo
  {pages} {035043} (\bibinfo {year} {2018})}\BibitemShut {NoStop}%
\bibitem [{\citenamefont {Okuma}\ \emph {et~al.}(2019)\citenamefont {Okuma},
  \citenamefont {Kawabata}, \citenamefont {Shiozaki},\ and\ \citenamefont
  {Sato}}]{okuma2019topological}%
  \BibitemOpen
  \bibfield  {author} {\bibinfo {author} {\bibfnamefont {N.}~\bibnamefont
  {Okuma}}, \bibinfo {author} {\bibfnamefont {K.}~\bibnamefont {Kawabata}},
  \bibinfo {author} {\bibfnamefont {K.}~\bibnamefont {Shiozaki}}, \ and\
  \bibinfo {author} {\bibfnamefont {M.}~\bibnamefont {Sato}},\ }\href@noop {}
  {\bibfield  {journal} {\bibinfo  {journal} {arXiv preprint arXiv:1910.02878}\
  } (\bibinfo {year} {2019})}\BibitemShut {NoStop}%
\bibitem [{\citenamefont {Zhang}\ \emph
  {et~al.}(2019{\natexlab{a}})\citenamefont {Zhang}, \citenamefont {Yang},\
  and\ \citenamefont {Fang}}]{zhang2019correspondence}%
  \BibitemOpen
  \bibfield  {author} {\bibinfo {author} {\bibfnamefont {K.}~\bibnamefont
  {Zhang}}, \bibinfo {author} {\bibfnamefont {Z.}~\bibnamefont {Yang}}, \ and\
  \bibinfo {author} {\bibfnamefont {C.}~\bibnamefont {Fang}},\ }\href@noop {}
  {\bibfield  {journal} {\bibinfo  {journal} {arXiv preprint arXiv:1910.01131}\
  } (\bibinfo {year} {2019}{\natexlab{a}})}\BibitemShut {NoStop}%
\bibitem [{\citenamefont {Lee}\ \emph {et~al.}(2019{\natexlab{a}})\citenamefont
  {Lee}, \citenamefont {Ahn}, \citenamefont {Zhou},\ and\ \citenamefont
  {Vishwanath}}]{lee2019topological}%
  \BibitemOpen
  \bibfield  {author} {\bibinfo {author} {\bibfnamefont {J.~Y.}\ \bibnamefont
  {Lee}}, \bibinfo {author} {\bibfnamefont {J.}~\bibnamefont {Ahn}}, \bibinfo
  {author} {\bibfnamefont {H.}~\bibnamefont {Zhou}}, \ and\ \bibinfo {author}
  {\bibfnamefont {A.}~\bibnamefont {Vishwanath}},\ }\href@noop {} {\bibfield
  {journal} {\bibinfo  {journal} {arXiv preprint arXiv:1906.08782}\ } (\bibinfo
  {year} {2019}{\natexlab{a}})}\BibitemShut {NoStop}%
\bibitem [{\citenamefont {Ghatak}\ and\ \citenamefont
  {Das}(2019)}]{ghatak2019new}%
  \BibitemOpen
  \bibfield  {author} {\bibinfo {author} {\bibfnamefont {A.}~\bibnamefont
  {Ghatak}}\ and\ \bibinfo {author} {\bibfnamefont {T.}~\bibnamefont {Das}},\
  }\href@noop {} {\bibfield  {journal} {\bibinfo  {journal} {Journal of
  Physics: Condensed Matter}\ }\textbf {\bibinfo {volume} {31}},\ \bibinfo
  {pages} {263001} (\bibinfo {year} {2019})}\BibitemShut {NoStop}%
\bibitem [{\citenamefont {Tserkovnyak}(2019)}]{tserkovnyak2019exceptional}%
  \BibitemOpen
  \bibfield  {author} {\bibinfo {author} {\bibfnamefont {Y.}~\bibnamefont
  {Tserkovnyak}},\ }\href@noop {} {\bibfield  {journal} {\bibinfo  {journal}
  {arXiv preprint arXiv:1911.01619}\ } (\bibinfo {year} {2019})}\BibitemShut
  {NoStop}%
\bibitem [{\citenamefont {Kawabata}\ \emph {et~al.}(2019)\citenamefont
  {Kawabata}, \citenamefont {Shiozaki}, \citenamefont {Ueda},\ and\
  \citenamefont {Sato}}]{kawabata2019symmetry}%
  \BibitemOpen
  \bibfield  {author} {\bibinfo {author} {\bibfnamefont {K.}~\bibnamefont
  {Kawabata}}, \bibinfo {author} {\bibfnamefont {K.}~\bibnamefont {Shiozaki}},
  \bibinfo {author} {\bibfnamefont {M.}~\bibnamefont {Ueda}}, \ and\ \bibinfo
  {author} {\bibfnamefont {M.}~\bibnamefont {Sato}},\ }\href@noop {} {\bibfield
   {journal} {\bibinfo  {journal} {Physical Review X}\ }\textbf {\bibinfo
  {volume} {9}},\ \bibinfo {pages} {041015} (\bibinfo {year}
  {2019})}\BibitemShut {NoStop}%
\bibitem [{\citenamefont {Gong}\ \emph {et~al.}(2018)\citenamefont {Gong},
  \citenamefont {Ashida}, \citenamefont {Kawabata}, \citenamefont {Takasan},
  \citenamefont {Higashikawa},\ and\ \citenamefont
  {Ueda}}]{gong2018topological}%
  \BibitemOpen
  \bibfield  {author} {\bibinfo {author} {\bibfnamefont {Z.}~\bibnamefont
  {Gong}}, \bibinfo {author} {\bibfnamefont {Y.}~\bibnamefont {Ashida}},
  \bibinfo {author} {\bibfnamefont {K.}~\bibnamefont {Kawabata}}, \bibinfo
  {author} {\bibfnamefont {K.}~\bibnamefont {Takasan}}, \bibinfo {author}
  {\bibfnamefont {S.}~\bibnamefont {Higashikawa}}, \ and\ \bibinfo {author}
  {\bibfnamefont {M.}~\bibnamefont {Ueda}},\ }\href@noop {} {\bibfield
  {journal} {\bibinfo  {journal} {Physical Review X}\ }\textbf {\bibinfo
  {volume} {8}},\ \bibinfo {pages} {031079} (\bibinfo {year}
  {2018})}\BibitemShut {NoStop}%
\bibitem [{\citenamefont
  {Longhi}(2019{\natexlab{a}})}]{PhysRevResearch.1.023013}%
  \BibitemOpen
  \bibfield  {author} {\bibinfo {author} {\bibfnamefont {S.}~\bibnamefont
  {Longhi}},\ }\href {\doibase 10.1103/PhysRevResearch.1.023013} {\bibfield
  {journal} {\bibinfo  {journal} {Phys. Rev. Research}\ }\textbf {\bibinfo
  {volume} {1}},\ \bibinfo {pages} {023013} (\bibinfo {year}
  {2019}{\natexlab{a}})}\BibitemShut {NoStop}%
\bibitem [{\citenamefont {Malzard}\ \emph {et~al.}(2015)\citenamefont
  {Malzard}, \citenamefont {Poli},\ and\ \citenamefont
  {Schomerus}}]{malzard2015topologically}%
  \BibitemOpen
  \bibfield  {author} {\bibinfo {author} {\bibfnamefont {S.}~\bibnamefont
  {Malzard}}, \bibinfo {author} {\bibfnamefont {C.}~\bibnamefont {Poli}}, \
  and\ \bibinfo {author} {\bibfnamefont {H.}~\bibnamefont {Schomerus}},\
  }\href@noop {} {\bibfield  {journal} {\bibinfo  {journal} {Physical review
  letters}\ }\textbf {\bibinfo {volume} {115}},\ \bibinfo {pages} {200402}
  (\bibinfo {year} {2015})}\BibitemShut {NoStop}%
\bibitem [{\citenamefont {Zhao}\ \emph {et~al.}(2019)\citenamefont {Zhao},
  \citenamefont {Chen}, \citenamefont {Fu},\ and\ \citenamefont
  {Yi}}]{zhao2019topological}%
  \BibitemOpen
  \bibfield  {author} {\bibinfo {author} {\bibfnamefont {X.}~\bibnamefont
  {Zhao}}, \bibinfo {author} {\bibfnamefont {L.}~\bibnamefont {Chen}}, \bibinfo
  {author} {\bibfnamefont {L.}~\bibnamefont {Fu}}, \ and\ \bibinfo {author}
  {\bibfnamefont {X.}~\bibnamefont {Yi}},\ }\href@noop {} {\bibfield  {journal}
  {\bibinfo  {journal} {arXiv preprint arXiv:1907.07924}\ } (\bibinfo {year}
  {2019})}\BibitemShut {NoStop}%
\bibitem [{\citenamefont {Yoshida}\ \emph
  {et~al.}(2019{\natexlab{a}})\citenamefont {Yoshida}, \citenamefont {Kudo},\
  and\ \citenamefont {Hatsugai}}]{yoshida2019non}%
  \BibitemOpen
  \bibfield  {author} {\bibinfo {author} {\bibfnamefont {T.}~\bibnamefont
  {Yoshida}}, \bibinfo {author} {\bibfnamefont {K.}~\bibnamefont {Kudo}}, \
  and\ \bibinfo {author} {\bibfnamefont {Y.}~\bibnamefont {Hatsugai}},\
  }\href@noop {} {\bibfield  {journal} {\bibinfo  {journal} {arXiv preprint
  arXiv:1907.07596}\ } (\bibinfo {year} {2019}{\natexlab{a}})}\BibitemShut
  {NoStop}%
\bibitem [{\citenamefont {Longhi}(2019{\natexlab{b}})}]{longhi2019topological}%
  \BibitemOpen
  \bibfield  {author} {\bibinfo {author} {\bibfnamefont {S.}~\bibnamefont
  {Longhi}},\ }\href@noop {} {\bibfield  {journal} {\bibinfo  {journal}
  {Physical Review Letters}\ }\textbf {\bibinfo {volume} {122}},\ \bibinfo
  {pages} {237601} (\bibinfo {year} {2019}{\natexlab{b}})}\BibitemShut
  {NoStop}%
\bibitem [{\citenamefont {Liu}\ and\ \citenamefont
  {Chen}(2019)}]{PhysRevB.100.144106}%
  \BibitemOpen
  \bibfield  {author} {\bibinfo {author} {\bibfnamefont {C.-H.}\ \bibnamefont
  {Liu}}\ and\ \bibinfo {author} {\bibfnamefont {S.}~\bibnamefont {Chen}},\
  }\href {\doibase 10.1103/PhysRevB.100.144106} {\bibfield  {journal} {\bibinfo
   {journal} {Phys. Rev. B}\ }\textbf {\bibinfo {volume} {100}},\ \bibinfo
  {pages} {144106} (\bibinfo {year} {2019})}\BibitemShut {NoStop}%
\bibitem [{\citenamefont {Bender}(2007)}]{bender2007making}%
  \BibitemOpen
  \bibfield  {author} {\bibinfo {author} {\bibfnamefont {C.~M.}\ \bibnamefont
  {Bender}},\ }\href@noop {} {\bibfield  {journal} {\bibinfo  {journal}
  {Reports on Progress in Physics}\ }\textbf {\bibinfo {volume} {70}},\
  \bibinfo {pages} {947} (\bibinfo {year} {2007})}\BibitemShut {NoStop}%
\bibitem [{\citenamefont {Wu}\ and\ \citenamefont
  {Hou}(2019)}]{PhysRevA.99.062107}%
  \BibitemOpen
  \bibfield  {author} {\bibinfo {author} {\bibfnamefont {Y.-J.}\ \bibnamefont
  {Wu}}\ and\ \bibinfo {author} {\bibfnamefont {J.}~\bibnamefont {Hou}},\
  }\href {\doibase 10.1103/PhysRevA.99.062107} {\bibfield  {journal} {\bibinfo
  {journal} {Phys. Rev. A}\ }\textbf {\bibinfo {volume} {99}},\ \bibinfo
  {pages} {062107} (\bibinfo {year} {2019})}\BibitemShut {NoStop}%
\bibitem [{\citenamefont {Chang}\ \emph {et~al.}(2019)\citenamefont {Chang},
  \citenamefont {You}, \citenamefont {Wen},\ and\ \citenamefont
  {Ryu}}]{chang2019entanglement}%
  \BibitemOpen
  \bibfield  {author} {\bibinfo {author} {\bibfnamefont {P.-Y.}\ \bibnamefont
  {Chang}}, \bibinfo {author} {\bibfnamefont {J.-S.}\ \bibnamefont {You}},
  \bibinfo {author} {\bibfnamefont {X.}~\bibnamefont {Wen}}, \ and\ \bibinfo
  {author} {\bibfnamefont {S.}~\bibnamefont {Ryu}},\ }\href@noop {} {\bibfield
  {journal} {\bibinfo  {journal} {arXiv preprint arXiv:1909.01346}\ } (\bibinfo
  {year} {2019})}\BibitemShut {NoStop}%
\bibitem [{\citenamefont {Yoshida}\ and\ \citenamefont
  {Hatsugai}(2019)}]{PhysRevB.100.054109}%
  \BibitemOpen
  \bibfield  {author} {\bibinfo {author} {\bibfnamefont {T.}~\bibnamefont
  {Yoshida}}\ and\ \bibinfo {author} {\bibfnamefont {Y.}~\bibnamefont
  {Hatsugai}},\ }\href {\doibase 10.1103/PhysRevB.100.054109} {\bibfield
  {journal} {\bibinfo  {journal} {Phys. Rev. B}\ }\textbf {\bibinfo {volume}
  {100}},\ \bibinfo {pages} {054109} (\bibinfo {year} {2019})}\BibitemShut
  {NoStop}%
\bibitem [{\citenamefont {Yoshida}\ \emph {et~al.}(2018)\citenamefont
  {Yoshida}, \citenamefont {Peters},\ and\ \citenamefont
  {Kawakami}}]{PhysRevB.98.035141}%
  \BibitemOpen
  \bibfield  {author} {\bibinfo {author} {\bibfnamefont {T.}~\bibnamefont
  {Yoshida}}, \bibinfo {author} {\bibfnamefont {R.}~\bibnamefont {Peters}}, \
  and\ \bibinfo {author} {\bibfnamefont {N.}~\bibnamefont {Kawakami}},\ }\href
  {\doibase 10.1103/PhysRevB.98.035141} {\bibfield  {journal} {\bibinfo
  {journal} {Phys. Rev. B}\ }\textbf {\bibinfo {volume} {98}},\ \bibinfo
  {pages} {035141} (\bibinfo {year} {2018})}\BibitemShut {NoStop}%
\bibitem [{\citenamefont {Yoshida}\ \emph
  {et~al.}(2019{\natexlab{b}})\citenamefont {Yoshida}, \citenamefont {Peters},
  \citenamefont {Kawakami},\ and\ \citenamefont
  {Hatsugai}}]{PhysRevB.99.121101}%
  \BibitemOpen
  \bibfield  {author} {\bibinfo {author} {\bibfnamefont {T.}~\bibnamefont
  {Yoshida}}, \bibinfo {author} {\bibfnamefont {R.}~\bibnamefont {Peters}},
  \bibinfo {author} {\bibfnamefont {N.}~\bibnamefont {Kawakami}}, \ and\
  \bibinfo {author} {\bibfnamefont {Y.}~\bibnamefont {Hatsugai}},\ }\href
  {\doibase 10.1103/PhysRevB.99.121101} {\bibfield  {journal} {\bibinfo
  {journal} {Phys. Rev. B}\ }\textbf {\bibinfo {volume} {99}},\ \bibinfo
  {pages} {121101} (\bibinfo {year} {2019}{\natexlab{b}})}\BibitemShut
  {NoStop}%
\bibitem [{\citenamefont {Borgnia}\ \emph {et~al.}(2019)\citenamefont
  {Borgnia}, \citenamefont {Kruchkov},\ and\ \citenamefont
  {Slager}}]{borgnia2019nonhermitian}%
  \BibitemOpen
  \bibfield  {author} {\bibinfo {author} {\bibfnamefont {D.~S.}\ \bibnamefont
  {Borgnia}}, \bibinfo {author} {\bibfnamefont {A.~J.}\ \bibnamefont
  {Kruchkov}}, \ and\ \bibinfo {author} {\bibfnamefont {R.-J.}\ \bibnamefont
  {Slager}},\ }\href@noop {} {\enquote {\bibinfo {title} {Non-hermitian
  boundary modes},}\ } (\bibinfo {year} {2019}),\ \Eprint
  {http://arxiv.org/abs/1902.07217} {arXiv:1902.07217 [cond-mat.mes-hall]}
  \BibitemShut {NoStop}%
\bibitem [{\citenamefont {Makris}\ \emph {et~al.}(2008)\citenamefont {Makris},
  \citenamefont {El-Ganainy}, \citenamefont {Christodoulides},\ and\
  \citenamefont {Musslimani}}]{makris2008beam}%
  \BibitemOpen
  \bibfield  {author} {\bibinfo {author} {\bibfnamefont {K.~G.}\ \bibnamefont
  {Makris}}, \bibinfo {author} {\bibfnamefont {R.}~\bibnamefont {El-Ganainy}},
  \bibinfo {author} {\bibfnamefont {D.}~\bibnamefont {Christodoulides}}, \ and\
  \bibinfo {author} {\bibfnamefont {Z.~H.}\ \bibnamefont {Musslimani}},\
  }\href@noop {} {\bibfield  {journal} {\bibinfo  {journal} {Physical Review
  Letters}\ }\textbf {\bibinfo {volume} {100}},\ \bibinfo {pages} {103904}
  (\bibinfo {year} {2008})}\BibitemShut {NoStop}%
\bibitem [{\citenamefont {Liu}\ \emph {et~al.}(2019)\citenamefont {Liu},
  \citenamefont {Zhang}, \citenamefont {Ai}, \citenamefont {Gong},
  \citenamefont {Kawabata}, \citenamefont {Ueda},\ and\ \citenamefont
  {Nori}}]{liu2019second}%
  \BibitemOpen
  \bibfield  {author} {\bibinfo {author} {\bibfnamefont {T.}~\bibnamefont
  {Liu}}, \bibinfo {author} {\bibfnamefont {Y.-R.}\ \bibnamefont {Zhang}},
  \bibinfo {author} {\bibfnamefont {Q.}~\bibnamefont {Ai}}, \bibinfo {author}
  {\bibfnamefont {Z.}~\bibnamefont {Gong}}, \bibinfo {author} {\bibfnamefont
  {K.}~\bibnamefont {Kawabata}}, \bibinfo {author} {\bibfnamefont
  {M.}~\bibnamefont {Ueda}}, \ and\ \bibinfo {author} {\bibfnamefont
  {F.}~\bibnamefont {Nori}},\ }\href@noop {} {\bibfield  {journal} {\bibinfo
  {journal} {Physical review letters}\ }\textbf {\bibinfo {volume} {122}},\
  \bibinfo {pages} {076801} (\bibinfo {year} {2019})}\BibitemShut {NoStop}%
\bibitem [{\citenamefont {Lee}\ \emph {et~al.}(2019{\natexlab{b}})\citenamefont
  {Lee}, \citenamefont {Li},\ and\ \citenamefont {Gong}}]{lee2019hybrid}%
  \BibitemOpen
  \bibfield  {author} {\bibinfo {author} {\bibfnamefont {C.~H.}\ \bibnamefont
  {Lee}}, \bibinfo {author} {\bibfnamefont {L.}~\bibnamefont {Li}}, \ and\
  \bibinfo {author} {\bibfnamefont {J.}~\bibnamefont {Gong}},\ }\href@noop {}
  {\bibfield  {journal} {\bibinfo  {journal} {Physical review letters}\
  }\textbf {\bibinfo {volume} {123}},\ \bibinfo {pages} {016805} (\bibinfo
  {year} {2019}{\natexlab{b}})}\BibitemShut {NoStop}%
\bibitem [{\citenamefont {Edvardsson}\ \emph {et~al.}(2019)\citenamefont
  {Edvardsson}, \citenamefont {Kunst},\ and\ \citenamefont
  {Bergholtz}}]{PhysRevB.99.081302}%
  \BibitemOpen
  \bibfield  {author} {\bibinfo {author} {\bibfnamefont {E.}~\bibnamefont
  {Edvardsson}}, \bibinfo {author} {\bibfnamefont {F.~K.}\ \bibnamefont
  {Kunst}}, \ and\ \bibinfo {author} {\bibfnamefont {E.~J.}\ \bibnamefont
  {Bergholtz}},\ }\href {\doibase 10.1103/PhysRevB.99.081302} {\bibfield
  {journal} {\bibinfo  {journal} {Phys. Rev. B}\ }\textbf {\bibinfo {volume}
  {99}},\ \bibinfo {pages} {081302} (\bibinfo {year} {2019})}\BibitemShut
  {NoStop}%
\bibitem [{\citenamefont {L{\'o}pez}\ \emph {et~al.}(2019)\citenamefont
  {L{\'o}pez}, \citenamefont {Zhang}, \citenamefont {Torrent},\ and\
  \citenamefont {Christensen}}]{lopez2019multiple}%
  \BibitemOpen
  \bibfield  {author} {\bibinfo {author} {\bibfnamefont {M.~R.}\ \bibnamefont
  {L{\'o}pez}}, \bibinfo {author} {\bibfnamefont {Z.}~\bibnamefont {Zhang}},
  \bibinfo {author} {\bibfnamefont {D.}~\bibnamefont {Torrent}}, \ and\
  \bibinfo {author} {\bibfnamefont {J.}~\bibnamefont {Christensen}},\
  }\href@noop {} {\bibfield  {journal} {\bibinfo  {journal} {Communications
  Physics}\ }\textbf {\bibinfo {volume} {2}},\ \bibinfo {pages} {1} (\bibinfo
  {year} {2019})}\BibitemShut {NoStop}%
\bibitem [{\citenamefont {Chong}\ \emph {et~al.}(2011)\citenamefont {Chong},
  \citenamefont {Ge},\ and\ \citenamefont {Stone}}]{chong2011p}%
  \BibitemOpen
  \bibfield  {author} {\bibinfo {author} {\bibfnamefont {Y.}~\bibnamefont
  {Chong}}, \bibinfo {author} {\bibfnamefont {L.}~\bibnamefont {Ge}}, \ and\
  \bibinfo {author} {\bibfnamefont {A.~D.}\ \bibnamefont {Stone}},\ }\href@noop
  {} {\bibfield  {journal} {\bibinfo  {journal} {Physical Review Letters}\
  }\textbf {\bibinfo {volume} {106}},\ \bibinfo {pages} {093902} (\bibinfo
  {year} {2011})}\BibitemShut {NoStop}%
\bibitem [{\citenamefont {Regensburger}\ \emph {et~al.}(2012)\citenamefont
  {Regensburger}, \citenamefont {Bersch}, \citenamefont {Miri}, \citenamefont
  {Onishchukov}, \citenamefont {Christodoulides},\ and\ \citenamefont
  {Peschel}}]{regensburger2012parity}%
  \BibitemOpen
  \bibfield  {author} {\bibinfo {author} {\bibfnamefont {A.}~\bibnamefont
  {Regensburger}}, \bibinfo {author} {\bibfnamefont {C.}~\bibnamefont
  {Bersch}}, \bibinfo {author} {\bibfnamefont {M.-A.}\ \bibnamefont {Miri}},
  \bibinfo {author} {\bibfnamefont {G.}~\bibnamefont {Onishchukov}}, \bibinfo
  {author} {\bibfnamefont {D.~N.}\ \bibnamefont {Christodoulides}}, \ and\
  \bibinfo {author} {\bibfnamefont {U.}~\bibnamefont {Peschel}},\ }\href@noop
  {} {\bibfield  {journal} {\bibinfo  {journal} {Nature}\ }\textbf {\bibinfo
  {volume} {488}},\ \bibinfo {pages} {167} (\bibinfo {year}
  {2012})}\BibitemShut {NoStop}%
\bibitem [{\citenamefont {Hodaei}\ \emph {et~al.}(2014)\citenamefont {Hodaei},
  \citenamefont {Miri}, \citenamefont {Heinrich}, \citenamefont
  {Christodoulides},\ and\ \citenamefont {Khajavikhan}}]{hodaei2014parity}%
  \BibitemOpen
  \bibfield  {author} {\bibinfo {author} {\bibfnamefont {H.}~\bibnamefont
  {Hodaei}}, \bibinfo {author} {\bibfnamefont {M.-A.}\ \bibnamefont {Miri}},
  \bibinfo {author} {\bibfnamefont {M.}~\bibnamefont {Heinrich}}, \bibinfo
  {author} {\bibfnamefont {D.~N.}\ \bibnamefont {Christodoulides}}, \ and\
  \bibinfo {author} {\bibfnamefont {M.}~\bibnamefont {Khajavikhan}},\
  }\href@noop {} {\bibfield  {journal} {\bibinfo  {journal} {Science}\ }\textbf
  {\bibinfo {volume} {346}},\ \bibinfo {pages} {975} (\bibinfo {year}
  {2014})}\BibitemShut {NoStop}%
\bibitem [{\citenamefont {Peng}\ \emph
  {et~al.}(2014{\natexlab{a}})\citenamefont {Peng}, \citenamefont
  {{\"O}zdemir}, \citenamefont {Lei}, \citenamefont {Monifi}, \citenamefont
  {Gianfreda}, \citenamefont {Long}, \citenamefont {Fan}, \citenamefont {Nori},
  \citenamefont {Bender},\ and\ \citenamefont {Yang}}]{peng2014parity}%
  \BibitemOpen
  \bibfield  {author} {\bibinfo {author} {\bibfnamefont {B.}~\bibnamefont
  {Peng}}, \bibinfo {author} {\bibfnamefont {{\c{S}}.~K.}\ \bibnamefont
  {{\"O}zdemir}}, \bibinfo {author} {\bibfnamefont {F.}~\bibnamefont {Lei}},
  \bibinfo {author} {\bibfnamefont {F.}~\bibnamefont {Monifi}}, \bibinfo
  {author} {\bibfnamefont {M.}~\bibnamefont {Gianfreda}}, \bibinfo {author}
  {\bibfnamefont {G.~L.}\ \bibnamefont {Long}}, \bibinfo {author}
  {\bibfnamefont {S.}~\bibnamefont {Fan}}, \bibinfo {author} {\bibfnamefont
  {F.}~\bibnamefont {Nori}}, \bibinfo {author} {\bibfnamefont {C.~M.}\
  \bibnamefont {Bender}}, \ and\ \bibinfo {author} {\bibfnamefont
  {L.}~\bibnamefont {Yang}},\ }\href@noop {} {\bibfield  {journal} {\bibinfo
  {journal} {Nature Physics}\ }\textbf {\bibinfo {volume} {10}},\ \bibinfo
  {pages} {394} (\bibinfo {year} {2014}{\natexlab{a}})}\BibitemShut {NoStop}%
\bibitem [{\citenamefont {Feng}\ \emph {et~al.}(2014)\citenamefont {Feng},
  \citenamefont {Wong}, \citenamefont {Ma}, \citenamefont {Wang},\ and\
  \citenamefont {Zhang}}]{feng2014single}%
  \BibitemOpen
  \bibfield  {author} {\bibinfo {author} {\bibfnamefont {L.}~\bibnamefont
  {Feng}}, \bibinfo {author} {\bibfnamefont {Z.~J.}\ \bibnamefont {Wong}},
  \bibinfo {author} {\bibfnamefont {R.-M.}\ \bibnamefont {Ma}}, \bibinfo
  {author} {\bibfnamefont {Y.}~\bibnamefont {Wang}}, \ and\ \bibinfo {author}
  {\bibfnamefont {X.}~\bibnamefont {Zhang}},\ }\href@noop {} {\bibfield
  {journal} {\bibinfo  {journal} {Science}\ }\textbf {\bibinfo {volume}
  {346}},\ \bibinfo {pages} {972} (\bibinfo {year} {2014})}\BibitemShut
  {NoStop}%
\bibitem [{\citenamefont {Peng}\ \emph
  {et~al.}(2014{\natexlab{b}})\citenamefont {Peng}, \citenamefont
  {{\"O}zdemir}, \citenamefont {Rotter}, \citenamefont {Yilmaz}, \citenamefont
  {Liertzer}, \citenamefont {Monifi}, \citenamefont {Bender}, \citenamefont
  {Nori},\ and\ \citenamefont {Yang}}]{peng2014loss}%
  \BibitemOpen
  \bibfield  {author} {\bibinfo {author} {\bibfnamefont {B.}~\bibnamefont
  {Peng}}, \bibinfo {author} {\bibfnamefont {{\c{S}}.}~\bibnamefont
  {{\"O}zdemir}}, \bibinfo {author} {\bibfnamefont {S.}~\bibnamefont {Rotter}},
  \bibinfo {author} {\bibfnamefont {H.}~\bibnamefont {Yilmaz}}, \bibinfo
  {author} {\bibfnamefont {M.}~\bibnamefont {Liertzer}}, \bibinfo {author}
  {\bibfnamefont {F.}~\bibnamefont {Monifi}}, \bibinfo {author} {\bibfnamefont
  {C.}~\bibnamefont {Bender}}, \bibinfo {author} {\bibfnamefont
  {F.}~\bibnamefont {Nori}}, \ and\ \bibinfo {author} {\bibfnamefont
  {L.}~\bibnamefont {Yang}},\ }\href@noop {} {\bibfield  {journal} {\bibinfo
  {journal} {Science}\ }\textbf {\bibinfo {volume} {346}},\ \bibinfo {pages}
  {328} (\bibinfo {year} {2014}{\natexlab{b}})}\BibitemShut {NoStop}%
\bibitem [{\citenamefont {Jing}\ \emph {et~al.}(2015)\citenamefont {Jing},
  \citenamefont {{\"O}zdemir}, \citenamefont {Geng}, \citenamefont {Zhang},
  \citenamefont {L{\"u}}, \citenamefont {Peng}, \citenamefont {Yang},\ and\
  \citenamefont {Nori}}]{jing2015optomechanically}%
  \BibitemOpen
  \bibfield  {author} {\bibinfo {author} {\bibfnamefont {H.}~\bibnamefont
  {Jing}}, \bibinfo {author} {\bibfnamefont {{\c{S}}.~K.}\ \bibnamefont
  {{\"O}zdemir}}, \bibinfo {author} {\bibfnamefont {Z.}~\bibnamefont {Geng}},
  \bibinfo {author} {\bibfnamefont {J.}~\bibnamefont {Zhang}}, \bibinfo
  {author} {\bibfnamefont {X.-Y.}\ \bibnamefont {L{\"u}}}, \bibinfo {author}
  {\bibfnamefont {B.}~\bibnamefont {Peng}}, \bibinfo {author} {\bibfnamefont
  {L.}~\bibnamefont {Yang}}, \ and\ \bibinfo {author} {\bibfnamefont
  {F.}~\bibnamefont {Nori}},\ }\href@noop {} {\bibfield  {journal} {\bibinfo
  {journal} {Scientific reports}\ }\textbf {\bibinfo {volume} {5}},\ \bibinfo
  {pages} {9663} (\bibinfo {year} {2015})}\BibitemShut {NoStop}%
\bibitem [{\citenamefont {Liu}\ \emph {et~al.}(2016)\citenamefont {Liu},
  \citenamefont {Zhang}, \citenamefont {{\"O}zdemir}, \citenamefont {Peng},
  \citenamefont {Jing}, \citenamefont {L{\"u}}, \citenamefont {Li},
  \citenamefont {Yang}, \citenamefont {Nori},\ and\ \citenamefont
  {Liu}}]{liu2016metrology}%
  \BibitemOpen
  \bibfield  {author} {\bibinfo {author} {\bibfnamefont {Z.-P.}\ \bibnamefont
  {Liu}}, \bibinfo {author} {\bibfnamefont {J.}~\bibnamefont {Zhang}}, \bibinfo
  {author} {\bibfnamefont {{\c{S}}.~K.}\ \bibnamefont {{\"O}zdemir}}, \bibinfo
  {author} {\bibfnamefont {B.}~\bibnamefont {Peng}}, \bibinfo {author}
  {\bibfnamefont {H.}~\bibnamefont {Jing}}, \bibinfo {author} {\bibfnamefont
  {X.-Y.}\ \bibnamefont {L{\"u}}}, \bibinfo {author} {\bibfnamefont {C.-W.}\
  \bibnamefont {Li}}, \bibinfo {author} {\bibfnamefont {L.}~\bibnamefont
  {Yang}}, \bibinfo {author} {\bibfnamefont {F.}~\bibnamefont {Nori}}, \ and\
  \bibinfo {author} {\bibfnamefont {Y.-x.}\ \bibnamefont {Liu}},\ }\href@noop
  {} {\bibfield  {journal} {\bibinfo  {journal} {Physical review letters}\
  }\textbf {\bibinfo {volume} {117}},\ \bibinfo {pages} {110802} (\bibinfo
  {year} {2016})}\BibitemShut {NoStop}%
\bibitem [{\citenamefont {Kawabata}\ \emph {et~al.}(2017)\citenamefont
  {Kawabata}, \citenamefont {Ashida},\ and\ \citenamefont
  {Ueda}}]{kawabata2017information}%
  \BibitemOpen
  \bibfield  {author} {\bibinfo {author} {\bibfnamefont {K.}~\bibnamefont
  {Kawabata}}, \bibinfo {author} {\bibfnamefont {Y.}~\bibnamefont {Ashida}}, \
  and\ \bibinfo {author} {\bibfnamefont {M.}~\bibnamefont {Ueda}},\ }\href@noop
  {} {\bibfield  {journal} {\bibinfo  {journal} {Physical review letters}\
  }\textbf {\bibinfo {volume} {119}},\ \bibinfo {pages} {190401} (\bibinfo
  {year} {2017})}\BibitemShut {NoStop}%
\bibitem [{\citenamefont {Jing}\ \emph {et~al.}(2017)\citenamefont {Jing},
  \citenamefont {{\"O}zdemir}, \citenamefont {L{\"u}},\ and\ \citenamefont
  {Nori}}]{jing2017high}%
  \BibitemOpen
  \bibfield  {author} {\bibinfo {author} {\bibfnamefont {H.}~\bibnamefont
  {Jing}}, \bibinfo {author} {\bibfnamefont {{\c{S}}.}~\bibnamefont
  {{\"O}zdemir}}, \bibinfo {author} {\bibfnamefont {H.}~\bibnamefont {L{\"u}}},
  \ and\ \bibinfo {author} {\bibfnamefont {F.}~\bibnamefont {Nori}},\
  }\href@noop {} {\bibfield  {journal} {\bibinfo  {journal} {Scientific
  reports}\ }\textbf {\bibinfo {volume} {7}},\ \bibinfo {pages} {3386}
  (\bibinfo {year} {2017})}\BibitemShut {NoStop}%
\bibitem [{\citenamefont {L{\"u}}\ \emph {et~al.}(2017)\citenamefont {L{\"u}},
  \citenamefont {{\"O}zdemir}, \citenamefont {Kuang}, \citenamefont {Nori},\
  and\ \citenamefont {Jing}}]{lu2017exceptional}%
  \BibitemOpen
  \bibfield  {author} {\bibinfo {author} {\bibfnamefont {H.}~\bibnamefont
  {L{\"u}}}, \bibinfo {author} {\bibfnamefont {S.}~\bibnamefont {{\"O}zdemir}},
  \bibinfo {author} {\bibfnamefont {L.-M.}\ \bibnamefont {Kuang}}, \bibinfo
  {author} {\bibfnamefont {F.}~\bibnamefont {Nori}}, \ and\ \bibinfo {author}
  {\bibfnamefont {H.}~\bibnamefont {Jing}},\ }\href@noop {} {\bibfield
  {journal} {\bibinfo  {journal} {Physical Review Applied}\ }\textbf {\bibinfo
  {volume} {8}},\ \bibinfo {pages} {044020} (\bibinfo {year}
  {2017})}\BibitemShut {NoStop}%
\bibitem [{\citenamefont {Ashida}\ \emph {et~al.}(2017)\citenamefont {Ashida},
  \citenamefont {Furukawa},\ and\ \citenamefont {Ueda}}]{ashida2017parity}%
  \BibitemOpen
  \bibfield  {author} {\bibinfo {author} {\bibfnamefont {Y.}~\bibnamefont
  {Ashida}}, \bibinfo {author} {\bibfnamefont {S.}~\bibnamefont {Furukawa}}, \
  and\ \bibinfo {author} {\bibfnamefont {M.}~\bibnamefont {Ueda}},\ }\href@noop
  {} {\bibfield  {journal} {\bibinfo  {journal} {Nature communications}\
  }\textbf {\bibinfo {volume} {8}},\ \bibinfo {pages} {15791} (\bibinfo {year}
  {2017})}\BibitemShut {NoStop}%
\bibitem [{\citenamefont {El-Ganainy}\ \emph {et~al.}(2018)\citenamefont
  {El-Ganainy}, \citenamefont {Makris}, \citenamefont {Khajavikhan},
  \citenamefont {Musslimani}, \citenamefont {Rotter},\ and\ \citenamefont
  {Christodoulides}}]{el2018non}%
  \BibitemOpen
  \bibfield  {author} {\bibinfo {author} {\bibfnamefont {R.}~\bibnamefont
  {El-Ganainy}}, \bibinfo {author} {\bibfnamefont {K.~G.}\ \bibnamefont
  {Makris}}, \bibinfo {author} {\bibfnamefont {M.}~\bibnamefont {Khajavikhan}},
  \bibinfo {author} {\bibfnamefont {Z.~H.}\ \bibnamefont {Musslimani}},
  \bibinfo {author} {\bibfnamefont {S.}~\bibnamefont {Rotter}}, \ and\ \bibinfo
  {author} {\bibfnamefont {D.~N.}\ \bibnamefont {Christodoulides}},\
  }\href@noop {} {\bibfield  {journal} {\bibinfo  {journal} {Nature Physics}\
  }\textbf {\bibinfo {volume} {14}},\ \bibinfo {pages} {11} (\bibinfo {year}
  {2018})}\BibitemShut {NoStop}%
\bibitem [{\citenamefont {Zhang}\ \emph
  {et~al.}(2018{\natexlab{a}})\citenamefont {Zhang}, \citenamefont {Peng},
  \citenamefont {{\"O}zdemir}, \citenamefont {Pichler}, \citenamefont {Krimer},
  \citenamefont {Zhao}, \citenamefont {Nori}, \citenamefont {Liu},
  \citenamefont {Rotter},\ and\ \citenamefont {Yang}}]{zhang2018phonon}%
  \BibitemOpen
  \bibfield  {author} {\bibinfo {author} {\bibfnamefont {J.}~\bibnamefont
  {Zhang}}, \bibinfo {author} {\bibfnamefont {B.}~\bibnamefont {Peng}},
  \bibinfo {author} {\bibfnamefont {{\c{S}}.~K.}\ \bibnamefont {{\"O}zdemir}},
  \bibinfo {author} {\bibfnamefont {K.}~\bibnamefont {Pichler}}, \bibinfo
  {author} {\bibfnamefont {D.~O.}\ \bibnamefont {Krimer}}, \bibinfo {author}
  {\bibfnamefont {G.}~\bibnamefont {Zhao}}, \bibinfo {author} {\bibfnamefont
  {F.}~\bibnamefont {Nori}}, \bibinfo {author} {\bibfnamefont {Y.-x.}\
  \bibnamefont {Liu}}, \bibinfo {author} {\bibfnamefont {S.}~\bibnamefont
  {Rotter}}, \ and\ \bibinfo {author} {\bibfnamefont {L.}~\bibnamefont
  {Yang}},\ }\href@noop {} {\bibfield  {journal} {\bibinfo  {journal} {Nature
  Photonics}\ }\textbf {\bibinfo {volume} {12}},\ \bibinfo {pages} {479}
  (\bibinfo {year} {2018}{\natexlab{a}})}\BibitemShut {NoStop}%
\bibitem [{\citenamefont {Lu}\ \emph {et~al.}(2014)\citenamefont {Lu},
  \citenamefont {Joannopoulos},\ and\ \citenamefont
  {Solja{\v{c}}i{\'c}}}]{lu2014topological}%
  \BibitemOpen
  \bibfield  {author} {\bibinfo {author} {\bibfnamefont {L.}~\bibnamefont
  {Lu}}, \bibinfo {author} {\bibfnamefont {J.~D.}\ \bibnamefont
  {Joannopoulos}}, \ and\ \bibinfo {author} {\bibfnamefont {M.}~\bibnamefont
  {Solja{\v{c}}i{\'c}}},\ }\href@noop {} {\bibfield  {journal} {\bibinfo
  {journal} {Nature photonics}\ }\textbf {\bibinfo {volume} {8}},\ \bibinfo
  {pages} {821} (\bibinfo {year} {2014})}\BibitemShut {NoStop}%
\bibitem [{\citenamefont {Ozawa}\ \emph {et~al.}(2019)\citenamefont {Ozawa},
  \citenamefont {Price}, \citenamefont {Amo}, \citenamefont {Goldman},
  \citenamefont {Hafezi}, \citenamefont {Lu}, \citenamefont {Rechtsman},
  \citenamefont {Schuster}, \citenamefont {Simon}, \citenamefont {Zilberberg}
  \emph {et~al.}}]{ozawa2019topological}%
  \BibitemOpen
  \bibfield  {author} {\bibinfo {author} {\bibfnamefont {T.}~\bibnamefont
  {Ozawa}}, \bibinfo {author} {\bibfnamefont {H.~M.}\ \bibnamefont {Price}},
  \bibinfo {author} {\bibfnamefont {A.}~\bibnamefont {Amo}}, \bibinfo {author}
  {\bibfnamefont {N.}~\bibnamefont {Goldman}}, \bibinfo {author} {\bibfnamefont
  {M.}~\bibnamefont {Hafezi}}, \bibinfo {author} {\bibfnamefont
  {L.}~\bibnamefont {Lu}}, \bibinfo {author} {\bibfnamefont {M.~C.}\
  \bibnamefont {Rechtsman}}, \bibinfo {author} {\bibfnamefont {D.}~\bibnamefont
  {Schuster}}, \bibinfo {author} {\bibfnamefont {J.}~\bibnamefont {Simon}},
  \bibinfo {author} {\bibfnamefont {O.}~\bibnamefont {Zilberberg}},  \emph
  {et~al.},\ }\href@noop {} {\bibfield  {journal} {\bibinfo  {journal} {Reviews
  of Modern Physics}\ }\textbf {\bibinfo {volume} {91}},\ \bibinfo {pages}
  {015006} (\bibinfo {year} {2019})}\BibitemShut {NoStop}%
\bibitem [{\citenamefont {Malzard}\ and\ \citenamefont
  {Schomerus}(2018)}]{malzard2018bulk}%
  \BibitemOpen
  \bibfield  {author} {\bibinfo {author} {\bibfnamefont {S.}~\bibnamefont
  {Malzard}}\ and\ \bibinfo {author} {\bibfnamefont {H.}~\bibnamefont
  {Schomerus}},\ }\href@noop {} {\bibfield  {journal} {\bibinfo  {journal}
  {Physical Review A}\ }\textbf {\bibinfo {volume} {98}},\ \bibinfo {pages}
  {033807} (\bibinfo {year} {2018})}\BibitemShut {NoStop}%
\bibitem [{\citenamefont {Longhi}(2018)}]{longhi2018parity}%
  \BibitemOpen
  \bibfield  {author} {\bibinfo {author} {\bibfnamefont {S.}~\bibnamefont
  {Longhi}},\ }\href@noop {} {\bibfield  {journal} {\bibinfo  {journal} {EPL
  (Europhysics Letters)}\ }\textbf {\bibinfo {volume} {120}},\ \bibinfo {pages}
  {64001} (\bibinfo {year} {2018})}\BibitemShut {NoStop}%
\bibitem [{\citenamefont {Zhen}\ \emph {et~al.}(2015)\citenamefont {Zhen},
  \citenamefont {Hsu}, \citenamefont {Igarashi}, \citenamefont {Lu},
  \citenamefont {Kaminer}, \citenamefont {Pick}, \citenamefont {Chua},
  \citenamefont {Joannopoulos},\ and\ \citenamefont
  {Solja{\v{c}}i{\'c}}}]{zhen2015spawning}%
  \BibitemOpen
  \bibfield  {author} {\bibinfo {author} {\bibfnamefont {B.}~\bibnamefont
  {Zhen}}, \bibinfo {author} {\bibfnamefont {C.~W.}\ \bibnamefont {Hsu}},
  \bibinfo {author} {\bibfnamefont {Y.}~\bibnamefont {Igarashi}}, \bibinfo
  {author} {\bibfnamefont {L.}~\bibnamefont {Lu}}, \bibinfo {author}
  {\bibfnamefont {I.}~\bibnamefont {Kaminer}}, \bibinfo {author} {\bibfnamefont
  {A.}~\bibnamefont {Pick}}, \bibinfo {author} {\bibfnamefont {S.-L.}\
  \bibnamefont {Chua}}, \bibinfo {author} {\bibfnamefont {J.~D.}\ \bibnamefont
  {Joannopoulos}}, \ and\ \bibinfo {author} {\bibfnamefont {M.}~\bibnamefont
  {Solja{\v{c}}i{\'c}}},\ }\href@noop {} {\bibfield  {journal} {\bibinfo
  {journal} {Nature}\ }\textbf {\bibinfo {volume} {525}},\ \bibinfo {pages}
  {354} (\bibinfo {year} {2015})}\BibitemShut {NoStop}%
\bibitem [{\citenamefont {Zhang}\ \emph
  {et~al.}(2018{\natexlab{b}})\citenamefont {Zhang}, \citenamefont {Hu},
  \citenamefont {Lin}, \citenamefont {Niu}, \citenamefont {Xia}, \citenamefont
  {Gong},\ and\ \citenamefont {Gong}}]{zhang2018thermal}%
  \BibitemOpen
  \bibfield  {author} {\bibinfo {author} {\bibfnamefont {S.}~\bibnamefont
  {Zhang}}, \bibinfo {author} {\bibfnamefont {Y.}~\bibnamefont {Hu}}, \bibinfo
  {author} {\bibfnamefont {G.}~\bibnamefont {Lin}}, \bibinfo {author}
  {\bibfnamefont {Y.}~\bibnamefont {Niu}}, \bibinfo {author} {\bibfnamefont
  {K.}~\bibnamefont {Xia}}, \bibinfo {author} {\bibfnamefont {J.}~\bibnamefont
  {Gong}}, \ and\ \bibinfo {author} {\bibfnamefont {S.}~\bibnamefont {Gong}},\
  }\href@noop {} {\bibfield  {journal} {\bibinfo  {journal} {Nature Photonics}\
  }\textbf {\bibinfo {volume} {12}},\ \bibinfo {pages} {744} (\bibinfo {year}
  {2018}{\natexlab{b}})}\BibitemShut {NoStop}%
\bibitem [{\citenamefont {Hofmann}\ \emph
  {et~al.}(2019{\natexlab{a}})\citenamefont {Hofmann}, \citenamefont {Helbig},
  \citenamefont {Lee}, \citenamefont {Greiter},\ and\ \citenamefont
  {Thomale}}]{hofmann2019chiral}%
  \BibitemOpen
  \bibfield  {author} {\bibinfo {author} {\bibfnamefont {T.}~\bibnamefont
  {Hofmann}}, \bibinfo {author} {\bibfnamefont {T.}~\bibnamefont {Helbig}},
  \bibinfo {author} {\bibfnamefont {C.~H.}\ \bibnamefont {Lee}}, \bibinfo
  {author} {\bibfnamefont {M.}~\bibnamefont {Greiter}}, \ and\ \bibinfo
  {author} {\bibfnamefont {R.}~\bibnamefont {Thomale}},\ }\href@noop {}
  {\bibfield  {journal} {\bibinfo  {journal} {Physical Review Letters}\
  }\textbf {\bibinfo {volume} {122}},\ \bibinfo {pages} {247702} (\bibinfo
  {year} {2019}{\natexlab{a}})}\BibitemShut {NoStop}%
\bibitem [{\citenamefont {Ezawa}(2019{\natexlab{a}})}]{PhysRevB.99.201411}%
  \BibitemOpen
  \bibfield  {author} {\bibinfo {author} {\bibfnamefont {M.}~\bibnamefont
  {Ezawa}},\ }\href {\doibase 10.1103/PhysRevB.99.201411} {\bibfield  {journal}
  {\bibinfo  {journal} {Phys. Rev. B}\ }\textbf {\bibinfo {volume} {99}},\
  \bibinfo {pages} {201411} (\bibinfo {year} {2019}{\natexlab{a}})}\BibitemShut
  {NoStop}%
\bibitem [{\citenamefont {Xu}\ \emph {et~al.}(2016)\citenamefont {Xu},
  \citenamefont {Mason}, \citenamefont {Jiang},\ and\ \citenamefont
  {Harris}}]{xu2016topological}%
  \BibitemOpen
  \bibfield  {author} {\bibinfo {author} {\bibfnamefont {H.}~\bibnamefont
  {Xu}}, \bibinfo {author} {\bibfnamefont {D.}~\bibnamefont {Mason}}, \bibinfo
  {author} {\bibfnamefont {L.}~\bibnamefont {Jiang}}, \ and\ \bibinfo {author}
  {\bibfnamefont {J.}~\bibnamefont {Harris}},\ }\href@noop {} {\bibfield
  {journal} {\bibinfo  {journal} {Nature}\ }\textbf {\bibinfo {volume} {537}},\
  \bibinfo {pages} {80} (\bibinfo {year} {2016})}\BibitemShut {NoStop}%
\bibitem [{\citenamefont {Tzortzakakis}\ \emph {et~al.}(2019)\citenamefont
  {Tzortzakakis}, \citenamefont {Makris},\ and\ \citenamefont
  {Economou}}]{tzortzakakis2019non}%
  \BibitemOpen
  \bibfield  {author} {\bibinfo {author} {\bibfnamefont {A.}~\bibnamefont
  {Tzortzakakis}}, \bibinfo {author} {\bibfnamefont {K.}~\bibnamefont
  {Makris}}, \ and\ \bibinfo {author} {\bibfnamefont {E.}~\bibnamefont
  {Economou}},\ }\href@noop {} {\bibfield  {journal} {\bibinfo  {journal}
  {arXiv preprint arXiv:1909.13816}\ } (\bibinfo {year} {2019})}\BibitemShut
  {NoStop}%
\bibitem [{\citenamefont {Liertzer}\ \emph {et~al.}(2012)\citenamefont
  {Liertzer}, \citenamefont {Ge}, \citenamefont {Cerjan}, \citenamefont
  {Stone}, \citenamefont {T{\"u}reci},\ and\ \citenamefont
  {Rotter}}]{liertzer2012pump}%
  \BibitemOpen
  \bibfield  {author} {\bibinfo {author} {\bibfnamefont {M.}~\bibnamefont
  {Liertzer}}, \bibinfo {author} {\bibfnamefont {L.}~\bibnamefont {Ge}},
  \bibinfo {author} {\bibfnamefont {A.}~\bibnamefont {Cerjan}}, \bibinfo
  {author} {\bibfnamefont {A.}~\bibnamefont {Stone}}, \bibinfo {author}
  {\bibfnamefont {H.~E.}\ \bibnamefont {T{\"u}reci}}, \ and\ \bibinfo {author}
  {\bibfnamefont {S.}~\bibnamefont {Rotter}},\ }\href@noop {} {\bibfield
  {journal} {\bibinfo  {journal} {Physical Review Letters}\ }\textbf {\bibinfo
  {volume} {108}},\ \bibinfo {pages} {173901} (\bibinfo {year}
  {2012})}\BibitemShut {NoStop}%
\bibitem [{\citenamefont {Bergholtz}\ \emph {et~al.}(2019)\citenamefont
  {Bergholtz}, \citenamefont {Budich},\ and\ \citenamefont
  {Kunst}}]{bergholtz2019exceptional}%
  \BibitemOpen
  \bibfield  {author} {\bibinfo {author} {\bibfnamefont {E.~J.}\ \bibnamefont
  {Bergholtz}}, \bibinfo {author} {\bibfnamefont {J.~C.}\ \bibnamefont
  {Budich}}, \ and\ \bibinfo {author} {\bibfnamefont {F.~K.}\ \bibnamefont
  {Kunst}},\ }\href@noop {} {\enquote {\bibinfo {title} {Exceptional topology
  of non-hermitian systems},}\ } (\bibinfo {year} {2019}),\ \Eprint
  {http://arxiv.org/abs/1912.10048} {arXiv:1912.10048 [cond-mat.mes-hall]}
  \BibitemShut {NoStop}%
\bibitem [{\citenamefont {Hofmann}\ \emph
  {et~al.}(2019{\natexlab{b}})\citenamefont {Hofmann}, \citenamefont {Helbig},
  \citenamefont {Schindler}, \citenamefont {Salgo}, \citenamefont
  {Brzezi{\'n}ska}, \citenamefont {Greiter}, \citenamefont {Kiessling},
  \citenamefont {Wolf}, \citenamefont {Vollhardt}, \citenamefont {Kaba{\v{s}}i}
  \emph {et~al.}}]{hofmann2019reciprocal}%
  \BibitemOpen
  \bibfield  {author} {\bibinfo {author} {\bibfnamefont {T.}~\bibnamefont
  {Hofmann}}, \bibinfo {author} {\bibfnamefont {T.}~\bibnamefont {Helbig}},
  \bibinfo {author} {\bibfnamefont {F.}~\bibnamefont {Schindler}}, \bibinfo
  {author} {\bibfnamefont {N.}~\bibnamefont {Salgo}}, \bibinfo {author}
  {\bibfnamefont {M.}~\bibnamefont {Brzezi{\'n}ska}}, \bibinfo {author}
  {\bibfnamefont {M.}~\bibnamefont {Greiter}}, \bibinfo {author} {\bibfnamefont
  {T.}~\bibnamefont {Kiessling}}, \bibinfo {author} {\bibfnamefont
  {D.}~\bibnamefont {Wolf}}, \bibinfo {author} {\bibfnamefont {A.}~\bibnamefont
  {Vollhardt}}, \bibinfo {author} {\bibfnamefont {A.}~\bibnamefont
  {Kaba{\v{s}}i}},  \emph {et~al.},\ }\href@noop {} {\bibfield  {journal}
  {\bibinfo  {journal} {arXiv preprint arXiv:1908.02759}\ } (\bibinfo {year}
  {2019}{\natexlab{b}})}\BibitemShut {NoStop}%
\bibitem [{\citenamefont {Helbig}\ \emph {et~al.}(2019)\citenamefont {Helbig},
  \citenamefont {Hofmann}, \citenamefont {Imhof}, \citenamefont {Abdelghany},
  \citenamefont {Kiessling}, \citenamefont {Molenkamp}, \citenamefont {Lee},
  \citenamefont {Szameit}, \citenamefont {Greiter},\ and\ \citenamefont
  {Thomale}}]{helbig2019observation}%
  \BibitemOpen
  \bibfield  {author} {\bibinfo {author} {\bibfnamefont {T.}~\bibnamefont
  {Helbig}}, \bibinfo {author} {\bibfnamefont {T.}~\bibnamefont {Hofmann}},
  \bibinfo {author} {\bibfnamefont {S.}~\bibnamefont {Imhof}}, \bibinfo
  {author} {\bibfnamefont {M.}~\bibnamefont {Abdelghany}}, \bibinfo {author}
  {\bibfnamefont {T.}~\bibnamefont {Kiessling}}, \bibinfo {author}
  {\bibfnamefont {L.~W.}\ \bibnamefont {Molenkamp}}, \bibinfo {author}
  {\bibfnamefont {C.~H.}\ \bibnamefont {Lee}}, \bibinfo {author} {\bibfnamefont
  {A.}~\bibnamefont {Szameit}}, \bibinfo {author} {\bibfnamefont
  {M.}~\bibnamefont {Greiter}}, \ and\ \bibinfo {author} {\bibfnamefont
  {R.}~\bibnamefont {Thomale}},\ }\href@noop {} {\bibfield  {journal} {\bibinfo
   {journal} {arXiv preprint arXiv:1907.11562}\ } (\bibinfo {year}
  {2019})}\BibitemShut {NoStop}%
\bibitem [{\citenamefont {Ezawa}(2019{\natexlab{b}})}]{ezawa2019electric}%
  \BibitemOpen
  \bibfield  {author} {\bibinfo {author} {\bibfnamefont {M.}~\bibnamefont
  {Ezawa}},\ }\href@noop {} {\bibfield  {journal} {\bibinfo  {journal}
  {Physical Review B}\ }\textbf {\bibinfo {volume} {100}},\ \bibinfo {pages}
  {165419} (\bibinfo {year} {2019}{\natexlab{b}})}\BibitemShut {NoStop}%
\bibitem [{\citenamefont {Xiao}\ \emph {et~al.}(2019)\citenamefont {Xiao},
  \citenamefont {Deng}, \citenamefont {Wang}, \citenamefont {Zhu},
  \citenamefont {Wang}, \citenamefont {Yi},\ and\ \citenamefont
  {Xue}}]{xiao2019observation}%
  \BibitemOpen
  \bibfield  {author} {\bibinfo {author} {\bibfnamefont {L.}~\bibnamefont
  {Xiao}}, \bibinfo {author} {\bibfnamefont {T.}~\bibnamefont {Deng}}, \bibinfo
  {author} {\bibfnamefont {K.}~\bibnamefont {Wang}}, \bibinfo {author}
  {\bibfnamefont {G.}~\bibnamefont {Zhu}}, \bibinfo {author} {\bibfnamefont
  {Z.}~\bibnamefont {Wang}}, \bibinfo {author} {\bibfnamefont {W.}~\bibnamefont
  {Yi}}, \ and\ \bibinfo {author} {\bibfnamefont {P.}~\bibnamefont {Xue}},\
  }\href@noop {} {\bibfield  {journal} {\bibinfo  {journal} {arXiv preprint
  arXiv:1907.12566}\ } (\bibinfo {year} {2019})}\BibitemShut {NoStop}%
\bibitem [{\citenamefont {Ezawa}(2019{\natexlab{c}})}]{ezawa2019non}%
  \BibitemOpen
  \bibfield  {author} {\bibinfo {author} {\bibfnamefont {M.}~\bibnamefont
  {Ezawa}},\ }\href@noop {} {\bibfield  {journal} {\bibinfo  {journal}
  {Physical Review B}\ }\textbf {\bibinfo {volume} {99}},\ \bibinfo {pages}
  {201411} (\bibinfo {year} {2019}{\natexlab{c}})}\BibitemShut {NoStop}%
\bibitem [{\citenamefont {Zhang}\ and\ \citenamefont
  {Franz}(2020)}]{PhysRevLett.124.046401}%
  \BibitemOpen
  \bibfield  {author} {\bibinfo {author} {\bibfnamefont {X.-X.}\ \bibnamefont
  {Zhang}}\ and\ \bibinfo {author} {\bibfnamefont {M.}~\bibnamefont {Franz}},\
  }\href {\doibase 10.1103/PhysRevLett.124.046401} {\bibfield  {journal}
  {\bibinfo  {journal} {Phys. Rev. Lett.}\ }\textbf {\bibinfo {volume} {124}},\
  \bibinfo {pages} {046401} (\bibinfo {year} {2020})}\BibitemShut {NoStop}%
\bibitem [{\citenamefont {Zhang}\ \emph
  {et~al.}(2019{\natexlab{b}})\citenamefont {Zhang}, \citenamefont
  {Rosendo~L\'opez}, \citenamefont {Cheng}, \citenamefont {Liu},\ and\
  \citenamefont {Christensen}}]{PhysRevLett.122.195501}%
  \BibitemOpen
  \bibfield  {author} {\bibinfo {author} {\bibfnamefont {Z.}~\bibnamefont
  {Zhang}}, \bibinfo {author} {\bibfnamefont {M.}~\bibnamefont
  {Rosendo~L\'opez}}, \bibinfo {author} {\bibfnamefont {Y.}~\bibnamefont
  {Cheng}}, \bibinfo {author} {\bibfnamefont {X.}~\bibnamefont {Liu}}, \ and\
  \bibinfo {author} {\bibfnamefont {J.}~\bibnamefont {Christensen}},\ }\href
  {\doibase 10.1103/PhysRevLett.122.195501} {\bibfield  {journal} {\bibinfo
  {journal} {Phys. Rev. Lett.}\ }\textbf {\bibinfo {volume} {122}},\ \bibinfo
  {pages} {195501} (\bibinfo {year} {2019}{\natexlab{b}})}\BibitemShut
  {NoStop}%
\bibitem [{\citenamefont {Jing}\ \emph {et~al.}(2014)\citenamefont {Jing},
  \citenamefont {{\"O}zdemir}, \citenamefont {L{\"u}}, \citenamefont {Zhang},
  \citenamefont {Yang},\ and\ \citenamefont {Nori}}]{jing2014pt}%
  \BibitemOpen
  \bibfield  {author} {\bibinfo {author} {\bibfnamefont {H.}~\bibnamefont
  {Jing}}, \bibinfo {author} {\bibfnamefont {S.}~\bibnamefont {{\"O}zdemir}},
  \bibinfo {author} {\bibfnamefont {X.-Y.}\ \bibnamefont {L{\"u}}}, \bibinfo
  {author} {\bibfnamefont {J.}~\bibnamefont {Zhang}}, \bibinfo {author}
  {\bibfnamefont {L.}~\bibnamefont {Yang}}, \ and\ \bibinfo {author}
  {\bibfnamefont {F.}~\bibnamefont {Nori}},\ }\href@noop {} {\bibfield
  {journal} {\bibinfo  {journal} {Physical review letters}\ }\textbf {\bibinfo
  {volume} {113}},\ \bibinfo {pages} {053604} (\bibinfo {year}
  {2014})}\BibitemShut {NoStop}%
\bibitem [{\citenamefont {Scheibner}\ \emph
  {et~al.}(2020{\natexlab{a}})\citenamefont {Scheibner}, \citenamefont
  {Souslov}, \citenamefont {Banerjee}, \citenamefont {Surowka}, \citenamefont
  {Irvine},\ and\ \citenamefont {Vitelli}}]{scheibner2020odd}%
  \BibitemOpen
  \bibfield  {author} {\bibinfo {author} {\bibfnamefont {C.}~\bibnamefont
  {Scheibner}}, \bibinfo {author} {\bibfnamefont {A.}~\bibnamefont {Souslov}},
  \bibinfo {author} {\bibfnamefont {D.}~\bibnamefont {Banerjee}}, \bibinfo
  {author} {\bibfnamefont {P.}~\bibnamefont {Surowka}}, \bibinfo {author}
  {\bibfnamefont {W.~T.}\ \bibnamefont {Irvine}}, \ and\ \bibinfo {author}
  {\bibfnamefont {V.}~\bibnamefont {Vitelli}},\ }\href@noop {} {\bibfield
  {journal} {\bibinfo  {journal} {Nature Physics}\ }\textbf {\bibinfo {volume}
  {16}},\ \bibinfo {pages} {475} (\bibinfo {year}
  {2020}{\natexlab{a}})}\BibitemShut {NoStop}%
\bibitem [{\citenamefont {Alberts}\ \emph {et~al.}(2002)\citenamefont
  {Alberts}, \citenamefont {Johnson}, \citenamefont {Lewis}, \citenamefont
  {Raff}, \citenamefont {Roberts},\ and\ \citenamefont
  {Walter}}]{alberts2002drosophila}%
  \BibitemOpen
  \bibfield  {author} {\bibinfo {author} {\bibfnamefont {B.}~\bibnamefont
  {Alberts}}, \bibinfo {author} {\bibfnamefont {A.}~\bibnamefont {Johnson}},
  \bibinfo {author} {\bibfnamefont {J.}~\bibnamefont {Lewis}}, \bibinfo
  {author} {\bibfnamefont {M.}~\bibnamefont {Raff}}, \bibinfo {author}
  {\bibfnamefont {K.}~\bibnamefont {Roberts}}, \ and\ \bibinfo {author}
  {\bibfnamefont {P.}~\bibnamefont {Walter}},\ }in\ \href@noop {} {\emph
  {\bibinfo {booktitle} {Molecular Biology of the Cell. 4th edition}}}\
  (\bibinfo  {publisher} {Garland Science},\ \bibinfo {year}
  {2002})\BibitemShut {NoStop}%
\bibitem [{\citenamefont {Brangwynne}\ \emph {et~al.}(2008)\citenamefont
  {Brangwynne}, \citenamefont {Koenderink}, \citenamefont {MacKintosh},\ and\
  \citenamefont {Weitz}}]{brangwynne2008cytoplasmic}%
  \BibitemOpen
  \bibfield  {author} {\bibinfo {author} {\bibfnamefont {C.~P.}\ \bibnamefont
  {Brangwynne}}, \bibinfo {author} {\bibfnamefont {G.~H.}\ \bibnamefont
  {Koenderink}}, \bibinfo {author} {\bibfnamefont {F.~C.}\ \bibnamefont
  {MacKintosh}}, \ and\ \bibinfo {author} {\bibfnamefont {D.~A.}\ \bibnamefont
  {Weitz}},\ }\href@noop {} {\bibfield  {journal} {\bibinfo  {journal} {The
  Journal of cell biology}\ }\textbf {\bibinfo {volume} {183}},\ \bibinfo
  {pages} {583} (\bibinfo {year} {2008})}\BibitemShut {NoStop}%
\bibitem [{\citenamefont {Joanny}\ and\ \citenamefont
  {Prost}(2009)}]{joanny2009active}%
  \BibitemOpen
  \bibfield  {author} {\bibinfo {author} {\bibfnamefont {J.-F.}\ \bibnamefont
  {Joanny}}\ and\ \bibinfo {author} {\bibfnamefont {J.}~\bibnamefont {Prost}},\
  }\href@noop {} {\bibfield  {journal} {\bibinfo  {journal} {HFSP journal}\
  }\textbf {\bibinfo {volume} {3}},\ \bibinfo {pages} {94} (\bibinfo {year}
  {2009})}\BibitemShut {NoStop}%
\bibitem [{\citenamefont {Foster}\ \emph {et~al.}(2015)\citenamefont {Foster},
  \citenamefont {F{\"u}rthauer}, \citenamefont {Shelley},\ and\ \citenamefont
  {Needleman}}]{foster2015active}%
  \BibitemOpen
  \bibfield  {author} {\bibinfo {author} {\bibfnamefont {P.~J.}\ \bibnamefont
  {Foster}}, \bibinfo {author} {\bibfnamefont {S.}~\bibnamefont
  {F{\"u}rthauer}}, \bibinfo {author} {\bibfnamefont {M.~J.}\ \bibnamefont
  {Shelley}}, \ and\ \bibinfo {author} {\bibfnamefont {D.~J.}\ \bibnamefont
  {Needleman}},\ }\href@noop {} {\bibfield  {journal} {\bibinfo  {journal}
  {Elife}\ }\textbf {\bibinfo {volume} {4}},\ \bibinfo {pages} {e10837}
  (\bibinfo {year} {2015})}\BibitemShut {NoStop}%
\bibitem [{\citenamefont {Wang}\ \emph {et~al.}(2015)\citenamefont {Wang},
  \citenamefont {Lu},\ and\ \citenamefont {Bertoldi}}]{wang2015topological}%
  \BibitemOpen
  \bibfield  {author} {\bibinfo {author} {\bibfnamefont {P.}~\bibnamefont
  {Wang}}, \bibinfo {author} {\bibfnamefont {L.}~\bibnamefont {Lu}}, \ and\
  \bibinfo {author} {\bibfnamefont {K.}~\bibnamefont {Bertoldi}},\ }\href@noop
  {} {\bibfield  {journal} {\bibinfo  {journal} {Physical review letters}\
  }\textbf {\bibinfo {volume} {115}},\ \bibinfo {pages} {104302} (\bibinfo
  {year} {2015})}\BibitemShut {NoStop}%
\bibitem [{\citenamefont {Mitchell}\ \emph
  {et~al.}(2018{\natexlab{b}})\citenamefont {Mitchell}, \citenamefont {Nash},\
  and\ \citenamefont {Irvine}}]{mitchell2018realization}%
  \BibitemOpen
  \bibfield  {author} {\bibinfo {author} {\bibfnamefont {N.~P.}\ \bibnamefont
  {Mitchell}}, \bibinfo {author} {\bibfnamefont {L.~M.}\ \bibnamefont {Nash}},
  \ and\ \bibinfo {author} {\bibfnamefont {W.~T.}\ \bibnamefont {Irvine}},\
  }\href@noop {} {\bibfield  {journal} {\bibinfo  {journal} {Physical Review
  B}\ }\textbf {\bibinfo {volume} {97}},\ \bibinfo {pages} {100302} (\bibinfo
  {year} {2018}{\natexlab{b}})}\BibitemShut {NoStop}%
\bibitem [{\citenamefont {Engheta}\ \emph {et~al.}(2005)\citenamefont
  {Engheta}, \citenamefont {Salandrino},\ and\ \citenamefont
  {Alu}}]{engheta2005circuit}%
  \BibitemOpen
  \bibfield  {author} {\bibinfo {author} {\bibfnamefont {N.}~\bibnamefont
  {Engheta}}, \bibinfo {author} {\bibfnamefont {A.}~\bibnamefont {Salandrino}},
  \ and\ \bibinfo {author} {\bibfnamefont {A.}~\bibnamefont {Alu}},\
  }\href@noop {} {\bibfield  {journal} {\bibinfo  {journal} {Physical Review
  Letters}\ }\textbf {\bibinfo {volume} {95}},\ \bibinfo {pages} {095504}
  (\bibinfo {year} {2005})}\BibitemShut {NoStop}%
\bibitem [{\citenamefont {Souslov}\ \emph {et~al.}(2019)\citenamefont
  {Souslov}, \citenamefont {Dasbiswas}, \citenamefont {Fruchart}, \citenamefont
  {Vaikuntanathan},\ and\ \citenamefont {Vitelli}}]{souslov2019topological}%
  \BibitemOpen
  \bibfield  {author} {\bibinfo {author} {\bibfnamefont {A.}~\bibnamefont
  {Souslov}}, \bibinfo {author} {\bibfnamefont {K.}~\bibnamefont {Dasbiswas}},
  \bibinfo {author} {\bibfnamefont {M.}~\bibnamefont {Fruchart}}, \bibinfo
  {author} {\bibfnamefont {S.}~\bibnamefont {Vaikuntanathan}}, \ and\ \bibinfo
  {author} {\bibfnamefont {V.}~\bibnamefont {Vitelli}},\ }\href@noop {}
  {\bibfield  {journal} {\bibinfo  {journal} {Physical review letters}\
  }\textbf {\bibinfo {volume} {122}},\ \bibinfo {pages} {128001} (\bibinfo
  {year} {2019})}\BibitemShut {NoStop}%
\bibitem [{\citenamefont {Salbreux}\ and\ \citenamefont
  {J\"ulicher}(2017)}]{PhysRevE.96.032404}%
  \BibitemOpen
  \bibfield  {author} {\bibinfo {author} {\bibfnamefont {G.}~\bibnamefont
  {Salbreux}}\ and\ \bibinfo {author} {\bibfnamefont {F.}~\bibnamefont
  {J\"ulicher}},\ }\href {\doibase 10.1103/PhysRevE.96.032404} {\bibfield
  {journal} {\bibinfo  {journal} {Phys. Rev. E}\ }\textbf {\bibinfo {volume}
  {96}},\ \bibinfo {pages} {032404} (\bibinfo {year} {2017})}\BibitemShut
  {NoStop}%
\bibitem [{\citenamefont {Ghatak}\ \emph {et~al.}(2019)\citenamefont {Ghatak},
  \citenamefont {Brandenbourger}, \citenamefont {van Wezel},\ and\
  \citenamefont {Coulais}}]{ghatak2019observation}%
  \BibitemOpen
  \bibfield  {author} {\bibinfo {author} {\bibfnamefont {A.}~\bibnamefont
  {Ghatak}}, \bibinfo {author} {\bibfnamefont {M.}~\bibnamefont
  {Brandenbourger}}, \bibinfo {author} {\bibfnamefont {J.}~\bibnamefont {van
  Wezel}}, \ and\ \bibinfo {author} {\bibfnamefont {C.}~\bibnamefont
  {Coulais}},\ }\href@noop {} {\bibfield  {journal} {\bibinfo  {journal} {arXiv
  preprint arXiv:1907.11619}\ } (\bibinfo {year} {2019})}\BibitemShut {NoStop}%
\bibitem [{\citenamefont {Schomerus}(2019)}]{schomerus2019nonreciprocal}%
  \BibitemOpen
  \bibfield  {author} {\bibinfo {author} {\bibfnamefont {H.}~\bibnamefont
  {Schomerus}},\ }\href@noop {} {\bibfield  {journal} {\bibinfo  {journal}
  {arXiv preprint arXiv:1908.06312}\ } (\bibinfo {year} {2019})}\BibitemShut
  {NoStop}%
\bibitem [{\citenamefont {Marchetti}\ \emph {et~al.}(2013)\citenamefont
  {Marchetti}, \citenamefont {Joanny}, \citenamefont {Ramaswamy}, \citenamefont
  {Liverpool}, \citenamefont {Prost}, \citenamefont {Rao},\ and\ \citenamefont
  {Simha}}]{RevModPhys.85.1143}%
  \BibitemOpen
  \bibfield  {author} {\bibinfo {author} {\bibfnamefont {M.~C.}\ \bibnamefont
  {Marchetti}}, \bibinfo {author} {\bibfnamefont {J.~F.}\ \bibnamefont
  {Joanny}}, \bibinfo {author} {\bibfnamefont {S.}~\bibnamefont {Ramaswamy}},
  \bibinfo {author} {\bibfnamefont {T.~B.}\ \bibnamefont {Liverpool}}, \bibinfo
  {author} {\bibfnamefont {J.}~\bibnamefont {Prost}}, \bibinfo {author}
  {\bibfnamefont {M.}~\bibnamefont {Rao}}, \ and\ \bibinfo {author}
  {\bibfnamefont {R.~A.}\ \bibnamefont {Simha}},\ }\href {\doibase
  10.1103/RevModPhys.85.1143} {\bibfield  {journal} {\bibinfo  {journal} {Rev.
  Mod. Phys.}\ }\textbf {\bibinfo {volume} {85}},\ \bibinfo {pages} {1143}
  (\bibinfo {year} {2013})}\BibitemShut {NoStop}%
\bibitem [{\citenamefont {Toner}\ \emph {et~al.}(2005)\citenamefont {Toner},
  \citenamefont {Tu},\ and\ \citenamefont
  {Ramaswamy}}]{toner2005hydrodynamics}%
  \BibitemOpen
  \bibfield  {author} {\bibinfo {author} {\bibfnamefont {J.}~\bibnamefont
  {Toner}}, \bibinfo {author} {\bibfnamefont {Y.}~\bibnamefont {Tu}}, \ and\
  \bibinfo {author} {\bibfnamefont {S.}~\bibnamefont {Ramaswamy}},\ }\href@noop
  {} {\bibfield  {journal} {\bibinfo  {journal} {Annals of Physics}\ }\textbf
  {\bibinfo {volume} {318}},\ \bibinfo {pages} {170} (\bibinfo {year}
  {2005})}\BibitemShut {NoStop}%
\bibitem [{\citenamefont {Geyer}\ \emph {et~al.}(2018)\citenamefont {Geyer},
  \citenamefont {Morin},\ and\ \citenamefont {Bartolo}}]{geyer2018sounds}%
  \BibitemOpen
  \bibfield  {author} {\bibinfo {author} {\bibfnamefont {D.}~\bibnamefont
  {Geyer}}, \bibinfo {author} {\bibfnamefont {A.}~\bibnamefont {Morin}}, \ and\
  \bibinfo {author} {\bibfnamefont {D.}~\bibnamefont {Bartolo}},\ }\href@noop
  {} {\bibfield  {journal} {\bibinfo  {journal} {Nature materials}\ }\textbf
  {\bibinfo {volume} {17}},\ \bibinfo {pages} {789} (\bibinfo {year}
  {2018})}\BibitemShut {NoStop}%
\bibitem [{\citenamefont {Petrie}\ and\ \citenamefont
  {Yamada}(2012)}]{petrie2012leading}%
  \BibitemOpen
  \bibfield  {author} {\bibinfo {author} {\bibfnamefont {R.~J.}\ \bibnamefont
  {Petrie}}\ and\ \bibinfo {author} {\bibfnamefont {K.~M.}\ \bibnamefont
  {Yamada}},\ }\href@noop {} {\bibfield  {journal} {\bibinfo  {journal} {J Cell
  Sci}\ }\textbf {\bibinfo {volume} {125}},\ \bibinfo {pages} {5917} (\bibinfo
  {year} {2012})}\BibitemShut {NoStop}%
\bibitem [{\citenamefont {Petrie}\ and\ \citenamefont
  {Yamada}(2016)}]{petrie2016multiple}%
  \BibitemOpen
  \bibfield  {author} {\bibinfo {author} {\bibfnamefont {R.~J.}\ \bibnamefont
  {Petrie}}\ and\ \bibinfo {author} {\bibfnamefont {K.~M.}\ \bibnamefont
  {Yamada}},\ }\href@noop {} {\bibfield  {journal} {\bibinfo  {journal}
  {Current opinion in cell biology}\ }\textbf {\bibinfo {volume} {42}},\
  \bibinfo {pages} {7} (\bibinfo {year} {2016})}\BibitemShut {NoStop}%
\bibitem [{\citenamefont {Benalcazar}\ \emph {et~al.}(2017)\citenamefont
  {Benalcazar}, \citenamefont {Bernevig},\ and\ \citenamefont
  {Hughes}}]{benalcazar2017electric}%
  \BibitemOpen
  \bibfield  {author} {\bibinfo {author} {\bibfnamefont {W.~A.}\ \bibnamefont
  {Benalcazar}}, \bibinfo {author} {\bibfnamefont {B.~A.}\ \bibnamefont
  {Bernevig}}, \ and\ \bibinfo {author} {\bibfnamefont {T.~L.}\ \bibnamefont
  {Hughes}},\ }\href@noop {} {\bibfield  {journal} {\bibinfo  {journal}
  {Physical Review B}\ }\textbf {\bibinfo {volume} {96}},\ \bibinfo {pages}
  {245115} (\bibinfo {year} {2017})}\BibitemShut {NoStop}%
\bibitem [{\citenamefont {Xiao}\ \emph
  {et~al.}(2010{\natexlab{a}})\citenamefont {Xiao}, \citenamefont {Chang},\
  and\ \citenamefont {Niu}}]{RevModPhys.82.1959}%
  \BibitemOpen
  \bibfield  {author} {\bibinfo {author} {\bibfnamefont {D.}~\bibnamefont
  {Xiao}}, \bibinfo {author} {\bibfnamefont {M.-C.}\ \bibnamefont {Chang}}, \
  and\ \bibinfo {author} {\bibfnamefont {Q.}~\bibnamefont {Niu}},\ }\href
  {\doibase 10.1103/RevModPhys.82.1959} {\bibfield  {journal} {\bibinfo
  {journal} {Rev. Mod. Phys.}\ }\textbf {\bibinfo {volume} {82}},\ \bibinfo
  {pages} {1959} (\bibinfo {year} {2010}{\natexlab{a}})}\BibitemShut {NoStop}%
\bibitem [{\citenamefont {K{\"o}pfler}\ \emph {et~al.}(2019)\citenamefont
  {K{\"o}pfler}, \citenamefont {Frenzel}, \citenamefont {Kadic}, \citenamefont
  {Schmalian},\ and\ \citenamefont {Wegener}}]{kopfler2019topologically}%
  \BibitemOpen
  \bibfield  {author} {\bibinfo {author} {\bibfnamefont {J.}~\bibnamefont
  {K{\"o}pfler}}, \bibinfo {author} {\bibfnamefont {T.}~\bibnamefont
  {Frenzel}}, \bibinfo {author} {\bibfnamefont {M.}~\bibnamefont {Kadic}},
  \bibinfo {author} {\bibfnamefont {J.}~\bibnamefont {Schmalian}}, \ and\
  \bibinfo {author} {\bibfnamefont {M.}~\bibnamefont {Wegener}},\ }\href@noop
  {} {\bibfield  {journal} {\bibinfo  {journal} {Physical Review Applied}\
  }\textbf {\bibinfo {volume} {11}},\ \bibinfo {pages} {034059} (\bibinfo
  {year} {2019})}\BibitemShut {NoStop}%
\bibitem [{\citenamefont {Wakao}\ \emph {et~al.}(2019)\citenamefont {Wakao},
  \citenamefont {Yoshida}, \citenamefont {Araki}, \citenamefont {Mizoguchi},\
  and\ \citenamefont {Hatsugai}}]{wakao2019higher}%
  \BibitemOpen
  \bibfield  {author} {\bibinfo {author} {\bibfnamefont {H.}~\bibnamefont
  {Wakao}}, \bibinfo {author} {\bibfnamefont {T.}~\bibnamefont {Yoshida}},
  \bibinfo {author} {\bibfnamefont {H.}~\bibnamefont {Araki}}, \bibinfo
  {author} {\bibfnamefont {T.}~\bibnamefont {Mizoguchi}}, \ and\ \bibinfo
  {author} {\bibfnamefont {Y.}~\bibnamefont {Hatsugai}},\ }\href@noop {}
  {\bibfield  {journal} {\bibinfo  {journal} {arXiv preprint arXiv:1909.02828}\
  } (\bibinfo {year} {2019})}\BibitemShut {NoStop}%
\bibitem [{\citenamefont {Alexandradinata}\ \emph {et~al.}(2014)\citenamefont
  {Alexandradinata}, \citenamefont {Dai},\ and\ \citenamefont
  {Bernevig}}]{PhysRevB.89.155114}%
  \BibitemOpen
  \bibfield  {author} {\bibinfo {author} {\bibfnamefont {A.}~\bibnamefont
  {Alexandradinata}}, \bibinfo {author} {\bibfnamefont {X.}~\bibnamefont
  {Dai}}, \ and\ \bibinfo {author} {\bibfnamefont {B.~A.}\ \bibnamefont
  {Bernevig}},\ }\href {\doibase 10.1103/PhysRevB.89.155114} {\bibfield
  {journal} {\bibinfo  {journal} {Phys. Rev. B}\ }\textbf {\bibinfo {volume}
  {89}},\ \bibinfo {pages} {155114} (\bibinfo {year} {2014})}\BibitemShut
  {NoStop}%
\bibitem [{\citenamefont {Scheibner}\ \emph
  {et~al.}(2020{\natexlab{b}})\citenamefont {Scheibner}, \citenamefont
  {Irvine},\ and\ \citenamefont {Vitelli}}]{scheibner2020non}%
  \BibitemOpen
  \bibfield  {author} {\bibinfo {author} {\bibfnamefont {C.}~\bibnamefont
  {Scheibner}}, \bibinfo {author} {\bibfnamefont {W.}~\bibnamefont {Irvine}}, \
  and\ \bibinfo {author} {\bibfnamefont {V.}~\bibnamefont {Vitelli}},\
  }\href@noop {} {\bibfield  {journal} {\bibinfo  {journal} {arXiv preprint
  arXiv:2001.04969}\ } (\bibinfo {year} {2020}{\natexlab{b}})}\BibitemShut
  {NoStop}%
\bibitem [{\citenamefont {Kato}(1950)}]{kato1950adiabatic}%
  \BibitemOpen
  \bibfield  {author} {\bibinfo {author} {\bibfnamefont {T.}~\bibnamefont
  {Kato}},\ }\href@noop {} {\bibfield  {journal} {\bibinfo  {journal} {Journal
  of the Physical Society of Japan}\ }\textbf {\bibinfo {volume} {5}},\
  \bibinfo {pages} {435} (\bibinfo {year} {1950})}\BibitemShut {NoStop}%
\bibitem [{\citenamefont {Messiah}(1962)}]{messiah1962quantum}%
  \BibitemOpen
  \bibfield  {author} {\bibinfo {author} {\bibfnamefont {A.}~\bibnamefont
  {Messiah}},\ }\href@noop {} {\bibfield  {journal} {\bibinfo  {journal}
  {Appedix C (Section IV)(North-Holland Publishing Company, Amsterdam, 1969)}\
  } (\bibinfo {year} {1962})}\BibitemShut {NoStop}%
\bibitem [{\citenamefont {Xiao}\ \emph
  {et~al.}(2010{\natexlab{b}})\citenamefont {Xiao}, \citenamefont {Chang},\
  and\ \citenamefont {Niu}}]{xiao2010berry}%
  \BibitemOpen
  \bibfield  {author} {\bibinfo {author} {\bibfnamefont {D.}~\bibnamefont
  {Xiao}}, \bibinfo {author} {\bibfnamefont {M.-C.}\ \bibnamefont {Chang}}, \
  and\ \bibinfo {author} {\bibfnamefont {Q.}~\bibnamefont {Niu}},\ }\href@noop
  {} {\bibfield  {journal} {\bibinfo  {journal} {Reviews of modern physics}\
  }\textbf {\bibinfo {volume} {82}},\ \bibinfo {pages} {1959} (\bibinfo {year}
  {2010}{\natexlab{b}})}\BibitemShut {NoStop}%
\end{thebibliography}
\end{document}